\documentclass[12pt,a4paper]{article}
\pdfoutput=1

\usepackage[english]{babel} 
\usepackage[utf8]{inputenc}
\usepackage[left=01in,right=1in,top=1.0in,bottom=1.1in]{geometry}
\usepackage[labelfont=bf]{caption}
\usepackage[hyperfootnotes=false]{hyperref}
\usepackage[normalem]{ulem}
\usepackage{dsfont}
\usepackage{fontawesome5}
\usepackage{amsfonts}
\usepackage{amsmath}
\usepackage{amssymb}
\usepackage{amsthm}
\usepackage{bbold}
\usepackage{calc}
\usepackage{cancel}
\usepackage{float}
\usepackage{slashed}
\usepackage{mathrsfs} 
\usepackage{pdfpages}
\usepackage{multirow}
\usepackage{mathtools}
\usepackage{braket}
\usepackage{cite}
\usepackage{tikz}
\usepackage{tikzsymbols}
\usepackage{caption}
\usepackage{subcaption}
\usepackage{array}
\usepackage{colortbl} 	
\definecolor{Gray}{gray}{0.95}
\definecolor{RGray}{gray}{0.90}
\definecolor{CGray}{gray}{0.92}
\usepackage{arydshln}
\usepackage{soul}
\usepackage{booktabs}

\usepackage[hang,flushmargin]{footmisc}
\usepackage{scalerel}
\usepackage{xcolor}

\definecolor{codegreen}{rgb}{0,0.6,0}
\definecolor{codegray}{rgb}{0.5,0.5,0.5}
\definecolor{codepurple}{rgb}{0.58,0,0.82}
\definecolor{backcolour}{rgb}{0.95,0.95,0.92}
\definecolor{lightgray}{rgb}{0.9,0.9,0.9}
\definecolor{niceblue}{rgb}{0.15,0.15,0.6}
\definecolor{nicegreen}{rgb}{0.1,0.5,0.1}
\definecolor{Red}{rgb}{1.,0.,0.}

\definecolor{Green}{rgb}{0.2,.7,0.2}

\usepackage{listings}
\lstdefinestyle{mystyle}{
    backgroundcolor=\color{backcolour},   
    commentstyle=\color{codegreen},  
    keywordstyle=\color{magenta},
    numberstyle=\tiny\color{codegray},
    stringstyle=\color{codepurple},
    basicstyle=\ttfamily\footnotesize,
    breakatwhitespace=false,         
    breaklines=true,                 
    captionpos=b,                    
    keepspaces=true,                 
    numbers=left,                    
    numbersep=5pt,                  
    showspaces=false,                
    showstringspaces=false,
    showtabs=false,                  
    tabsize=2
}
\hypersetup{
    colorlinks=true,
    linkcolor=niceblue,
    citecolor=cadmiumgreen,
    filecolor=cadmiumgreen,      
    urlcolor=cadmiumgreen,
    linktoc=all
}
\usepackage{lipsum}
\numberwithin{equation}{section}
\numberwithin{figure}{section}
\numberwithin{table}{section}

\renewcommand\thefootnote{\textcolor{cadmiumgreen}{\arabic{footnote}}}

\allowdisplaybreaks

\usetikzlibrary{decorations}

\pgfdeclaredecoration{complete sines}{initial}
{
    \state{initial}[
        width=+0pt,
        next state=sine,
        persistent precomputation={\pgfmathsetmacro\matchinglength{
            \pgfdecoratedinputsegmentlength / int(\pgfdecoratedinputsegmentlength/\pgfdecorationsegmentlength)}
            \setlength{\pgfdecorationsegmentlength}{\matchinglength pt}
        }] {}
    \state{sine}[width=\pgfdecorationsegmentlength]{
        \pgfpathsine{\pgfpoint{0.25\pgfdecorationsegmentlength}{0.5\pgfdecorationsegmentamplitude}}
        \pgfpathcosine{\pgfpoint{0.25\pgfdecorationsegmentlength}{-0.5\pgfdecorationsegmentamplitude}}
        \pgfpathsine{\pgfpoint{0.25\pgfdecorationsegmentlength}{-0.5\pgfdecorationsegmentamplitude}}
        \pgfpathcosine{\pgfpoint{0.25\pgfdecorationsegmentlength}{0.5\pgfdecorationsegmentamplitude}}
}
    \state{final}{}
}

\tikzset{vector/.style={decorate, decoration={complete sines, amplitude=8pt, segment length=10pt}}}

\tikzset{
wc/.style = {circle, fill, minimum size=#1,
              inner sep=0pt, outer sep=0pt},
wc/.default = 6pt % size of the circle diameter 
}

\graphicspath{{./Figures/}{./figures/}}

\newcommand{\be}{\begin{equation}}
\newcommand{\ee}{\end{equation}}
\newcommand{\bea}{\begin{eqnarray}}
\newcommand{\eea}{\end{eqnarray}}  

\newcommand{\gsim}{\lower.7ex\hbox{$\;\stackrel{\textstyle>}{\sim}\;$}}
\newcommand{\lsim}{\lower.7ex\hbox{$\;\stackrel{\textstyle<}{\sim}\;$}}

\newcommand{\cC}{{\mathcal C}}

\definecolor{cadmiumgreen}{rgb}{0.0, 0.42, 0.24}

\definecolor{applegreen}{rgb}{0.55, 0.71, 0.0}

\newcommand\blfootnote[1]{%
  \begingroup
  \renewcommand\thefootnote{}\footnote{#1}%
  \addtocounter{footnote}{-1}%
  \endgroup
}
\makeatletter
\g@addto@macro\bfseries{\boldmath}
\makeatother

\begin{document}
 \begin{flushright}
\,\\% ZU-TH-28/22 \\
 \end{flushright}

\begin{center}
%\vspace{5.7cm}
{\Large\bf Flavor constraints from $pp\to Vh$ and $pp\to VW$ \\[0.3em] at the LHC}
\\[1.0cm]
{\sc O.J.P.~\'Eboli$^{\,a}$\blfootnote{\href{mailto:eboli@if.usp.br}{eboli@if.usp.br}}, 
L.P.S.~Leal$^{\,a}$\blfootnote{\href{mailto:luighi.leal@usp.br}{luighi.leal@usp.br}},
M.~Martines$^{\,a,b}$\blfootnote{\href{mailto:matheus.martines.silva@usp.br}{matheus.martines.silva@usp.br}},
O.~Sumensari    $^{\,b}$\blfootnote{\href{mailto:olcyr.sumensari@ijclab.in2p3.fr}{olcyr.sumensari@ijclab.in2p3.fr}}}
\vspace{0.7cm}

{\em\small ${}^{\ a}$Instituto de Física, Universidade de São Paulo, São Paulo–SP, Brazil}\\[0.2em]
{\em \small${}^{\ b}$IJCLab, P\^ole Th\'eorie (Bât.~210), CNRS/IN2P3 et Univ.~Paris-Saclay, 91405 Orsay, France}\\[0.2em]
\end{center}
\vspace{0.5 cm}

\centerline{\large\bf Abstract}
\begin{quote}
We investigate the potential of associated Higgs $pp\to Vh$ and diboson production $pp\to VW$ channels at the LHC (with $V=W,Z$) to constrain flavor-physics operators. Within a general Effective Field Theory (EFT) framework, we derive the helicity amplitudes for these processes at high energies and identify the leading contributions from dimension-six operators involving different quark flavors. Using available
LHC data, we show that these processes are sensitive to non-trivial flavor structures and provide complementary constraints to those from low-energy observables. We illustrate this synergy through an explicit comparison between our LHC bounds with electroweak precision data, and with flavor limits derived from charged-current pion and kaon decays. In particular, we show that our HL-LHC projections can probe viable EFT scenarios proposed to accommodate the discrepancies in the extraction of the Cabibbo angle.
\end{quote}
\thispagestyle{empty}
\clearpage
\setcounter{page}{1}

 \clearpage

 {\small\tableofcontents}

 \newpage

%%%%%%%%%%%%%%%%%%%%%%%%%%%%%%%%%%%%%%%%%%%%%%%%%%%%%%
%%%%%%%%%%%%%%%%%%%%%%%%%%%%%%%%%%%%%%%%%%%%%%%%%%%%%%
\section{Introduction}

The discovery of the Higgs boson at the LHC has made the Standard Model (SM) of particle physics complete~\cite{CMS:2012qbp}. However, several fundamental questions remain unanswered within the SM, such as the flavor and hierarchy problems, which motivate a vast experimental program to explore the effects of physics beyond the SM (BSM). The absence of heavy resonances in current direct searches at the LHC indicates a mass gap separating the electroweak scale from the scale of New Physics. Therefore, Effective Field Theories (EFTs) have become the most suitable approach for describing BSM effects in modern-day experiments. Within this framework, the most relevant BSM signals at the LHC are characterized by deviations in the tails of kinematical distributions.

The study of diboson production, $pp\to WV$, and associated Higgs production with a vector boson, $pp\to Vh$, with $V=W,Z$, offers powerful probes of EFTs at the LHC. These processes have received attention over the past decades, as they directly probe the interactions of Goldstone bosons at high energies via the longitudinal polarizations of electroweak bosons~\cite{Hagiwara:1986vm,Hagiwara:1993ck,Degrande:2012wf,Corbett:2012dm,Greljo:2015sla,Ethier:2021ydt,Butter:2016cvz,Falkowski:2015jaa,Baglio:2017bfe,Corbett:2017qgl,Franceschini:2017xkh,Banerjee:2018bio,Grojean:2018dqj,Biekotter:2014gup,Liu:2018pkg,deBlas:2025xhe}. The contributions from specific dimension-six operators to the $WV$ and $Vh$ production cross-sections are expected to grow with energy at parton level, with their amplitudes related by the Goldstone equivalence theorem~\cite{Cornwall:1974km}. Interestingly, this energy enhancement allows us to derive LHC constraints that are competitive and that can even supersede the LEP-I and LEP-II bounds for certain operators~\cite{Grojean:2018dqj, Franceschini:2017xkh, Liu:2018pkg, Baglio:2017bfe, Falkowski:2015jaa, Banerjee:2018bio, Biekotter:2014gup, Butter:2016cvz,deBlas:2025xhe}.

Previous studies of the $pp\to WV$ and $pp\to Vh$ processes at the LHC were based on very restrictive flavor assumptions to reduce the number of Wilson coefficients~\cite{Grojean:2018dqj,Biekotter:2014gup}, such as imposing the full $U(3)^5$ flavor symmetry or Minimal Flavor Violation (MFV)~\cite{DAmbrosio:2002vsn}. With these assumptions, the $WV$ and $Vh$ productions are driven by the contributions from operators with light quarks. While these contributions are dominant in a flavor-blind scenario due to the sizable valence quark Particle Distribution Functions (PDFs), the current and future precisions of the LHC and HL-LHC also open the possibility to probe contributions from heavy quarks. The latter contributions can be dominant for models with non-universal couplings to quarks, as motivated by several BSM constructions aiming to address the SM and New Physics flavor problems~\cite{Barbieri:2011ci,Faroughy:2020ina}.

In this paper, we use the available LHC data on the diboson and associated Higgs production to constrain the Standard Model Effective Field Theory (SMEFT) Wilson coefficients with a general flavor structure~\cite{Buchmuller:1985jz,Grzadkowski:2010es}. Most importantly, we do not impose a specific flavor ansatz on the operator basis, obtaining results that apply not only to the usually considered flavor-universal scenario but also to models with a non-trivial flavor structure. This analysis will complement the ongoing effort of building a general SMEFT likelihood of high-energy observables that could be combined with low-energy flavor constraints in full generality. The first steps of this program have been performed for Drell-Yan processes at high-$p_T$ in Ref.~\cite{Allwicher:2022gkm,Allwicher:2022mcg}, see also~Ref.~\cite{deBlas:2025xhe,Grunwald:2023nli}.

\begin{figure*}[!t]
    \includegraphics[width=1\textwidth]{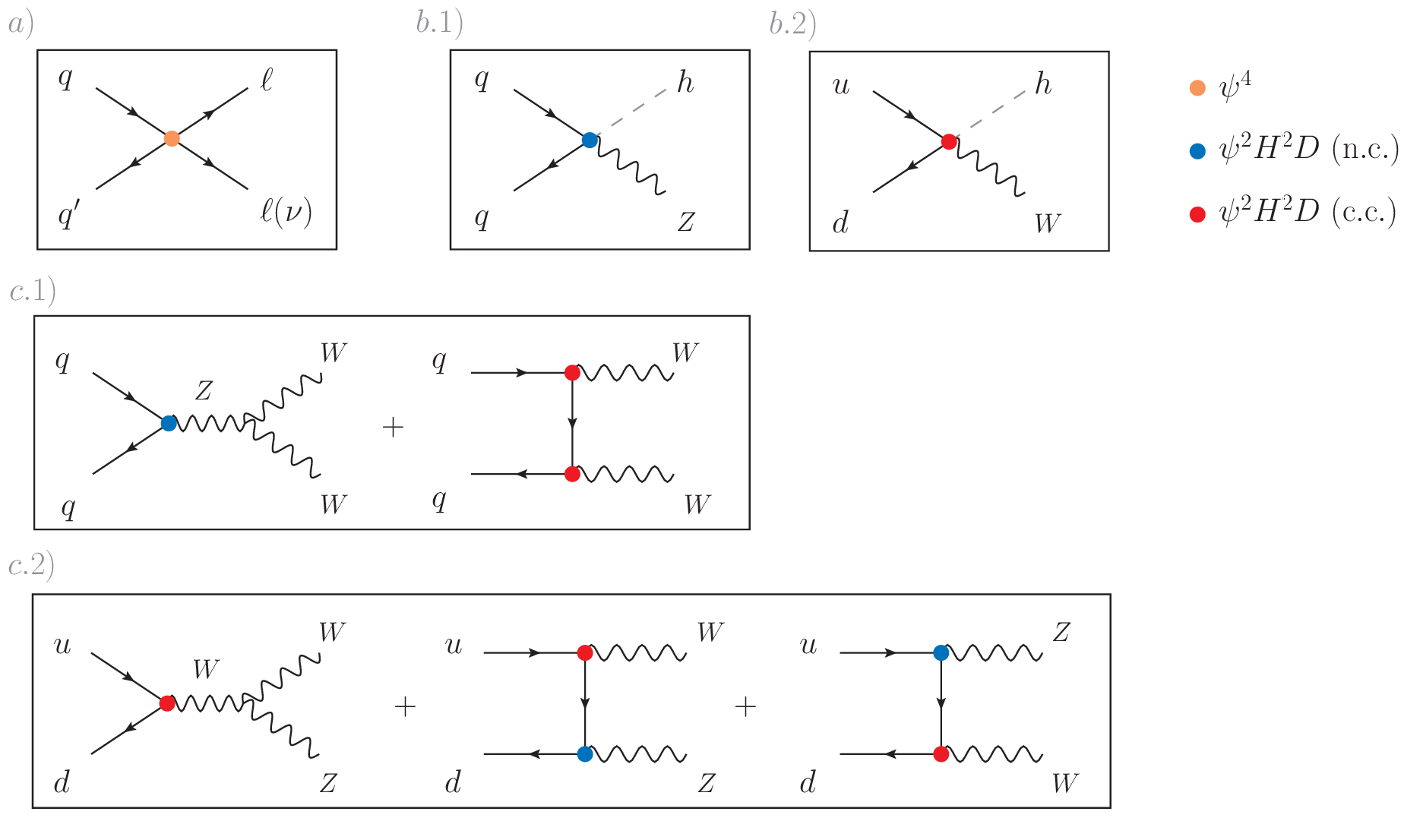}
    \caption{\small \sl Illustration of energy-enhanced EFTcontributions to a) $pp\to\ell \ell, \ell\nu$ b) $pp\to Vh$ and c) $pp\to VW$, with $V=W,Z$, from flavor-dependent operators. The insertions of $\psi^4$ semileptonic operators are denoted in orange, and those from Higgs-current $\psi^2 H^2 D$ operators to quarks in blue and red for neutral- and charged-current interactions, respectively. See Table~\ref{tab:SMEFT-ope-energy} and details in the main text.}
    \label{fig:lhc-illustration}
\end{figure*}

A subtle feature of the $pp\to WW$ process is that the unitarity of the CKM matrix ($V\equiv V_{\mathrm{CKM}}$) plays an essential role in guaranteeing that this process respects unitarity when different initial quark flavors are
considered. The leading contributions to the $q_i \bar{q}_j\to W W$ amplitude within the SM are the photon and  $Z$-boson exchanges in the $s$- and the $t$-channels, respectively. The latter are proportional to $(V\cdot V^\dagger)_{ij}$ at high energies, as depicted in the center panel of ~Fig.~\ref{fig:lhc-illustration}. While these individual contributions grow with the center-of-mass energy, their sum is well behaved at high energies provided that $V$ is unitary. Therefore, departures from the CKM unitarity in low-energy data should, in principle, manifest themselves in the kinematical tails of these processes~\cite{Gabrielli:2024bjw}. Such effects must be consistently parameterized by the relevant dimension-six operators~\cite{Buchmuller:1985jz}. The gauge-invariant EFT operators exhibit correlations between the modifications of the $W$- and $Z$-couplings to quarks with contact terms of the type $q\bar{q} hV$, which contribute to the $Vh$ production cross section. In fact, the latter processes provide the most stringent constraints on these operators due to the smaller SM background at high energies, as will be discussed in Sec.~\ref{sec:lhc}.

To illustrate the relevance of the constraints derived in this study, and based on the above discussion, we will consider the tests of the first row CKM unitarity as an example. The current determinations of the $|V_{ud}|$ and $|V_{us}|$ matrix elements from $\beta$-decay, and leptonic and semileptonic kaon decays show an apparent departure from CKM unitarity, with $\Delta_{\mathrm{CKM}}\equiv |V_{ud}|^2+|V_{us}|^2+|V_{ub}|^2-1$ differing from zero at the $\approx 3\sigma$ level; see Ref.~\cite{Cirigliano:2022yyo} and references therein. Although the significance of these discrepancies is relatively mild and improvements are still required -- particularly in fully addressing the hadronic uncertainties of these processes at the percent level -- they provide an excellent example for demonstrating that data from the LHC on Higgs and electroweak interactions can be competitive with flavor observables for charged-current transitions. We should also mention that the processes $pp\to Vh$ were considered for a similar purpose before~\cite{Alioli:2017ces}, setting useful constraints on right-handed currents with early LHC data on the $Wh$ signal strength. We will extend these results to both neutral and charged channels, using the latest LHC data that is nowadays also given at the differential level~\cite{CMS:2023vzh
,ATLAS:2024yzu}. Importantly, this will allow us to improve the previous bounds by roughly an order of magnitude due to the sensitivity of the differential distribution to the energy growth of the $Vh$ amplitude.

%%%%%%%%%%%%%
%%%%%%%%%%%%%

The remainder of this paper is organized as follows. In Sec.~\ref{sec:eft}, we set our notation and formulate the EFT Lagrangian used in our phenomenological analysis. In Sec.~\ref{sec:amp-energy}, we study the high-energy behaviour of the helicity amplitudes for the $Vh$ and $VW$ production channels for a general flavor structure. 
In Sec.~\ref{sec:lhc}, we present our numerical results using current LHC data. In Sec.~\ref{sec:EWPO}, we compare our collider constraints with those derived from electroweak precision observables at LEP for operators with different quark flavors. Finally, we illlustrate the relevance of our results by applying these LHC constraints to an explicit BSM example in Sec.~\ref{sec:illustration} which is motivated by the discrepancies in the extraction of the Cabibbo angle. We summarize our findings in Sec.~\ref{sec:conclusion}.

%%%%%%%%%%%%%%%%%%%%%%%%%%%%%%%%%%%%%%%%%%%%%%%%%%%%%%
%%%%%%%%%%%%%%%%%%%%%%%%%%%%%%%%%%%%%%%%%%%%%%%%%%%%%%
\section{Theoretical framework}
\label{sec:eft}

We assume that the scale of New Physics lies well above the electroweak scale, allowing us to use the Standard Model Effective Field Theory (SMEFT) to describe flavor and LHC data~\cite{Buchmuller:1985jz,Grzadkowski:2010es},
%%%%%%%%%%%%%%%%
\begin{equation}
    \label{eq:smeft}
        \mathcal{L}_{\mathrm{SMEFT}}^{(d=6)} \supset \sum_I \dfrac{\mathcal{C}_I}{\Lambda^2}\,\mathcal{O}_I \,,
\end{equation}
%%%%%%%%%%%%%%%%
where $\mathcal{O}_I$ are dimension-six operators invariant under the $SU(3)_c\times SU(2)_L \times U(1)_Y$ gauge symmetry, $\mathcal{C}_I$ are the corresponding Wilson coefficients and $\Lambda$ is the EFT cutoff. Our convention for the SMEFT operators and flavor indices follows Ref.~\cite{Allwicher:2022gkm}, with the assumption that down-type quark Yukawas are diagonal, {i.e.},~the quark weak doublet reads $q_i=[(V^\dagger u)_{Li}\,,d_{Li}]^T$ with a flavor index $i$, where we remind that $V$ denotes the CKM matrix. 

In the electroweak sector, we consider the $\lbrace \alpha_{\mathrm{em}},\,G_F,\,m_Z\rbrace$ input parameter scheme  and we express the $W$- and $Z$-couplings to quarks, for convenience, as follows~\footnote{Our convention for the covariant derivative acting on a generic field $\eta$ is $D_\mu \eta = (\partial_\mu +i g_3 G_\mu^A T^A+i g_2 W_\mu^I t^I+i g_1 B_\mu y) \,\eta $, 
where $g_3$, $g_2$ and $g_1$ are the gauge couplings of $SU(3)_c$, $SU(2)_L$ and $U(1)_Y$, the corresponding generators are labeled $T^A$, $t^I$ and $y$, and the associated gauge fields are $G_\mu^A$, $W_\mu^I$ and $B_\mu$, respectively.}
%%%%%%%%%%%%%
\begin{align}
    \label{eq:Leff-EW}
    \mathcal{L}_{V\bar{q}q} \supset &-\dfrac{g_2}{c_W} Z^\mu \sum_{\psi \in \lbrace u,d \rbrace} \sum_{ij} \bar{\psi}_i \gamma_\mu \Big{(} g^{Z\psi}_{\substack{L\\ij}} P_L +g^{Z\psi}_{\substack{R\\ij}} P_R  \Big{)} \psi_j\\
    &-\dfrac{g_2}{\sqrt{2}}\sum_{ij}\bigg{[}W_\mu^+ \,\bar{u}_{i}\gamma^\mu \Big{(} g^{Wq}_{\substack{L\\ij}} P_L +g^{Wq}_{\substack{R\\ij}} P_R  \Big{)} d_{j}+\mathrm{h.c.}\bigg{]}\,, \nonumber
\end{align}
%%%%%%%%%%%%

\noindent where
%%%%%%%%%%%%%
\begin{align}
    g^{Wq}_{\substack{L\\ij}}&\equiv V_{ij}+\delta g^{Wq}_{\substack{L\\ij}}\,, \qquad  \quad g^{Wq}_{\substack{R\\ij}} \equiv  \delta g^{Wq}_{\substack{R\\ij}}\,,\qquad \quad 
    g^{Z\psi}_{\substack{X\\ij}} \equiv  g^{Z\psi}_X\,\delta_{ij}+\delta g^{Z\psi}_{\substack{X\\ij}}\,,
\end{align}
%%%%%%%%%%%%

\noindent with $g_L^{Z\psi}=T^3_\psi-Q_\psi\, s^2_W$ and $g_R^{Z\psi}=-Q_\psi\, s^2_W$, where $Q_\psi$ and $T^3_\psi$ are the electric charge and the third-component of the weak isospin of $\psi$. The shifts of the SM couplings are induced by dimension-six operators (see discussion below), which respect the following relation due to gauge invariance~\cite{Efrati:2015eaa}, 
%%%%%%%%%%%%%
\begin{align}
    \label{eq:SU2-relation}
    \delta g_{L}^{W q} = \delta g_{L}^{Z u} \cdot V - 
 V\cdot \delta g_{L}^{Z d}\,,
\end{align}
%%%%%%%%%%%%
which is written as a matrix multiplication in flavor space. We have not included shifts of leptonic couplings in the above expressions, nor shifts of the $W$-boson mass, as they are irrelevant for the processes we consider and are tightly constrained by electroweak data~\cite{Efrati:2015eaa}. In principle, there should also be dipole interactions of the electroweak bosons with quarks in Eq.~\eqref{eq:Leff-EW}, however these are loop suppressed in concrete models of New Physics. Therefore, we neglect dipole operators in our analysis.

%%%%%%%%%%%%%%%%%
\begin{table}[!t]
\renewcommand{\arraystretch}{1.8}
\centering
\begin{tabular}{c||ccc|c}
\multirow{2}{*}{Operator} & \multicolumn{3}{c|}{Amplitude scaling} &   \multirow{2}{*}{Parameters} \\ \cline{2-4}
& Drell-Yan & $pp\to VW$ & $pp\to Vh$ & \\ \hline\hline
$\psi^4$ & $E^2/\Lambda^2$ & -- & -- & 456 (399) \\ 
$\psi^2 H^2 D$ & $v^2/\Lambda^2$ & $E^2/\Lambda^2$ & $E^2/\Lambda^2$ & 27 (16) \\
$\psi^2 X H$ & $v E/\Lambda^2$ & $E^2/\Lambda^2$ & $v^2/\Lambda^2$ &  30 (30)\\
\end{tabular}
\vspace{0.2cm}
\caption{\small \sl Energy scaling for $\sqrt{\hat{s}}\gg v$, at amplitude level, for Drell-Yan processes, diboson production and associated Higgs production for the SMEFT operators relevant for low-energy flavor processes in the quark sector. The number of CP-even (CP-odd) operators for each class of operators contributing at tree level to the mentioned LHC processes is given in the last column~\cite{Faroughy:2020ina}. For completeness, we also display the dipole operators ($\psi^2 X H$), which are not included in our numerical analysis, as they are generated at loop-level in weakly coupled UV scenarios.}
\label{tab:SMEFT-ope-energy} 
\end{table}
%%%%%%%%%%%%%%%%%

\subsection{Operator basis}

In this study, we are interested in the operators that contribute at \emph{tree level} to \emph{quark-flavor transitions} and that could also lead to significant contributions at the tails of the kinematical distributions of LHC observables. We will restrict our analysis to operators in the Warsaw basis~\cite{Grzadkowski:2010es} that can be generated at \emph{tree level}~\cite{Arzt:1994gp} by \emph{weakly coupled} and \emph{renormalizable ultraviolet} (UV) scenarios~\cite{deBlas:2017xtg}. This restriction applies, for instance, to dipole operators ($\psi^2 X H$), which can in principle lead to energy-enhanced effects in the observables discussed above, cf.~Table~\ref{tab:SMEFT-ope-energy}, but with loop-suppressed Wilson coefficients once a specific UV completion is imposed. The same conclusion applies to operators generating anomalous Triple Gauge Couplings (TGCs) in the Warsaw basis ($X^3$)~\cite{Grzadkowski:2010es}, which would be loop suppressed, thus being disregarded in our analysis~\cite{Arzt:1994gp}.~\footnote{See Ref.~\cite{Bobeth:2015zqa} for a study of flavor-changing neutral-current constraints on TGCs in the HISZ basis~\cite{Hagiwara:1986vm}.}

Given the above assumptions, the only relevant SMEFT operators in our analyses are the four-fermion operators ($\psi^4$) and the Higgs current ones ($\psi^2 H^2 D$). These operators can induce energy-enhanced effects in Drell-Yan processes, and in the $pp\to VW$ and $pp \to Vh$ channels, respectively, as illustrated in Fig.~\ref{fig:lhc-illustration}. The energy scaling at the amplitude level of the relevant operators is summarized in Table~\ref{tab:SMEFT-ope-energy}. While Drell-Yan processes and their connection to flavor observables have been extensively studied,  see Ref.~\cite{Allwicher:2022gkm,Allwicher:2022mcg} and references therein, similar studies have not yet been performed for the electroweak and Higgs-related processes discussed in this paper.~\footnote{Notice that four-quark operators also lead to LHC signatures such a dijet processes, with a potential connection to flavor observables, see e.g.~Ref.~\cite{Bordone:2021cca}. However, the main obstacles to combining low- and high-energy searches for these operators are the large hadronic uncertainties affecting non-leptonic meson decays.}

%%%%%%%%%%%%%%%%%
%%%%%%%%%%%%%%%%%
\begin{table}[!t]
    \renewcommand{\arraystretch}{1.9}
    \centering
        \centering
        \begin{tabular}{c|c}
        $\mathcal{O}_{\psi^2 H^2 D}$ &   Operator\\\hline\hline
        $\mathcal{O}_{Hq}^{(1)}$ & $\big{(}H^\dagger i {\overleftrightarrow{D}}_\mu  H\big{)}\big{(}\bar{q}_i \gamma^\mu  q_j\big{)}$ \\
        $\mathcal{O}_{Hq}^{(3)}$ & $\big{(}H^\dagger i {\overleftrightarrow{D}}_\mu^I  H\big{)}\big{(}\bar{q}_i \gamma^\mu \tau^I q_j\big{)}$\\
        $\mathcal{O}_{Hu}$ & $\big{(}H^\dagger i {\overleftrightarrow{D}}_\mu  H\big{)}\big{(}\bar{u}_i \gamma^\mu  u_j\big{)}$ \\
        $\mathcal{O}_{Hd}$ & $\big{(}H^\dagger i {\overleftrightarrow{D}}_\mu  H\big{)}\big{(}\bar{d}_i \gamma^\mu  d_j\big{)}$ \\
        $\mathcal{O}_{Hud}$ & $\big{(}\widetilde{H}^\dagger i {{D}}_\mu  H\big{)}\big{(}\bar{u}_i \gamma^\mu  d_j\big{)}+\mathrm{h.c.}$
        \end{tabular}
    \caption{\small \sl 
   Dimension-six Higgs-current operators ($\psi^2 H^2D$) with quark fields in the Warsaw basis~\cite{Grzadkowski:2010es}. Quark doublets are denoted by $q$, and quark singlets by $u$ and $d$,  with flavor indices represented by Latin letters. The conjugated Higgs doublet is defined as $\widetilde{H} = i \tau_2 H^\ast$. The Pauli matrices are denoted by~$\tau^I$ with ${I\in\{1,2,3\}}$, and we use the shorthand notation $H^\dagger i {\overleftrightarrow{D}}_\mu  H = i H^\dagger (D_\mu H)- i(D_\mu H^\dagger)H$ and $\smash{H^\dagger i {\overleftrightarrow{D}}_\mu^I  H = i H^\dagger \tau^I (D_\mu H)- i(D_\mu H^\dagger)\tau^I H}$.
    }
    \label{tab:SMEFT-Higgs-ope} 
\end{table}

The main operators that we consider in our analysis are the Higgs-current ones, which are defined in Table~\ref{tab:SMEFT-Higgs-ope} and which enter low-energy observables through shifts of the $W$- and $Z$-couplings to fermions, as defined in Eq.~\eqref{eq:Leff-EW}. These operators contribute not only to $pp\to VW$ at high energies (cf.~Fig.~\ref{fig:lhc-illustration}), but also to $pp\to Vh$ through contact interactions $q_i\bar{q}_j h V$ that are induced by the same gauge-invariant operators. For instance, the charged-current right-handed interactions are written in the unitary gauge,
%%%%%%%%%%%%%
\begin{align}
\mathcal{L}_{\mathrm{SMEFT}}^{(d=6)} &\supset  \dfrac{1}{\Lambda^2}\mathcal{C}_{\substack{Hud\\ij}}\sum_{ij}\big{(}\widetilde{H}^\dagger i {{D}}_\mu  H\big{)}\big{(}\bar{u}_i \gamma^\mu  d_j\big{)} +\mathrm{h.c.} \\*
&\to-\dfrac{g_2}{2\sqrt{2}}W_{\mu}^+\dfrac{1}{\Lambda^2}\sum_{ij}\mathcal{C}_{\substack{Hud\\ij}} \,  \big{(}\bar{u}_{Ri}\gamma^\mu d_{Rj}\big{)}\, \bigg{(}1+\dfrac{h}{v}\bigg{)}^2  + \,\dots\, +\mathrm{h.c.}\,,\nonumber
\end{align}
%%%%%%%%%%%%%

\noindent which affects the Higgs boson production in association with the $W$-boson, in addition to the induced $W$-boson couplings to right-handed quarks. Similar relations can be obtained for the neutral currents and for the left-handed charged currents with the matching between the SMEFT and Eq.~\eqref{eq:Leff-EW} at dimension six given by
%%%%%%%%%%%%%
\begin{align}
    \label{eq:Vqq-matching}
    \delta g_{\substack{L\\ij}}^{Wq} &= \dfrac{v^2}{\Lambda^2} \sum_{a} V_{ia} \, \mathcal{C}^{(3)}_{\substack{Hq\\aj}}\,, & \delta g_{\substack{R\\ij}}^{Wq} &= \dfrac{v^2}{2\Lambda^2}\mathcal{C}_{\substack{Hud\\ij}}\,,\\*
     \delta g_{\substack{L\\ij}}^{Zd} &= - \dfrac{v^2}{2\Lambda^2} \mathcal{C}_{\substack{Hq\\ij}}^{(1+3)}\,, & \delta g_{\substack{R\\ij}}^{Zd} &= - \dfrac{v^2}{2\Lambda^2} \mathcal{C}_{\substack{Hd\\ij}}\,, \nonumber \\*
    \delta g_{\substack{L \\ ij}}^{Zu} &= - \dfrac{v^2}{2\Lambda^2} \sum_{ab} V_{ia} \,\mathcal{C}_{\substack{Hq\\ab}}^{(1-3)}\, V_{jb}^\ast\,, & \delta g_{\substack{R \\ ij}}^{Zu} &= - \dfrac{v^2}{2\Lambda^2} \mathcal{C}_{\substack{Hu\\ij}}\,, \nonumber 
\end{align}
%%%%%%%%%%%%%
where we use the shorthand notation $\mathcal{C}_{Hq}^{(1\pm 3)}\equiv \mathcal{C}_{Hq}^{(1)}\pm \mathcal{C}_{Hq}^{(3)}$\,. From the above equation, it is straightforward to derive the $SU(2)_L$ relation in Eq.~\eqref{eq:SU2-relation}. 

In the following, we will only assume that the EFT provides a valid description of the LHC processes that we consider, without committing to a specific ultraviolet model. In other words, we require that the EFT cutoff $\Lambda$ is sufficiently larger than the typical energy scale $E$ of the LHC processes, i.e. $\Lambda\gg E$.

%%%%%%%%%%%%%%%%%%%%%%%%%%%%%%%%%%%%%%%%%%%%%%%%%%%%%%
%%%%%%%%%%%%%%%%%%%%%%%%%%%%%%%%%%%%%%%%%%%%%%%%%%%%%%
\section{Helicity amplitudes at high energies}
\label{sec:helicity}

In this Section, we present the analytic expressions for the dominant helicity amplitudes for the $Vh$ and $VW$ production channels at high energies. For illustration, we start our discussion with the $WW$ channel, which receives tree-level contributions in the SM from the $\gamma$- and $Z$-exchange in the $s$-channel, as well as from the $t$-channel diagram with two insertions of the CKM matrix, cf.~center panel in Fig.~\ref{fig:lhc-illustration}. While each of these diagrams induces energy-enhanced contributions for amplitudes with longitudinally polarized final states, 
their sum vanishes identically within the SM~\cite{Hagiwara:1986vm, Grojean:2018dqj}, 
\label{sec:amp-energy}
%%%%%%%%%%%%%
\begin{align}
\mathcal{M}(q_i \bar{q}_j&\to W_0W_0) \overset{\mathrm{SM}}{=} i \hat{s} \dfrac{e^2 \sin\theta}{2 m_W^2} \bigg{[} Q_q \,\delta_{ij}+\dfrac{1}{s^2_W}(T_q^3-s^2_W Q_q)\,\delta_{ij}-\dfrac{T_q^3}{s_W^2}\Delta_{ij}\bigg{]}+\mathcal{O}(\hat{s}^0)\,,
\label{eq:WWproductionSM}
\end{align}
%%%%%%%%%%%%%

\noindent where the angle $\theta$ is defined between $W^+$ and the beam axis in the center-of-mass frame, $\sqrt{\hat{s}}$ denotes the partonic center-of-mass energy, and quark masses have been neglected. We defined $\Delta\equiv V\cdot V^\dagger$ for $q=d$ and $\Delta \equiv V^\dagger\cdot V$ for $q=u$. Notice that $\Delta_{ij}=\delta_{ij}$ from the CKM matrix unitarity, ensuring that the total cross section
does not violate unitarity in the SM. Therefore, in the presence of modifications of $Z$- and $W$-boson couplings to fermions as we consider, or corrections to the TGCs, the SM cancellation can be disrupted. This would lead to energy-enhanced effects in the tails of kinematic distributions, as summarized in Table~\ref{tab:SMEFT-ope-energy}. The same conclusion applies to the $WZ$ channel, but with the difference that the CKM matrix only appears as an overall factor in this case.~\footnote{The energy-enhancement induced by these operators can be better understood in terms of the Goldstone equivalence theorem, as will be discussed below.}

In the presence of $\psi^2 H^2 D$ effective operators, 
the amplitude for the process with longitudinally polarized vector bosons, $u_i \overline{u}_j \rightarrow W_0 W_0$, receives energy-enhanced contribution due to  SMEFT corrections to the $Z$ coupling to left-handed down quarks,
%%%%%%%%%%%%%%%%
\begin{equation}
    \mathcal{M}(u_{Li} \bar{u}_{Lj}\to  W_0 W_0)= \dfrac{2i\, \hat{s} \,\sin\theta}{v^2}\, \, \sum_{ab}\,V^*_{ia}\,\,\delta g^{Zd *}_{\substack{L\\ab}} \,\,V_{jb} + \mathcal{O}(\hat{s}^0)\, \,,
    \label{eq:WWSMEFT}
\end{equation}
%%%%%%%%%%%%%%%
where we used Eq.~\eqref{eq:SU2-relation}.
As mentioned in Sec.~\ref{sec:eft}, the 
corrections to the $Z$- and $W$-boson
couplings to quark originate from effective operators that respect the $SU(2)_L \times U(1)_Y$ 
gauge invariance, which can also be probed in processes where a Higgs 
boson is produced in association with an electroweak boson, 
vide second panel in Fig.~\ref{fig:lhc-illustration}. In particular, the process 
$d_i \bar{d}_j \to Z h$ is sensitive to the same corrections to the $Z$ 
coupling that contribute to Eq.~\eqref{eq:WWSMEFT},
\begin{align}
\label{eq:amp-Zh}
& \mathcal{M}(d_{Li} \bar{d}_{Lj} \to Z_0 h )=-\dfrac{2i\, \hat{s} \,\sin\theta}{v^2}\,\delta g_{\substack{L \\ ij}}^{Zd\,\ast}+\mathcal{O}(\hat{s}^0)\,.
\end{align}

\noindent As pointed out, for instance, in Ref. \cite{Franceschini:2017xkh}, the fact that both diboson and
Higgs-associated production processes probe the same couplings, as illustrated 
in Eqs.\eqref{eq:WWSMEFT} and \eqref{eq:amp-Zh}, is a consequence of the Goldstone 
Boson Equivalence Theorem~\cite{Cornwall:1974km, Lee:1977eg, Gounaris:1986cr, Chanowitz:1987vj, Wulzer:2013mza}.
This theorem states that, in the high-energy limit ($\sqrt{\hat{s}} \gg m_W$), amplitudes with
external longitudinally polarized gauge bosons can be approximated by those where 
the gauge bosons are replaced by their corresponding Goldstone bosons, up to $\mathcal{O}(m_W / \sqrt{\hat{s}})$ corrections.
As a result, the high-energy behavior of the $pp \rightarrow Vh$ and $pp \rightarrow WV$ 
processes with longitudinal gauge bosons can be traced back to interactions among fermions, the Higgs and Goldstone bosons. For the effective operators 
listed in Table~\ref{tab:SMEFT-Higgs-ope}, these interactions
correspond to four-point contact terms. Therefore, these processes are related by the $SU(2)_L$ gauge symmetry at high energies and the energy growth of the cross section can be easily understood from the contact interactions.

\subsection{General expressions} We provide below the general expressions for the dominant helicity amplitudes of the $pp \rightarrow Vh$ and $pp \rightarrow WV$ processes at high energies. Using Eq.~\eqref{eq:SU2-relation}, we find that the $WW$ and $Zh$ production amplitudes depend on the effective couplings $\delta g_L^{Zd}$, $\delta g_L^{Zu}$,  $\delta g_R^{Zd}$ and $\delta g_R^{Zu}$, as given below, 
%%%%%%%%%%%%%
\begin{align}
    \label{eq:amp-Zh1}
    {\cal M}(u_{Li} \overline{u}_{Lj}\rightarrow Z_0 h) 
    &\simeq  -\, \sum_{ab}\, V^*_{ia} \, \mathcal{M}(d_{La} \overline{d}_{Lb} \rightarrow W_0 W_0) \, V_{jb} 
   \simeq -\dfrac{2i\, \hat{s} \,\sin\theta}{v^2} \,\, \delta g^{Zu *}_{\substack{L\\ij}}\,, \\[0.4em]
    {\cal M}(d_{Li} \overline{d}_{Lj}\rightarrow Z_0 h) 
    &\simeq  -\, \sum_{ab}\, V_{ai} \, \mathcal{M}(u_{La} \overline{u}_{Lb} \rightarrow W_0 W_0) \, V_{bj}^{*}
    \simeq  -\dfrac{2i\, \hat{s} \,\sin\theta}{v^2}\, \delta g^{Zd *}_{\substack{L\\ij}}\,, \nonumber \\[0.4em]
    {\cal M}(f_{Ri} \overline{f}_{Rj}\rightarrow Z_0 h) 
    &\simeq  - \, \mathcal{M}(f_{Ri} \overline{f}_{Rj} \rightarrow W_0 W_0)
    \simeq - \dfrac{2i\, \hat{s} \,\sin\theta}{v^2}\, \delta g^{Zf *}_{\substack{R\\ij}}\,. \nonumber
\end{align}
%%%%%%%%%%%%
where $f=u,d$. The above results are valid up to contributions from dimension-six operators and up to $\mathcal{O}(v/\sqrt{\hat{s}})$ corrections. Similarly, the $WZ$ and $Wh$ production
channels are sensitive to $\delta g_L^{Wq}$ and $\delta g_R^{Wq}$,
%%%%%%%%%%%%
\begin{align}
    \mathcal{M}(u_{Li} \overline{d}_{Lj} \rightarrow W_0 h ) 
    &\simeq - \, \mathcal{M}(u_{Li} \overline{d}_{Lj} \rightarrow W_0 Z_0 ) 
    \simeq -\dfrac{i\sqrt{2}\, \hat{s} \,\sin\theta}{v^2} \, \delta g^{Wq*}_{\substack{L\\ij}}\,,
   \label{eq:amp-Wh1}
   \\[0.4em]
    \mathcal{M}(u_{Ri} \overline{d}_{Rj} \rightarrow W_0 h ) 
    &\simeq + \, \mathcal{M}(u_{Ri} \overline{d}_{Rj} \rightarrow W_0 Z_0 ) 
    \simeq -\dfrac{i\sqrt{2}\, \hat{s} \,\sin\theta}{v^2} \, \delta g^{Wq*}_{\substack{R\\ij}}\,. \nonumber
    \label{eq:amp-Wh2}
\end{align}
%%%%%%%%%%%%
The main difference between our results and those in the 
literature~\cite{Grojean:2018dqj, Franceschini:2017xkh, Corbett:2017qgl} 
is the explicit inclusion of CKM matrix elements in the above expressions. Given the constraint from gauge invariance
in Eq.~\eqref{eq:SU2-relation}, the high-energy amplitudes can be described in terms of five different types of effective couplings. 
These are the modifications to the $Z$-boson couplings to quarks, $\delta g^{Zq}_L$ and $\delta g^{Zq}_R$ (with $q=u,d$), and the coupling that controls right-handed charged currents, $\delta g_R^{Wq}$. 
The former parameterizes the contributions that can interfere with the SM contributions and are
denoted as high-energy primaries in Ref.~\cite{Franceschini:2017xkh}. The coupling $\delta g^{Wq}_R$ is often neglected in collider studies, either because it does not interfere with the SM, or because it is suppressed under more restrictive flavor assumptions such as MFV. However, these couplings can arise in concrete scenarios of New Physics, as will be discussed, for instance, in~Sec.~\ref{sec:illustration}.
For this reason, we retain these contributions in the following discussion, in particular, exploring the complementarity of LHC and flavor limits on right-handed charged currents.

Finally, we note that there is an asymptotic $SU(2)_L$ relation between the $WZ$ and $WW$ process amplitudes for left-handed quarks,
%%%%%%%%%%%%%
\begin{align}
\mathcal{M}(u_{Li}\bar{d}_{Li} \to W_0 Z_0) = \dfrac{1}{\sqrt{2}} \sum_{a} \Big{[} V_{ia}^*\,&\mathcal{M}(d_{La} \bar{d}_{Lj} \to W_0 W_0) \\*
- &\mathcal{M}(u_{Li} \bar{u}_{La} \to W_0 W_0)\,V_{aj}^*\,\Big{]} +\mathcal{O}(\hat{s}^0) \,,\nonumber
\end{align}
%%%%%%%%%%%%%
which was already observed disregarding flavor~\cite{Franceschini:2017xkh,Grojean:2018dqj}, and can be extended for different quark flavors by including the CKM factors spelled out in the above equation. For right-handed quarks, it is not possible to derive a similar relation, given that the effective couplings associated with each process are independent; see e.g.~Eq.~\eqref{eq:Vqq-matching}.

%%%%%%%%%%%%%%%%%%%%%%%%%%%%%%%%%%%%%%%%%%%%%%%%%%%%%%
%%%%%%%%%%%%%%%%%%%%%%%%%%%%%%%%%%%%%%%%%%%%%%%%%%%%%%
\section{LHC constraints}
\label{sec:lhc}

%%%%%%%%%%%%%%%%%%%%%%%%%%%%%%%%%%%%%%%%%%%%%%%%%%%%%%
\subsection{LHC data and statistical analysis}
\label{sec:lhc-data}

To constrain the Wilson coefficients that modify the $Z$- and $W$-
couplings quarks in Eq.~\eqref{eq:Leff-EW}, 
we have considered the available LHC Run 2 data on the $WV$ and $Vh$ production channels, as
summarized in Table~\ref{tab:LHC-WV} with the experimental details of each analysis.~\footnote{Recently, the ATLAS Collaboration published measurements of the $WW$ production cross-section with 140 fb$^{-1}$ \cite{ATLAS:2025dhf}. These measurements have not been included in our analysis, since no information on the correlations between different bins of the kinematic distributions was made publicly available.} 
Most experimental analyses provide differential cross-section measurements at particle level, 
allowing us to avoid the inclusion of detector effects in our simulations, with the exception of the CMS $WW$ analysis reported in Ref.~\cite{CMS:2020mxy} which requires a specific recast that will be described below.

In our simulations, we use \textsc{MadGraph5\_aMC@NLO}~\cite{Frederix:2018nkq} to extract 
predictions for the 
effective operator contributions at leading order (LO) in QCD with the 
UFO files for the effective Lagrangian generated via \textsc{FeynRules}\cite{Christensen:2008py}. 
Parton showering and hadronization were accounted using \textsc{Pythia8}\cite{Sjostrand:2007gs}, 
while detector effects for the CMS $WW$ analysis were simulated
using \textsc{Delphes}\cite{deFavereau:2013fsa}.

We parameterize our observables at the cross-section level using the following expansion
%%%%%%%%%%%%%%%%
\begin{equation}
\sigma = \sigma_\text{SM} + \dfrac{1}{\Lambda^2} \sum_{i} 
\mathrm{Re}\big{(}\cC_i\big{)} \, A^{i}_{\text{NP}} + \dfrac{1}{\Lambda^4} \sum_{i,j} \cC_i \, \cC_j^\ast \, B^{ij}_{\text{NP}^2}\,,
\end{equation}
%%%%%%%%%%%%%%%%
%
where the sums run over all relevant Wilson coefficients for each channel. The  linear and dimension-six squared contributions
 $A^{i}_{\text{NP}}$ and $B^{ij}_{\text{NP}^2}$
are obtained following the simulation procedure outlined above. 
For the SM predictions, $\sigma_\text{SM}$, we used the 
state-of-the-art results reported in the experimental analyses listed in Table~\ref{tab:LHC-WV}.
In the following, we describe in more detail the statistical analysis
performed for each of the channels.

%%%%%%%%%%%%%%%%%
\begin{table*}[t!]
\renewcommand{\arraystretch}{1.7}
\centering
\begin{tabular}{c||c|c|c|c}
Channel & Distribution & Collaboration & $N_{\text{obs}}$ & $\mathcal{L}$ \\[1ex] \hline\hline 
\multirow{2}{*}[-1.5ex]{$pp\rightarrow WW$} 
  & $\dfrac{d\sigma}{dp_T^{\ell^\text{lead}}}$  & ATLAS  & 14 & 36.1 fb$^{-1}$ \cite{ATLAS:2019rob}\\[2ex]
  & $\dfrac{dN_{\text{ev}}}{d m_{e\mu}}$ & CMS & 11 & 35.9 fb$^{-1}$ \cite{CMS:2020mxy}\\[1.5ex]\hline
\multirow{2}{*}[-1.5ex]{$pp\rightarrow WZ$} & $\dfrac{d \sigma}{d m_T^{WZ}}$ & ATLAS & 12 & 140 fb$^{-1}$ \cite{ATLAS:2025edf}\\[2ex]
& $\dfrac{1}{\sigma}\dfrac{d\sigma}{dm_{WZ}}$ & CMS & 5 & 137 fb$^{-1}$ \cite{CMS:2021icx}\\[1.5ex]\hline\hline
\multirow{2}{*}[-1.5ex]{$pp\rightarrow Zh$} & \multirow{2}{*}[-0.8ex]{$\dfrac{d\sigma}{d p_T^Z}$} & ATLAS & 5 & 140 fb$^{-1}$ \cite{ATLAS:2024yzu}\\
&  & CMS & 3 & 138 fb$^{-1}$   \cite{CMS:2023vzh} \\[1.ex]\hline
\multirow{2}{*}[-1.5ex]{$pp\rightarrow Wh$} &  \multirow{2}{*}[-0.8ex]{$\dfrac{d\sigma}{d p_T^W}$} & ATLAS & 5 & 140 fb$^{-1}$ \cite{ATLAS:2024yzu}\\
& &CMS & 3 & 138 fb$^{-1}$ \cite{CMS:2023vzh}
\end{tabular}
\vspace{0.2cm}
\caption{\small \sl LHC Run 2 data on $pp\rightarrow WV$ and $pp\to Vh$ processes (with $V=W,Z$) considered in our analyses. The relevant experimental distribution, the number of observed bins $(N_{\mathrm{obs}})$ and the integrated luminosity ($\mathcal{L}$) are given for each search channel.}
\label{tab:LHC-WV} 
\end{table*}
%%%%%%%%%%%%%%%%%

\subsubsection{$pp\rightarrow WV$}

For the $WV$ analyses reported in Refs.~\cite{ATLAS:2019rob, ATLAS:2025edf,CMS:2021icx}, we use the available bin-by-bin cross-section measurements and uncertainties, 
along with the correlations between different bins provided in the experimental papers. These inputs are used to build the usual $\chi^2$-test statistic for each channel,
%%%%%%%%%%%%%%%%
\begin{equation}
\chi^2 = \sum_{i, j} ({\cal O}_i^\text{dat} - {\cal O}_i^\text{th}) \,\mathrm{cov}^{-1}_{ij}\, ({\cal O}_j^\text{dat} - {\cal O}_j^\text{th})\,,
\label{eq:gauss-chi}
\end{equation}
%%%%%%%%%%%%%%%
where ${\cal O}_i^\text{dat}$ and ${\cal O}_i^\text{th}$ denote the 
measured and predicted cross sections in the bin $i$, respectively, 
and $\mathrm{cov}_{ij}$ is the covariance matrix that incorporates the experimental 
and theoretical uncertainties reported by the corresponding 
experimental analyses. For simplicity, we have not include theoretical uncertainties 
associated with PDFs and factorization scale variations 
in the simulation of the EFT contributions, as they have a negligible impact on the results of our fit.

The CMS $WW$ analysis~\cite{CMS:2020mxy} only provide measurements 
of the event yields as a function of the invariant mass 
of the final-state electron and muon.~\footnote{In this search, 
the signal region requires two oppositely charged leptons 
of different flavors in the final state.} Given the small statistics in the high-energy bins, in this case we used a Poisson chi-square distribution defined as
\begin{equation}
    \chi^2 = \min_{\vec{\xi}}\,\Bigg\{ 2 \sum_{i=1}^3 \Bigg[N_i^\text{th}(\vec{\xi}) - N_i^\text{dat}(\vec{\xi}) + N_i^\text{dat}(\vec{\xi}) \log{\dfrac{N_i^\text{dat}(\vec{\xi})}{N_i^\text{th}(\vec{\xi})}} \Bigg] + \sum_{i=1}^3 \xi_i^2\,\Bigg\}\,,
    \label{eq:poisson-chi}
\end{equation}
where we have included three pulls ($\xi_i$, with $i=1,2,3$) to account for theoretical and/or systematic 
uncertainties on the signal, background and observed number of events, respectively,
\begin{align}
    N_i^\text{th} &= \Big(1 + \sigma^\text{th}_i \, \xi_1 \Big)\, N_i^\text{signal} + \Big(1 + \sigma^\text{backg}_i\, \xi_2 \Big)\, N_i^\text{backg}\,, \\*[0.3em]
    N_i^\text{dat} &= \Big(1 + \sigma^\text{dat}_i \,\xi_3 \Big)\, N_i^\text{events} \, . \nonumber
\end{align}
The values chosen for $\sigma^\text{th}_i$ and $\sigma^\text{backg}_i$ range from 0.1 to 0.2, 
while the value for $\sigma_i^\text{dat}$ was taken to be 0.02 across all bins. We validated our $\chi^2$ implementation for this channel by comparing the confidence interval
obtained for the same Wilson coefficients constrained by experimental
collaborations and our results are in good agreement with those reported in Ref.~\cite{CMS:2020mxy}. 

\subsubsection{$pp\rightarrow Vh$} 

For the associated Higgs-production channels reported in Refs. \cite{ATLAS:2024yzu,CMS:2023vzh},
we used the differential cross-section measurements in the 
format of Simplified Template Cross Sections (STXS) \cite{LHCHiggsCrossSectionWorkingGroup:2016ypw}. 
The STXS framework separates the measurements into distinct bins based on the Higgs production and decay modes, and further subdivided into  mutually exclusive kinematic regions. In our analysis, we included 
the most recent ATLAS \cite{ATLAS:2024yzu} and CMS \cite{CMS:2023vzh} STXS measurements for the processes 
$pp\rightarrow Zh$, with $Z \rightarrow \ell^+ \ell^-,\, \nu\overline{\nu}$, and  $pp\rightarrow W h$, with
$W \rightarrow \ell \nu$, including decays to $\tau$ leptons, with the Higgs decaying into a pair of $b$-quarks in both cases.
We have only considered distributions inclusive in the number of jets. 

%%%%%%%%%%%%%%%
\begin{figure*}[!p]
\begin{center}
    \includegraphics[width=0.49\textwidth]{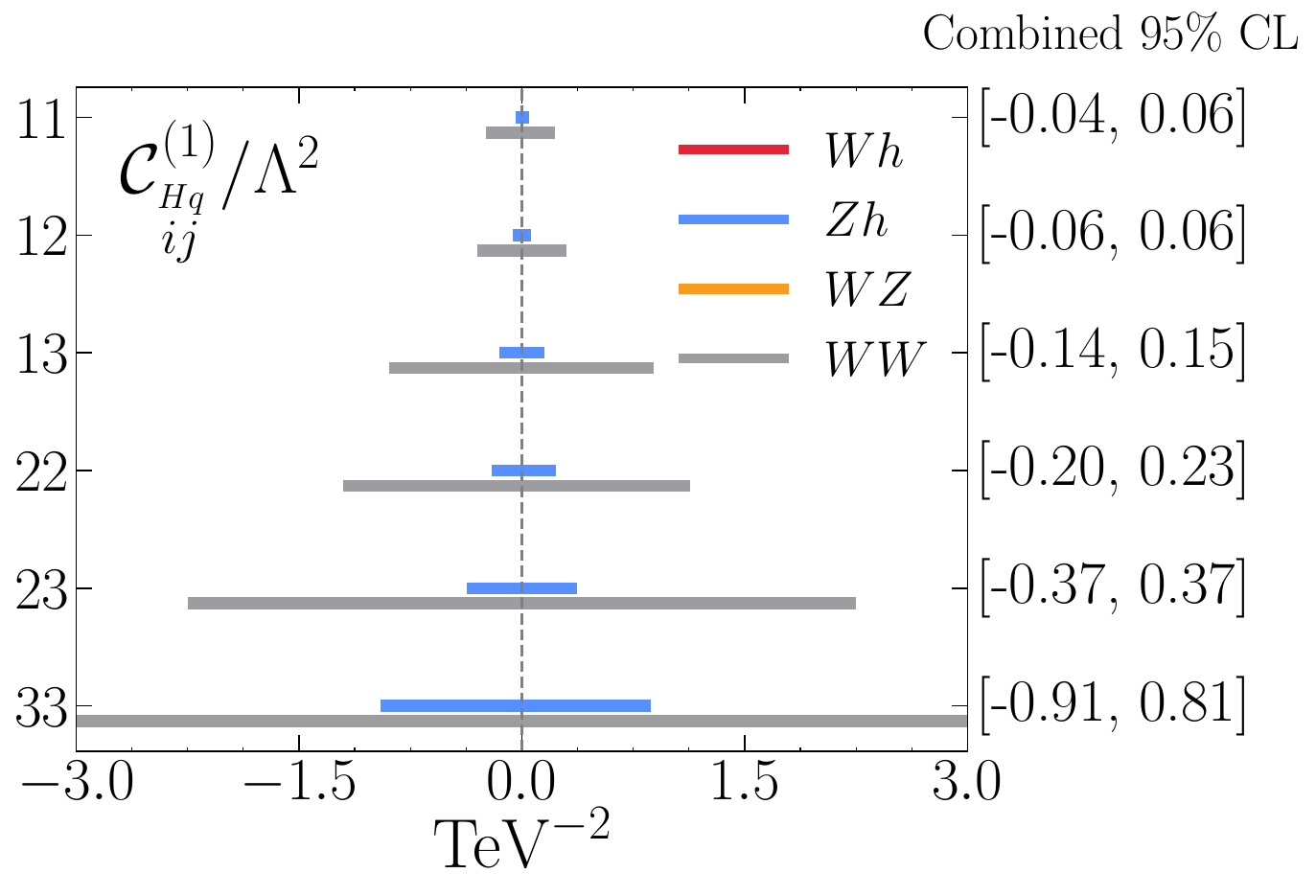}~\includegraphics[width=0.49\textwidth]{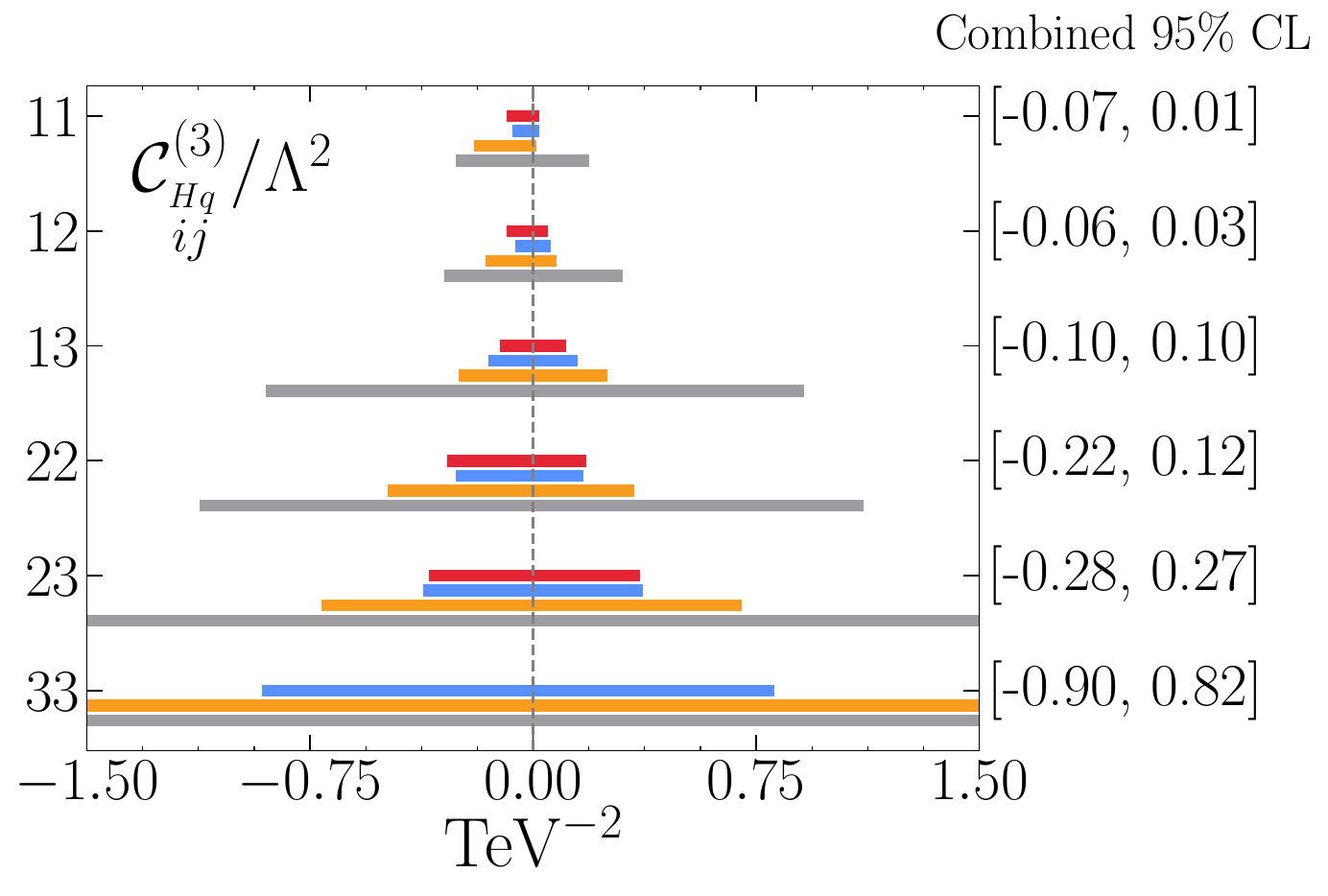}\\[0.85em]
    \includegraphics[width=0.49\textwidth]{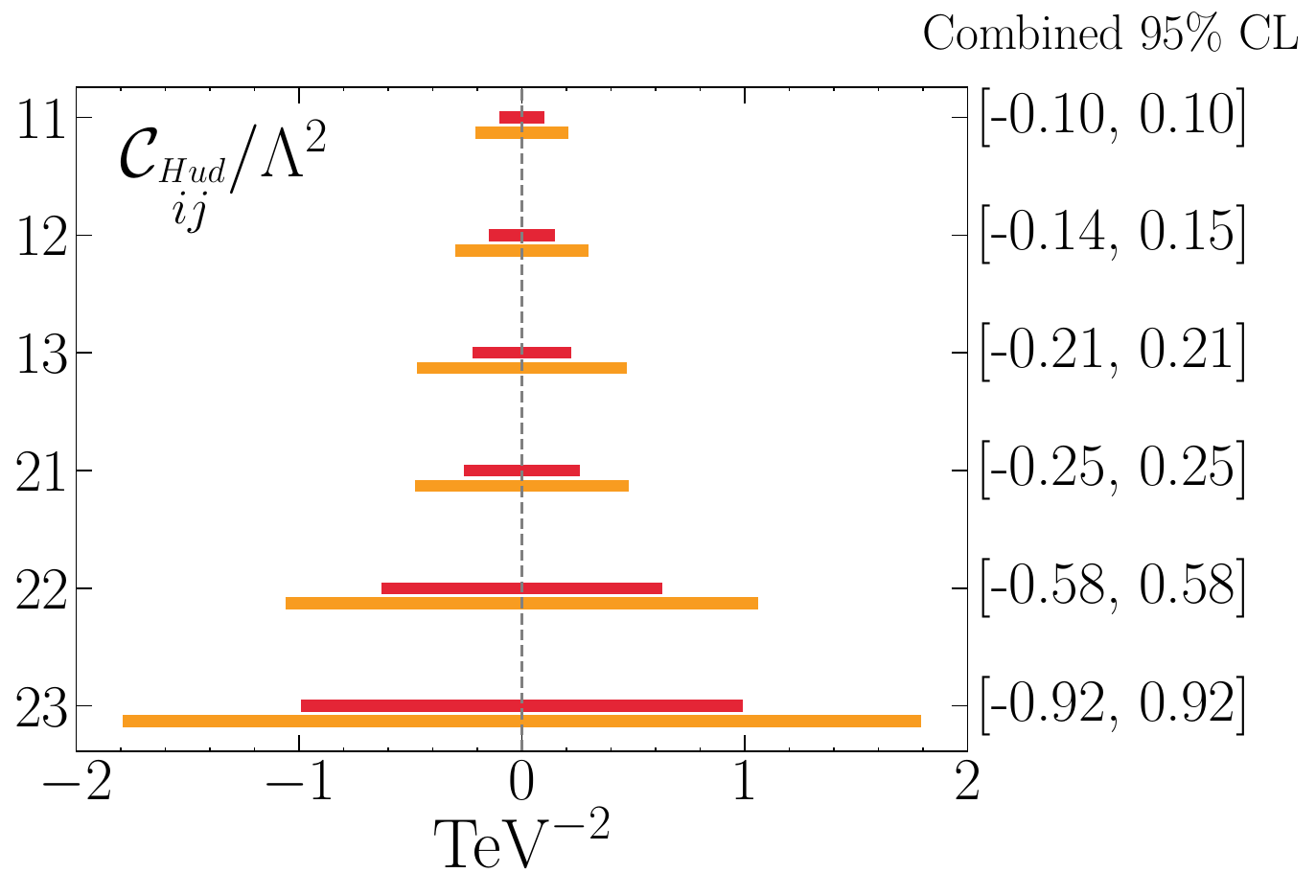}~\includegraphics[width=0.49\textwidth]{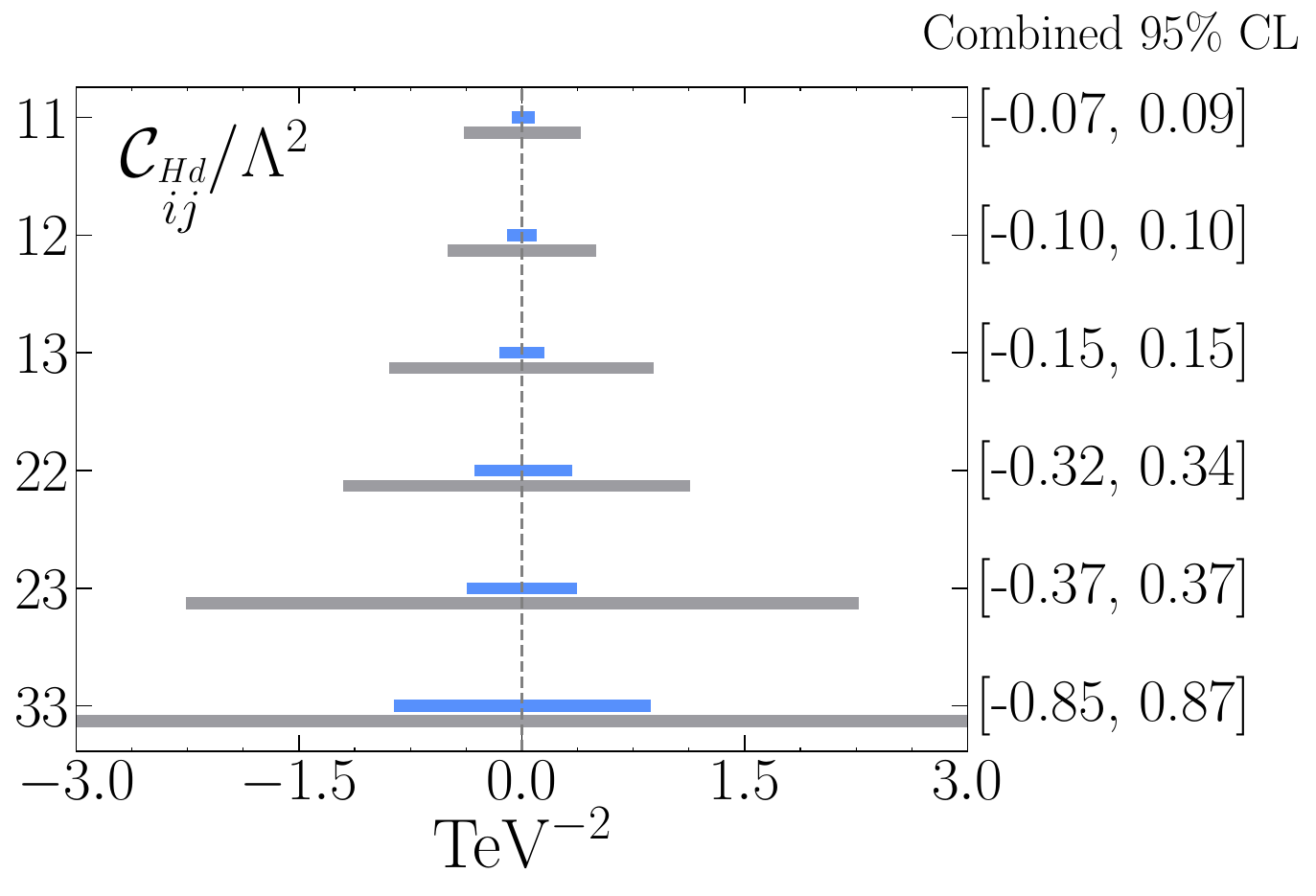}\\[0.85em]
    \includegraphics[width=0.49\textwidth]{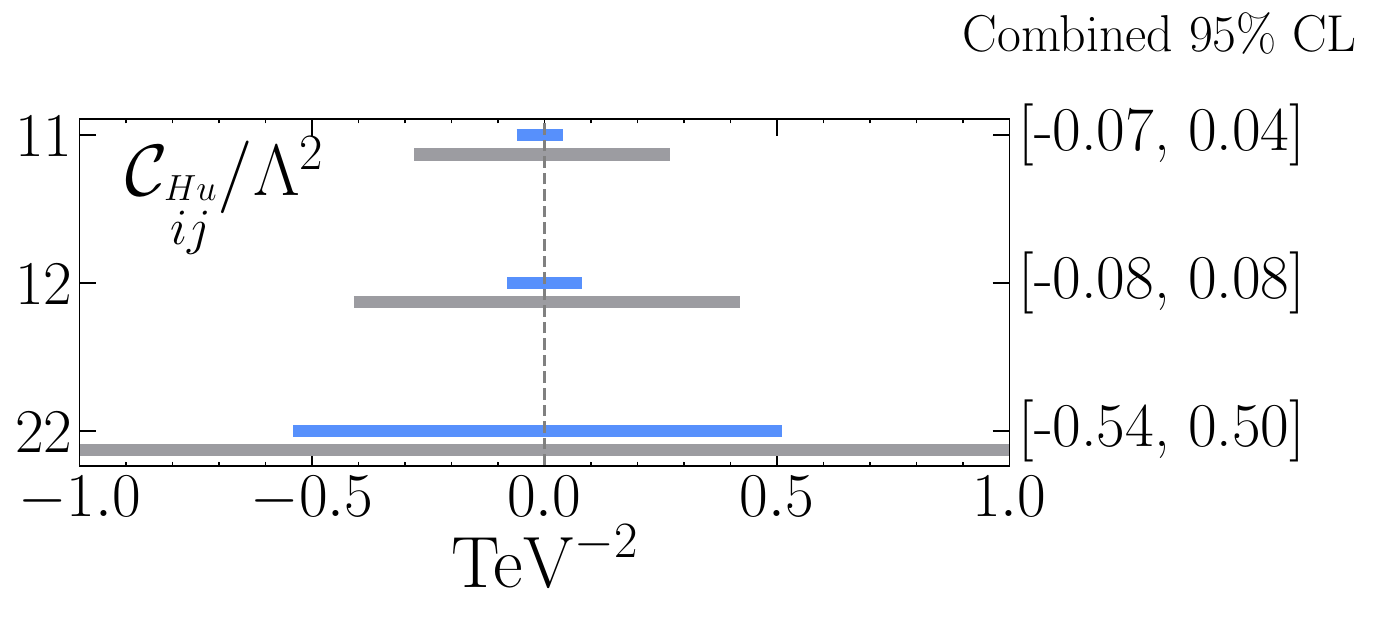}
\end{center}
\caption{\small \sl~Constraints on the dimension-six Higgs-current Wilson coefficients, considering a single non-vanishing coefficient at a time for the $Vh$ and $WV$ production channels. Quark flavor-indices  $ij$ are specified on the left-hand side of each plot with the 95\% confidence interval on the right panels. All coefficients are assumed to be real  and contributions to the cross-section include EFT contributions up to the quadratic terms.} 
    \label{fig:limits-Wcouplings}
\end{figure*}
%%%%%%%%%%%%%%%

In these channels, the cross-section is measured as a function
of the transverse momentum of the gauge boson in the final state, $p_T^V$.
The uncertainties and correlations for each of the bins are made available by the experimental collaborations, which allows us to write a Gaussian $\chi^2$-function as presented in Eq.~\eqref{eq:gauss-chi}. Currently, the theoretical uncertainties are small compared to the experimental ones,  which are unfolded in each bin, and can therefore be neglected.

\subsubsection{Summary}

In summary, the analysis of the LHC data can be described by the combined $\chi^2$ function
\begin{equation}
\chi^2_\text{LHC}(\vec{\cC}) \equiv \chi^2_{WV}(\vec{\cC}) + \chi^2_{Vh}(\vec{\cC})\,,
\end{equation}
where $\vec{\cC}$ denotes the Wilson coefficients of the Higgs current operators.
The terms $\chi^2_{WV}$ and $\chi^2_{Vh}$ account for
the diboson and Higgs-associated production channels, respectively, which are collected in Table~\ref{tab:LHC-WV} and which will be used in the numerical analysis in the following Section.

%%%%%%%%%%%%%%%%%%%%%%%%%%%%%%%%%%%%%%%%%%%%%%%%%%%%%%

%%%%%%%%%%%%%%%%%%%%%%%%%%%%%%%%%%%%%%%%%%%%%%%%%%%%%%
\subsection{Numerical results}
\label{sec:numerical}

In Fig.~\ref{fig:limits-Wcouplings}, we present the single-parameter 
bounds on the coefficients $\smash{\cC^{(1)}_{Hq}}$, $\smash{\cC^{(3)}_{Hq}}$, $\cC_{Hd}$, $\cC_{Hu}$, and $\cC_{Hud}$ with all possible flavor indices, which are derived from the LHC processes discussed in Sec.~\ref{sec:lhc-data}. We assume the Wilson coefficients to be real, and we only display the LHC limits arising from energy-enhanced EFT effects, cf.~discussion in Sec.~\ref{sec:helicity}. 
Several comments can be made by comparing the limits for different flavor indices:
\begin{itemize}
    \item[{\it i})] The limits given in Fig.~\ref{fig:limits-Wcouplings} are weaker for heavy quark flavors due to PDF suppression, as expected. More specifically, the bounds for operators with $b$-quarks are an order of magnitude weaker 
    than those for first-generation quarks for the three types of operators. Since our analysis retains the full CKM matrix 
    dependence (cf. Eq.~\eqref{eq:Vqq-matching}), the coefficients $\smash{\cC^{(1)}_{Hq}}$ and $\smash{\cC^{(3)}_{Hq}}$ 
    with specific flavor indices can also be constrained via processes involving initial state 
    quarks of different flavors.  However, these effects are small, 
    as they are suppressed by off-diagonal CKM matrix elements.
    \item[{\it ii})] The impact of interference terms is visible for the 
    coefficient $\cC^{(3)}_{Hq}$ for light and/or same-flavor quarks. This can be seen from the 
    asymmetric confidence intervals for the flavor indices 11, 12, 22, and 33. The interference terms for the flavor indices 13 and 23 are strongly suppressed within the SM, 
    resulting in symmetric limits.  Moreover, the limits for the coefficient $\cC_{Hud}$ are completely
    symmetric around zero since its energy-growing amplitude has a different helicity structure
    than the SM amplitude, resulting in an 
    interference term suppressed by the fermion masses.

%%%%%%%%%%%%%
\begin{figure*}[hbtp]
    ~\includegraphics[width=1.\textwidth]{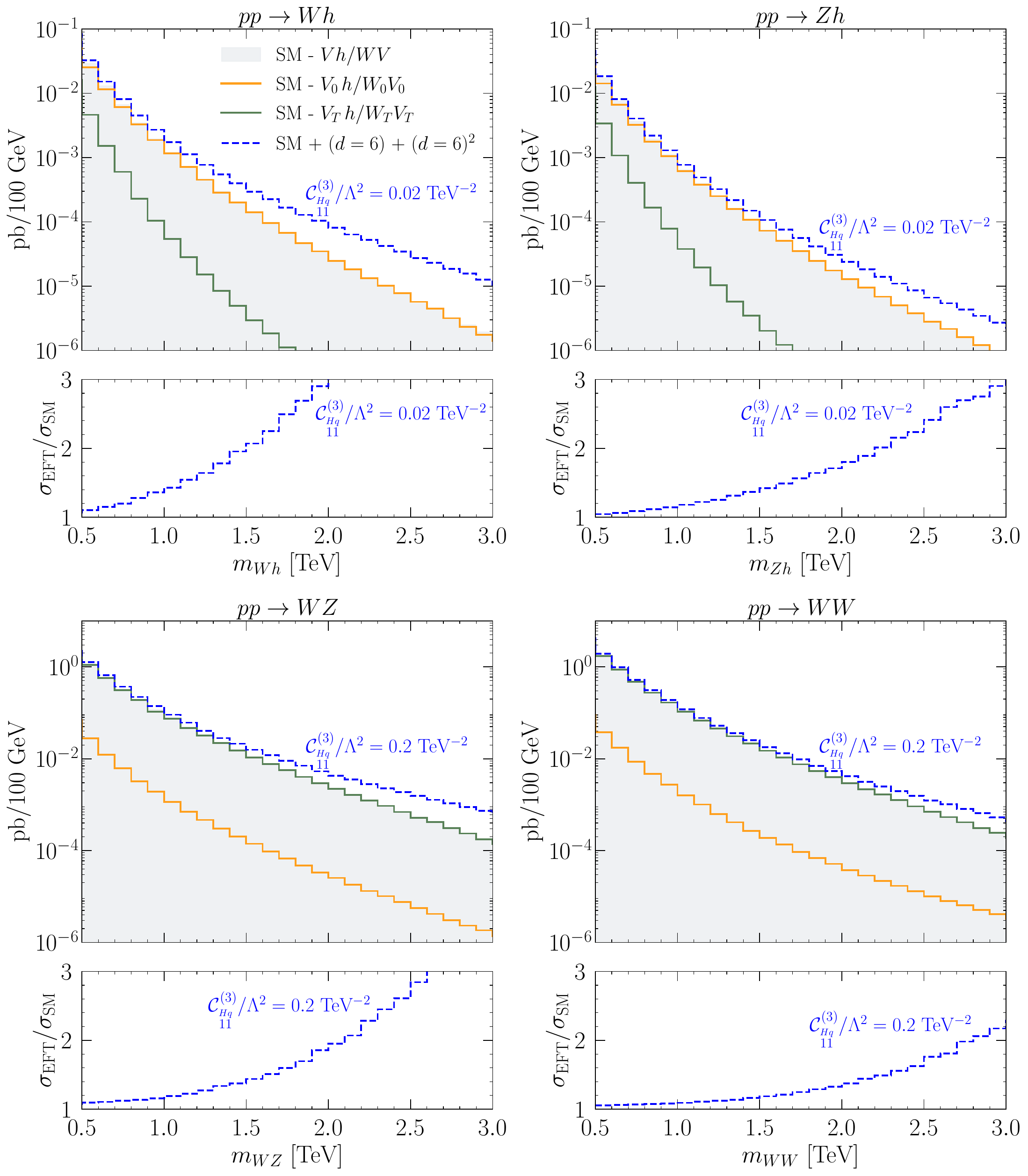}
    \caption{ \small \sl
    The parton-level distributions for the $Vh$ (top) and $WV$ (bottom) processes are plotted as functions of the process center-of-mass energy. 
    In the large panels, we present the SM predictions, including 
    all polarizations of the final-state vector bosons (shaded gray area), as well as the separate contributions from the 
    longitudinal (solid orange line) and transverse (solid green line) components. We also show the contribution 
    from New Physics, in addition to the SM background, originating 
    from the operator $\smash{\mathcal{O}^{(3)}_{Hq}}$. 
    In the smaller panels, we display the ratio between the EFT and SM cross-sections in each bin. Notice, in particular, that we consider different values of the Wilson coefficient in the upper and lower panels, as the diboson channels are less sensitive to New Physics effects.}
    \label{fig:WHvsWZ}
\end{figure*}
%%%%%%%%%%%%%
    \item[{\it iii})] For all Wilson coefficients displayed in Fig.~\ref{fig:limits-Wcouplings}, the limits from Higgs-associated production channels are roughly twice as strong as those from diboson production channels.  This can be understood qualitatively from Fig.~\ref{fig:WHvsWZ}, which shows the predictions for the 
    parton-level cross section as a function of the diboson invariant mass for the $Vh$ and $WV$ channels, within the SM (shaded gray) and for a representative EFT scenario with $\smash{\mathcal{C}_{Hq}^{(3)}\neq 0}$ for first-generation quarks (dashed blue). In Fig.~\ref{fig:WHvsWZ}, the SM contributions are also shown separately for longitudinal (orange) and transverse (green) polarizations of the electroweak bosons.
    Comparing these predictions, we observe a larger deviation from the SM values in the $Vh$ production channels for smaller values of the effective coefficients. This can be traced back to the size of the respective SM backgrounds, which are dominated by the transverse modes in the diboson channels, and by the longitudinal modes in the associated Higgs production~\cite{Franceschini:2017xkh}.
\end{itemize}

\noindent There are other observables that could be used to probe the modifications
of the $W$- and $Z$-boson couplings to SM quarks at the LHC. 
In Ref.~\cite{Alioli:2017ces}, the impact of $\delta g^{Wq}_R$ on other probes has been carefully studied, 
including vector-boson scattering at the LHC, among other high-energy processes. Their findings were that the early LHC constraints from the signal 
strength measurements of $pp\to Vh$ were largely dominant over the 
other available observables. We expect this conclusion to be further strengthened, since the latest LHC data on 
$pp\to Vh$~\cite{ATLAS:2024yzu,CMS:2023vzh}, 
as well as the related process $pp\to VW$~\cite{ATLAS:2019rob,CMS:2020mxy,ATLAS:2025edf,CMS:2021icx} that we consider here, are given at the differential level, thus largely benefiting from the energy enhancement of the EFT contributions.

%%%%%%%%%%%%%%%%%%%%%%%%%%%%%%%%%%%%%%%%%%%%%%%%%%%%%%
%%%%%%%%%%%%%%%%%%%%%%%%%%%%%%%%%%%%%%%%%%%%%%%%%%%%%%
\section{Confronting LHC and electroweak observables}
\label{sec:EWPO}

In this Section, we compare the LHC constraints derived in Sec.~\ref{sec:numerical} for Higgs-current operators, containing quarks of different flavors, with those obtained from Electroweak Precision Observables (EWPO). This complementarity has been extensively 
explored in the literature in the flavor-blind scenario~\cite{Grojean:2018dqj, Franceschini:2017xkh, Liu:2018pkg, Baglio:2017bfe, Falkowski:2015jaa, Banerjee:2018bio, Biekotter:2014gup, Butter:2016cvz,deBlas:2025xhe}. In the following, we focus on the flavor-conserving operators with quarks $q=u,d,c,s,b$, which simultaneously contribute to $pp\to Vh$ and $pp\to VW$ proceses, as well as to electroweak observables.~\footnote{Flavor-violating operators are not considered in this section, since those contributing to neutral currents are tightly constrained by $\Delta F=1$ and $\Delta F=2$ processes. Furthermore, operators with the top-quark are not considered as they do not contribute at tree level to the LHC processes studied in Sec.~\ref{sec:numerical}, although they can impact EWPO at loop level, see Ref.~\cite{Coy:2019rfr} and references therein.}

In our numerical analysis, we consider the $Z$- and $W$-pole observables listed in Table~\ref{tab:EWPO}, along with their experimental values and SM predictions. We computed the SMEFT contributions to these observables, including the leading-logarithm contributions due to the running of the Wilson coefficients from the scale $\mu = 1$ TeV to $\mu_\text{ew} = m_Z$~\cite{Jenkins:2013zja}, as provided in the {\tt DsixTools} package~\cite{Fuentes-Martin:2020zaz}. For simplicity, we neglected small theory uncertainties, only considering the experimental ones together with the correlations between different measurements, whenever available.

%%%%%%%%%%%%%%%%%
\begin{table*}[t!]
\renewcommand{\arraystretch}{1.4}
\centering
\begin{tabular}{c|c|c}
Observable & Measurement & SM prediction  \\[1ex] \hline\hline 
$\Gamma_Z$ [GeV] & $2.4955 \pm 0.0023$~\cite{Janot:2019oyi} & 2.4941~\cite{Breso-Pla:2021qoe} \\
$\sigma_\text{had}$ [nb] & $41.4807 \pm 0.0325$~\cite{Janot:2019oyi} & 41.4923~\cite{deBlas:2022hdk} \\
$R_e$ & $20.8038 \pm 0.0497$~\cite{Janot:2019oyi} & 20.734~\cite{Breso-Pla:2021qoe}\\
$R_\mu$ & $20.7842 \pm 0.0335$~\cite{Janot:2019oyi} & 20.734~\cite{Breso-Pla:2021qoe}\\
$R_\tau$ & $20.7644 \pm 0.0448$~\cite{Janot:2019oyi} & 20.781~\cite{Breso-Pla:2021qoe}\\
$R_b$ & $0.21629 \pm 0.00066$~\cite{ALEPH:2005ab} & 0.21591~\cite{deBlas:2022hdk}\\
$R_c$ & $0.1721 \pm 0.0030$~\cite{ALEPH:2005ab} & 0.1722~\cite{deBlas:2022hdk}\\
$A_\text{FB}^{0,\, b}$ & $0.0992 \pm 0.0016$~\cite{ALEPH:2005ab} & 0.1029~\cite{deBlas:2022hdk}\\
$A_\text{FB}^{0,\, c}$ & $0.0707 \pm 0.0035$~\cite{ALEPH:2005ab} & 0.0735~\cite{deBlas:2022hdk}\\
${\cal A}_b$ & $0.923 \pm 0.020$~\cite{ALEPH:2005ab} & 0.935~\cite{deBlas:2022hdk}\\
${\cal A}_c$ & $0.670 \pm 0.027$~\cite{ALEPH:2005ab} & 0.668~\cite{deBlas:2022hdk}\\
${\cal A}_s$ & $0.895 \pm 0.091$~\cite{ALEPH:2005ab} & 0.936~\cite{deBlas:2022hdk}\\
$R_{uc}$ & $0.166 \pm 0.009$~\cite{ParticleDataGroup:2024cfk} & 0.172~\cite{deBlas:2022hdk}\\\hline
$\Gamma_W$ [GeV] & $2.085 \pm 0.042$~\cite{ParticleDataGroup:2024cfk} & 2.087~\cite{deBlas:2022hdk}\\
$\text{Br}(W\rightarrow e\nu)$ &  $0.1071 \pm 0.0016$~\cite{ALEPH:2013dgf} & 0.1082~\cite{deBlas:2022hdk}\\
$\text{Br}(W\rightarrow \mu\nu)$ &  $0.1063 \pm 0.0015$~\cite{ALEPH:2013dgf} & 0.1082~\cite{deBlas:2022hdk}\\
$\text{Br}(W\rightarrow \tau\nu)$ &  $0.1138 \pm 0.0021$~\cite{ALEPH:2013dgf} & 0.1081~\cite{deBlas:2022hdk}
\end{tabular}
\vspace{0.2cm}
\caption{\small \sl $Z$- and $W$-pole observables considered in our 
analysis, along with the corresponding experimental values and SM predictions. We present only the central values for the SM predictions, as their uncertainties are significantly smaller than the experimental ones, thus being neglected in our analysis. Whenever available, correlations between the 
observables were considered in the fit. The experimental correlation matrices 
can be found in the respective experimental papers.}
\label{tab:EWPO} 
\end{table*}
%%%%%%%%%%%%%%%%%

\begin{figure*}[!t]
\begin{center}
    \includegraphics[width=1.\textwidth]{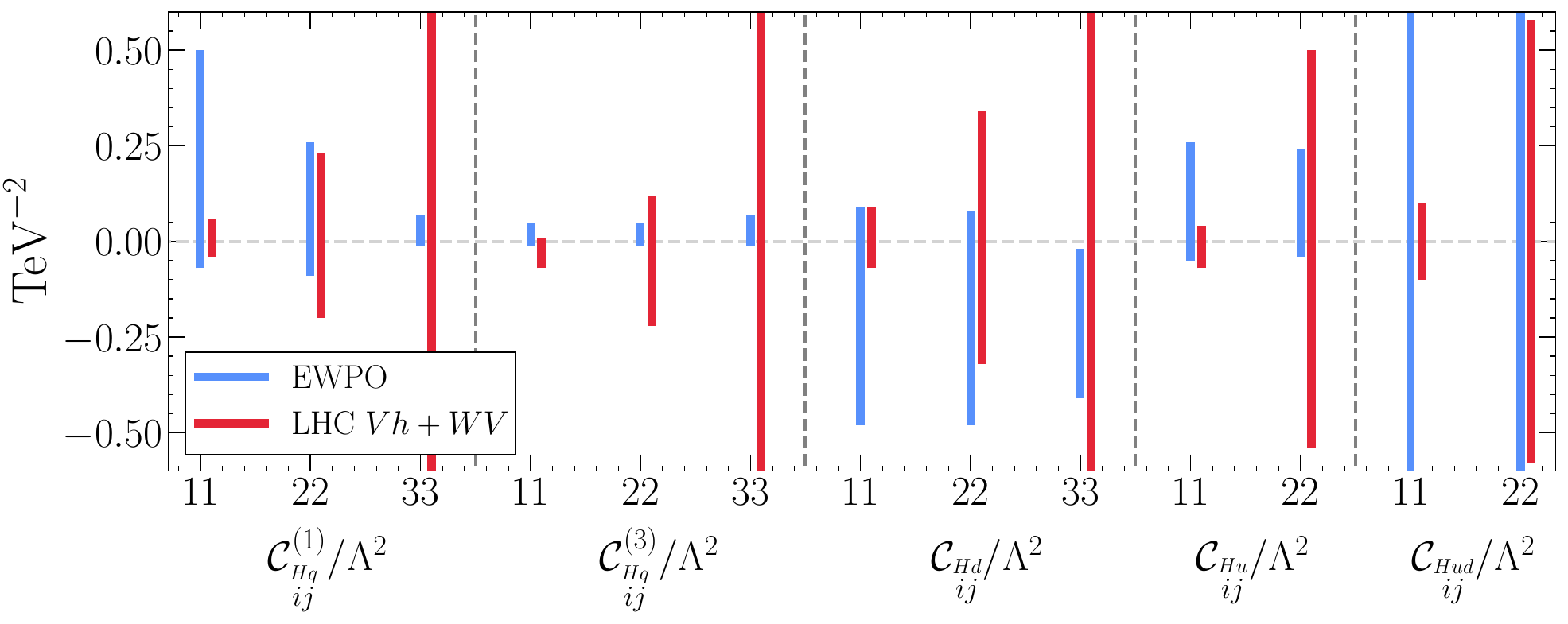}
    \caption{\small \sl Comparison of the limits at 95\% CL on the coefficients 
    $\cC^{(1)}_{Hq}$, $\cC^{(3)}_{Hq}$, $\cC_{Hd}$, $\cC_{Hu}$  and $\cC_{Hud}$, for different flavor-diagonal indices, which are extracted from EWPO (blue), and LHC diboson and Higgs associated production data (red). The EFT cutoff is fixed 
    to $\Lambda=1$~TeV in the EWPO analysis.}
    \label{fig:EWPOvsLHC}
\end{center}
\end{figure*}

The comparison between LHC processes and EWPO is provided in Fig.~\ref{fig:EWPOvsLHC}, considering a single non-vanishing coefficient at a time. We find that the LHC limits from diboson and Higgs associated production on light-flavor quark couplings (i.e., $11$ and $22$) are competitive with those from EWPO, while constraints on third-generation quarks are significantly weaker, as expected due to PDF suppression. Notice, also, that EWPO bounds on $\smash{\cC^{(1)}_{Hq}}$ for the $11$ and $22$ indices are weaker than those on $\smash{\cC^{(3)}_{Hq}}$ for the same flavors. This occurs because the contributions from up- and down-type quarks from the singlet operator to the total $Z$-boson width ($\Gamma_Z$) approximately cancel at linear order. No such accidental cancellation occurs for the triplet operator, as it induces contributions with up- and down-quarks with the same sign. Notice, also, that such feature is absent in the case of $\smash{\cC^{(1)}_{\substack{Hq}}}$ coupled to third-generatio quarks, since only the bottom-quark contributes to $\Gamma_Z$ in this scenario.

%%%%%%%%%%%%%
%%%%%%%%%%%%%

\begin{figure}[t!]
\begin{center}
    \includegraphics[width=1.\textwidth]{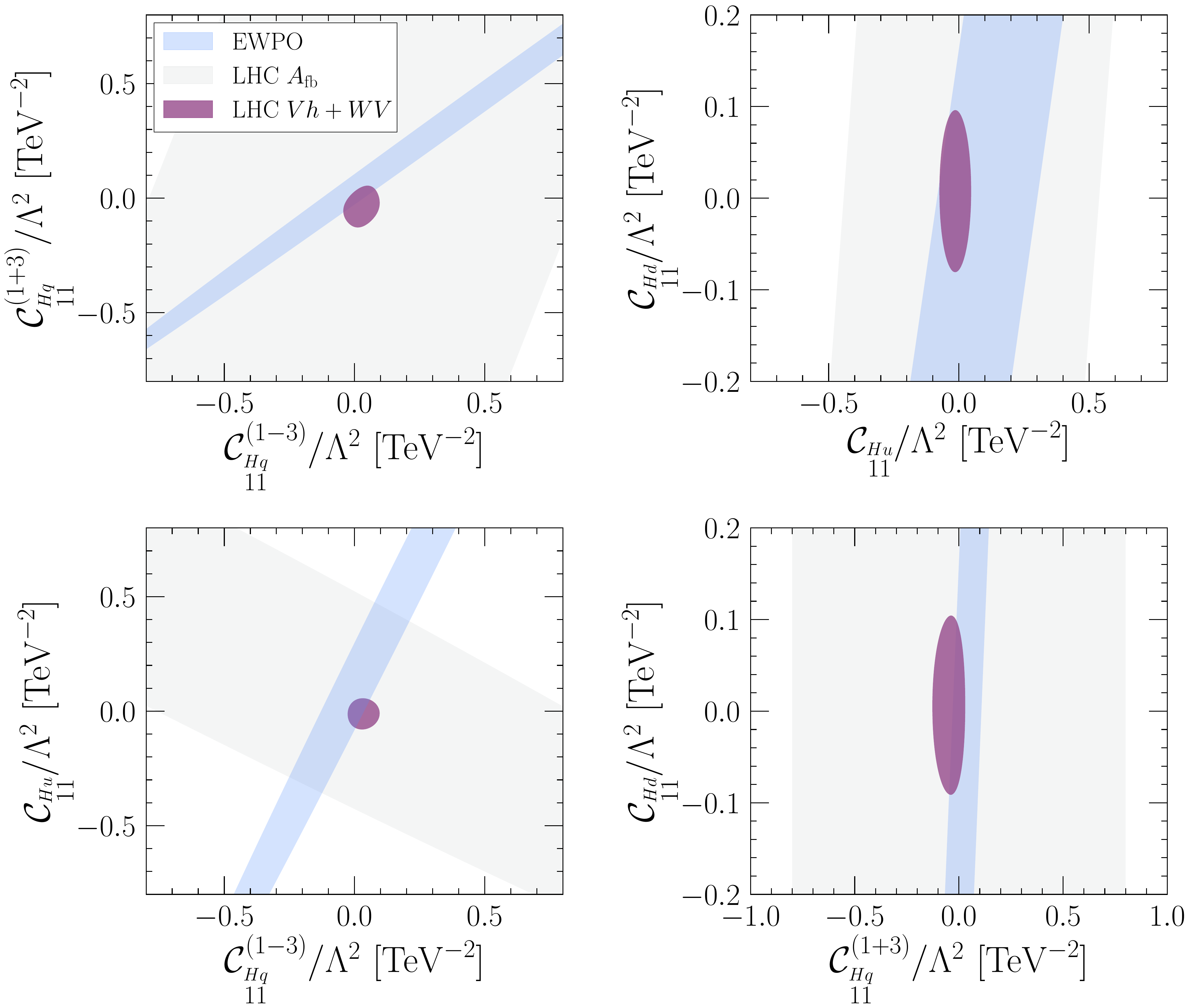}
    \caption{\small \sl Constraints on Higgs-current operators coupled to first-generation quarks derived from EWPO in Sec.~\ref{sec:EWPO} (blue), and $pp \to Vh$ and $pp\to VW$ data in Sec.~\ref{sec:numerical} (purple). For comparison, we also display the constraints from Drell-Yan $A_{\mathrm{fb}}$ taken from Ref.~\cite{Breso-Pla:2021qoe}, which are superseded by associated Higgs production. The two-dimensional constraints are provided at 95\% CL, setting the other effective coefficients to zero. The EFT cutoff is fixed 
    to $\Lambda=1$~TeV in the EWPO analysis.}
    \label{fig:EWPO-Regions}
\end{center}
\end{figure}

As argued in Ref.~\cite{Breso-Pla:2021qoe}, the electroweak data alone cannot fully constrain the corrections to the electroweak boson couplings to light quarks due to flat directions in the SMEFT likelihood at $\mathcal{O}(\Lambda^{-2})$. To lift these unconstrained directions, the LHC measurement of the forward-backward asymmetry ($A_{\mathrm{fb}}$) of Drell-Yan processes $pp\to\ell\ell$ (with $\ell=e,\mu$) at the $Z$-pole was considered in Ref.~\cite{Breso-Pla:2021qoe}. These observables can probe an orthogonal combination of Wilson coefficients than EWPO. In Fig.~\ref{fig:EWPO-Regions}, we show that
LHC data on Higgs-associated production can also remove these flat directions, providing much stronger constraints than $A_{\mathrm{fb}}$ with current LHC data, thanks to energy enhancement of the EFT contributions.
%%%%%%%%%%%%%
%%%%%%%%%%%%%

Finally, we stress that although EWPO observables can provide
meaningful constraints on modifications to the $Z$ couplings
and left-handed corrections to the $W$ couplings, they are not capable of limiting right-handed charged currents to light quarks. This can be attributed to two main reasons. Firstly, right-handed modifications to the $W$ couplings
to fermions do not interfere with the SM amplitudes, leading to changes to the $W$ width only at order ${\cal O}(\Lambda^{-4})$. Secondly,  the measurement of $\Gamma_W$ is less precise than that of $\Gamma_Z$. 
As an example, the limit on the coefficient 
$\cC_{Hud}$ from $W$ pole observables is 
\begin{equation}
   \Big|\cC_{\substack{Hud\\11}}\Big|/\Lambda^2 \le 6.6 \, \text{TeV}^{-2}\, , 
\end{equation}
which is more than one order of magnitude weaker than the one 
extracted using diboson measurements (cf.~Fig.~\ref{fig:limits-Wcouplings}). 
This highlights the importance of LHC data 
in constraining right-handed charged currents, as previously studied in Ref.~\cite{Alioli:2017ces}. In the following section, we examine the complementarity of LHC and flavor observables using a specific example that is motivated by discrepancies in low-energy $d(s)\to u \ell\nu$ data.

%%%%%%%%%%%%%%%%%%%%%%%%%%%%%%%%%%%%%%%%%%%%%%%%%%%%%%
%%%%%%%%%%%%%%%%%%%%%%%%%%%%%%%%%%%%%%%%%%%%%%%%%%%%%%
\section{Testing the first-row CKM unitarity}
\label{sec:illustration}

In this Section, we confront our LHC constraints with the ones stemming from low-energy flavor-physics observables within a specific example. For illustration, we  focus on the observables based on the $d(s)\to u \ell \nu$ transitions, which are notably used to determine the Cabibbo angle in the SM. Our motivation is the mild discrepancy in the unitarity tests of the first-row of the CKM unitarity,
%%%%%%%%%%%%%%%%
\begin{equation}
    \Delta_{\mathrm{CKM}}\equiv |V_{ud}|^2+|V_{us}|^2+|V_{ub}|^2-1\,,
\end{equation}
%%%%%%%%%%%%%%%%

\noindent which seem to differ from zero at the $2\sigma$ to $3\sigma$ level, depending on the set of low-energy observables considered~\cite{Cirigliano:2022yyo}. Although updated experimental measurements and refined SM predictions are still necessary to clarify the situation, it is worthwhile to explore whether a minimalistic New Physics model could account for the observed deviations while remaining consistent with available high-energy constraints. As will be discussed below, the preferred EFT scenarios identified in the literature require a modification of the $W\bar{u}d$ and/or $W\bar{u}s$ couplings~\cite{Crivellin:2022rhw,Cirigliano:2023nol}, which could have implications for $pp\to Vh$ and $pp\to VW$ processes at the LHC.

In the following, we briefly revisit the main constraints on the first-row CKM unitarity, and we show within the EFT approach that the LHC constraints derived in this paper will become competitive with flavor constraints in the HL-LHC. Finally, we apply these constraints to a concrete model with TeV-scale vector-like quarks that can generate the viable operators at tree level, which 
predict specific correlations between the different Higgs-current 
operators.

%%%%%%%%%%%%%%%%%%%%%%%%%%%%%%%%%%%%%%%%%%%%%%%%%%%%%%
%%%%%%%%%%%%%%%%%%%%%%%%%%%%%%%%%%%%%%%%%%%%%%%%%%%%%%
\subsection{Low-energy EFT}

%%%%%%%%%%%%%%%%
\begin{table*}[!t]
\renewcommand{\arraystretch}{2.}
\centering
\begin{tabular}{cc}
\centering
\begin{tabular}{c|c}
$\psi^4$ &   Operator\\ \hline\hline
$\mathcal{O}_{lq}^{(1)}$ & $\big{(}\bar{l}_i \gamma^\mu l_j\big{)}\big{(}\bar{q}_k \gamma_\mu q_l\big{)}$ \\ 
$\mathcal{O}_{lq}^{(3)}$ & $\big{(}\bar{l}_i \gamma^\mu \tau^I l_j\big{)}\big{(}\bar{q}_k \gamma_\mu \tau^I q_l\big{)}$\\ 
$\mathcal{O}_{lu}$ & $\big{(}\bar{l}_i \gamma^\mu l_j\big{)}\big{(}\bar{u}_k \gamma_\mu u_l\big{)}$\\ 
$\mathcal{O}_{ld}$ & $\big{(}\bar{l}_i \gamma^\mu l_j\big{)}\big{(}\bar{d}_k \gamma_\mu d_l\big{)}$\\ 
$\mathcal{O}_{eq}$ & $\big{(}\bar{e}_i \gamma^\mu e_j\big{)}\big{(}\bar{q}_k \gamma_\mu q_l\big{)}$\\ 
$\mathcal{O}_{eu}$ & $\big{(}\bar{e}_i \gamma^\mu e_j\big{)}\big{(}\bar{u}_k \gamma_\mu u_l\big{)}$\\ 
$\mathcal{O}_{ed}$ & $\big{(}\bar{e}_i \gamma^\mu e_j\big{)}\big{(}\bar{d}_k \gamma_\mu d_l\big{)}$\\ 
\end{tabular}
&
\hspace{1.2cm}
\begin{tabular}{c|c}
$\psi^4$ &   Operator $+\mathrm{h.c.}$\\ \hline\hline
$\mathcal{O}_{ledq}$ & $\big{(}\bar{l}^a_i e_j\big{)}\big{(}\bar{d}_k q^a_l\big{)}$ \\ 
$\mathcal{O}_{lequ}^{(1)}$ & $\big{(}\bar{l}^a_i e_j \big{)}\varepsilon_{ab}\big{(}\bar{q}^b_k u_l\big{)}$\\ 
$\mathcal{O}_{lequ}^{(3)}$ & $\big{(}\bar{l}^a_i \sigma^{\mu\nu}e_j \big{)}\varepsilon_{ab}\big{(}\bar{q}^b_k \sigma_{\mu\nu}u_l\big{)}$\\
\end{tabular}
\end{tabular}
\vspace{0.2cm}
\caption{\small \sl 
Hermitian (left panel) and non-Hermitian (right panel) dimension-six operators that contribute to semileptonic processes at tree level. See caption of Table~\ref{tab:SMEFT-Higgs-ope}.}
\label{tab:SMEFT-psi4-ope} 
\end{table*}
%%%%%%%%%%%%%%%%

Firstly, we introduce the most general low-energy Lagrangian to describe the $d(s)\to u \ell \nu$ with operators up to dimension six, which is normalized as follows,
%%%%%%%%%%%%%%%%
\begin{align}
    \label{eq:left}  
        \mathcal{L}_{\mathrm{LEFT}}&\supset -2\sqrt{2}G_F V_{uq} \sum_I g^{q\ell}_I\, O_I+\mathrm{h.c.}\,,
\end{align}
%%%%%%%%%%%%%%%%
where the relevant operators are given by
%%%%%%%%%%%%%%%%
\begin{align}
\label{eq:LEFT-lag}
O_{V_L}^{q\ell} &=\big{(}\bar{u}_L\gamma_{\mu}q_L\big{)}\big{(}\bar{\ell}_L\gamma^{\mu}\nu_L\big{)}\,, &
O_{S_L}^{q\ell} &=\big{(}\bar{u}_R q_L\big{)}\big{(}\bar{\ell}_R\nu_L\big{)}\,, \\*[0.3em]
O_{V_R}^{q\ell} &=\big{(}\bar{u}_R\gamma_{\mu}q_R\big{)}\big{(}\bar{\ell}_L\gamma^{\mu}\nu_L\big{)}\,, &
O_{S_R}^{q\ell} &=\big{(}\bar{u}_Lq_R\big{)}\big{(}\bar{\ell}_R\nu_L\big{)}\,,\nonumber\\*[0.3em]
O_T^{q\ell} &= \big{(}\bar{u}_R\sigma_{\mu\nu}q_L\big{)}\big{(}\bar{\ell}_R\sigma^{\mu\nu}\nu_L\big{)}\,,\nonumber
\end{align}
%%%%%%%%%%%%%%%%
where $q=d,s$, and $g_{I}^{q\ell}$ are the low-energy effective coefficients, which are defined at the renormalization scale $\mu_{\mathrm{low}}=2~\mathrm{GeV}$. We define $g_{V_L}^{q\ell}\equiv 1 +\delta g_{V_L}^{q\ell}$, and $g_{I}^{q\ell} \equiv \delta g_{I}^{q\ell}$ for $I\in \lbrace V_R,\,S_L,\,S_R,\,T \rbrace$, so that the SM case corresponds to $\smash{\delta g_I^{q\ell}=0}$ for all operators. For convenience, we also define $g_{S(P)} \equiv g_{S_R}\pm g_{S_L}$ and $g_{V(A)} \equiv g_{V_R}\pm g_{V_L}$, which will directly enter the leptonic and semileptonic processes discussed below.

The low-energy EFT coefficients can be evolved up to the electroweak scale, in which they can be matched to the SMEFT operators defined in Eq.~\eqref{eq:smeft} (see Table~\ref{tab:SMEFT-psi4-ope}),
%%%%%%%%%%%%%%%%
\begin{align}
    \label{eq:left-smeft-matching}
    \delta g_{V_L}^{q\ell} &= -\dfrac{v^2}{\Lambda^2} \sum_k \dfrac{V_{uk}}{V_{uq}}\Big{[}\mathcal{C}_{\substack{lq\\ \ell\ell kq}}^{(3)}- \mathcal{C}_{\substack{Hq\\ kq}}^{(3)}\Big{]}+\dfrac{v^2}{\Lambda^2}\mathcal{C}_{\substack{Hl\\ \ell\ell}}^{(3)}\,,\\*[0.35em]
    \delta g_{V_R}^{q\ell} &= \dfrac{v^2}{2\Lambda^2}\dfrac{1}{V_{uq}}\, \mathcal{C}_{\substack{Hud\\1q}}\,, \nonumber\\[0.35em]
    \delta g_{S_L}^{q\ell} &= -\dfrac{v^2}{2\Lambda^2}\dfrac{1}{V_{uq}}\,\mathcal{C}_{\substack{lequ\\ \ell\ell q1}}^{(1)\ast}\,, \nonumber\\*[0.35em]
    \delta g_{S_R}^{q\ell} &= -\dfrac{v^2}{2\Lambda^2} \sum_k \dfrac{V_{uk}}{V_{uq}}\,\mathcal{C}_{\substack{ledq\\ \ell\ell qk}}^\ast\,,  \nonumber\\*[0.35em]
    \delta g_{T}^{q\ell} &= -\dfrac{v^2}{2\Lambda^2}\dfrac{1}{V_{uq}}\,\mathcal{C}_{\substack{lequ\\ \ell\ell q1}}^{(3)\ast}\,,\nonumber
\end{align}
%%%%%%%%%%%%%%%%
where we have kept $\mathcal{O}(\Lambda^{-2})$, with all coefficients set at the renormalization scale $\mu=\mu_{\mathrm{ew}}$. 
From these expressions, we find that the right-handed vector coefficient is only induced at $d=6$ via Higgs-current operators, which are necessarily lepton flavor universal, i.e.,~$\delta g_{V_R}^{q\ell}\equiv \delta g_{V_R}^{q}$. Instead, the purely left-handed one ($\delta g_{V_L}$) receives contributions from four-fermion operators and both Higgs-current operators, including the purely leptonic one, $\smash{\mathcal{O}_{Hl}^{(3)} = \big{(}H^\dagger i {\overleftrightarrow{D}}_\mu^I  H\big{)}\big{(}\bar{l}_i \gamma^\mu \tau^I l_j\big{)}}$\,, which we disregard in the following as it is irrelevant for the LHC observables that we consider (see Ref.~\cite{Coutinho:2019aiy}). Finally, scalar and tensor coefficients only appear through four-fermion interactions.

%%%%%%%%%%%%%%%%%%%%%%%%%%%%%%%%%%%%%%%%%%%%%%%%%%%%%%
%%%%%%%%%%%%%%%%%%%%%%%%%%%%%%%%%%%%%%%%%%%%%%%%%%%%%%
\subsection{Flavor constraints}

In this Section, we discuss the most relevant low-energy observables that are used to constrain the $d\to u\ell \nu$ and $s\to u \ell \nu$ effective coefficients defined in Eq.~\eqref{eq:left}. For simplicity, we focus on the most robust and precise observables related to these transitions, cf.~Ref.~\cite{Cirigliano:2023nol,Crivellin:2022rhw} for more comprehensive EFT analyses.

\subsubsection{$K_{\ell2}$ and $\pi_{\ell 2}$} 

Among the cleanest probes of the Cabibbo angle are the leptonic decays $P\to \mu \nu$ with $P\in \lbrace \pi,K \rbrace$. These observables are usually combined in the following ratio that allows for a direct extraction of $|V_{us}|/|V_{ud}|$ within the SM, 
%%%%%%%%%%%%%%%%
\begin{equation}
\label{eq:Klnu-Pilnu}
 \dfrac{\Gamma(K\to\mu\nu)}{\Gamma(\pi\to\mu\nu)} \bigg{|}_{\mathrm{SM}} = \dfrac{|V_{us}|^2}{|V_{ud}|^2}\dfrac{f_K^2}{f_\pi^2}\dfrac{m_K}{m_\pi}\dfrac{(1-m_\mu^2/m_K^2)^2}{(1-m_\mu^2/m_\pi^2)^2}\,,
\end{equation}
%%%%%%%%%%%%%%% 

\noindent where $m_P$ denotes the $P$-meson mass, and $f_P$ stands for the $P$-meson decay constant defined as follows,
%%%%%%%%%%%%%%%%
\begin{equation}
\langle 0 | \bar{u}\gamma^\mu\gamma_5 q| P(p) \rangle = i p^\mu f_P \,,
\end{equation}
%%%%%%%%%%%%%%% 

\noindent which must be determined through numerical simulations of QCD on the lattice~\cite{FlavourLatticeAveragingGroupFLAG:2021npn}. Given the current theoretical and experimental precisions, isospin-breaking and QED structure-dependent corrections also have to be taken into account in Eq.~\eqref{eq:Klnu-Pilnu}~\cite{Giusti:2017dwk,DiCarlo:2019thl}. We consider the isospin-limit of the ratio $f_K/f_\pi=1.1978(22)$~\cite{Cirigliano:2022yyo} obtained through the average of the lattice QCD determinations with $N_f=2+1+1$ dynamical flavors~\cite{Dowdall:2013rya}. This value is combined with the QED corrections reported in Ref.~\cite{DiCarlo:2019thl}, allowing us to extract
%%%%%%%%%%%%%%%%
\begin{equation}
    \label{eq:Vus-Kl2}
    \dfrac{|V_{us}|}{|V_{ud}|}_{K_{\mu2}/\pi_{\mu 2}}=0.2312(6)\,,
\end{equation}
%%%%%%%%%%%%%%%

\noindent which leads to $|V_{us}|_{K_{\mu2}/\pi_{\mu 2}}=0.2252(6)$ after applying the CKM unitarity and replacing $|V_{ub}|=3.75(26)\times 10^{-3}$ from Ref.~\cite{UTfit:2022hsi} (see also Ref.~\cite{Charles:2004jd}). We stress that the value of $|V_{ub}|$ has no impact on this extraction given the  experimental and theoretical precisions. 

By using the effective Lagrangian \eqref{eq:left}, it is straightforward to compute the New Physics contributions to the $P\to\ell \nu$ branching fractions, 
%%%%%%%%%%%%%%%%
\begin{equation}
    \mathcal{B}(P\to \ell\bar{\nu})= \tau_P\, \dfrac{G_F^2 |V_{uq}|^2 f_P^2 m_{P} m_\ell^2}{8 \pi}\Bigg(1 - \dfrac{m_\ell^2}{m_P^2}\Bigg)^2\Bigg{|}g_A^{q\ell}-g_P^{q\ell}\dfrac{m_P^2}{m_\ell (m_u+m_q)}\Bigg{|}^2\,,    
\end{equation}
%%%%%%%%%%%%%%%
where $\tau_P$ denotes the $P$-meson lifetime, and the Wilson coefficients are evaluated at the scale $\mu =2$~GeV. These contributions would amount to a shift of the CKM values extracted by assuming the SM alone. Useful constraints on these effective couplings can also be obtained from the $e/\mu$ ratios,
%%%%%%%%%%%%%%%%
\begin{equation}
    \label{eq:LFU-pion}
    r_P^{(e/\mu)} \equiv \dfrac{\mathcal{B}(P\to e\nu)}{\mathcal{B}(P\to \mu\nu)}\,,
\end{equation}
%%%%%%%%%%%%%%%
which are determined to be $\smash{r_{\pi,\,\mathrm{exp}}^{(e/\mu)} =1.233(2)\times 10^{-4}}$ and $\smash{r_{K,\,\mathrm{exp}}^{(e/\mu)} =2.488(9)\times 10^{-5}}$~\cite{ParticleDataGroup:2024cfk}, in agreement with the SM predictions, namely, $r_{\pi,\,\mathrm{SM}}^{(e/\mu)} =1.2352(2)\times 10^{-4}$~\cite{Bryman:2011zz} and $r_{K,\,\mathrm{SM}}^{(e/\mu)} =2.477(1)\times 10^{-5}$~\cite{Bryman:2011zz}, implying stringent constraints on New Physics contributions that would potentially break lepton flavor universality. Further constraints can be derived on operators with the $\tau$-flavor, which we disregard in this study~\cite{Cirigliano:2018dyk}.

\subsubsection{$K_{\ell 3}$} 

The semileptonic processes $K\to \pi \ell \nu$ (with $\ell=e,\mu$) allow for a direct determination of $|V_{us}|$ in the SM, as well as complementary constraints on the $s\to u \ell \nu$ effective coefficients. The SM branching fraction can be written as~\cite{FlaviaNetWorkingGrouponKaonDecays:2010lot}
%%%%%%%%%%%%%
\begin{align}
\mathcal{B}(K\to \pi \ell\bar{\nu})_{\mathrm{SM}} = \tau_K\, \dfrac{G_F^2 m_K^5}{192 \pi^3} C_K^2 S_\mathrm{EW} |V_{us}|^2 \Big{(}1+\delta_{\mathrm{em}}^{K\ell}+\delta_{SU(2)}^{K\pi}\Big{)}\, I_{K\ell} \,,
\end{align}
%%%%%%%%%%%%%
where $C_K$ is the Clebsch-Gordan coefficient ($1$ for decays into $\pi^\pm$ and $1/\sqrt{2}$ for those to $\pi^0$), $S_\mathrm{EW}=1.0232(3)$ accounts for the short-distand electroweak corrections~\cite{Marciano:1993sh}, while $\delta_{\mathrm{em}}^{K\ell}$ and $\delta_{SU(2)}^{K\pi}$ stand for the channel-dependent electromagnetic and isospin-breaking corrections, respectively, which can be found in Ref.~\cite{Seng:2022wcw}. The phase-space integrals $I_{K\ell}$ are defined by
%%%%%%%%%%%%%
\begin{align}
I_{K\ell} = \int_{m_\ell^2}^{q^2_{\mathrm{max}}} \mathrm{d}q^2 \, \dfrac{\lambda_K^{1/2}}{m_K^4}\bigg{(}1-\dfrac{m_\ell^2}{ q^2}\bigg{)}^2\Big{[}|f_+(q^2)|^2\,\dfrac{\lambda_K}{m_K^4}\bigg{(}1+\dfrac{m_\ell^2}{2q^2}\bigg{)}+|f_0(q^2)|^2\, \dfrac{3 m_\ell^2}{2 q^2}\bigg{(}1- \dfrac{m_\pi^2}{m_K^2}\bigg{)}^2\bigg{]}\,,
\end{align}
%%%%%%%%%%%%%

\noindent where $q^2_\mathrm{max}\equiv (m_K-m_\pi)^2$, $\lambda_K\equiv (q^2-(m_K-m_\pi)^2)(q^2-(m_K+m_\pi)^2)$, and the scalar ($f_0$) and vector ($f_+$) form factors are defined as follows,
%%%%%%%%%%%%%
\begin{align}
\langle \pi^-(k) | \bar{s}\gamma^\mu u | K^0(p) \rangle = \Big{[}(p+k)^\mu -\dfrac{m_K^2-m_\pi^2}{q^2}\, q^\mu\Big{]}f_+(q^2) + \dfrac{m_K^2-m_\pi^2}{q^2}\,q^\mu f_0(q^2)\,,
\end{align}
%%%%%%%%%%%%%
where $q^2=(p-k)^2$ is the dilepton invariant mass. In our analysis, we use the $q^2$-shapes of $f_0$ and $f_+$ reported in Ref.~\cite{Carrasco:2016kpy} with $N_f = 2 + 1 + 1$ dynamical flavors. For the overall form-factor normalization, we use the FLAG average $f_+(0)=f_0(0)=0.9698(17)$~\cite{FlavourLatticeAveragingGroupFLAG:2024oxs}, which is dominated by the results from ETMC~\cite{Carrasco:2016kpy} and MILC/Fermilab~\cite{FermilabLattice:2018zqv}. By using these theoretical inputs and the experimental determination of the $K_{\ell 3}$ branching fractions summarized in Ref.~\cite{Moulson:2017ive}, we obtain~\footnote{Notice, in particular, that $K^- \to \pi^0 \ell\nu$ data (with $\ell=e,\mu$) has several inconsistencies, as noted in Ref.~\cite{Becirevic:2020rzi}. Indeed, the fitted value reported in Ref.~\cite{Moulson:2017ive} for these modes are considerably larger than the PDG averages, although their ratio is consistent with the SM~\cite{ParticleDataGroup:2024cfk}. We believe that further discussion and updated experimental analyses are essential in assessing the correct values of the $K_{\ell 3}$ branching fractions. }
%%%%%%%%%%%%%%%%%
\begin{equation}
\label{eq:Vus-Kl3}
|V_{us}|_{K_{\ell3}} =0.2233(8)\,,
\end{equation}
%%%%%%%%%%%%%%%%
which is about ${2}\sigma$ smaller than the results from $K_{\ell 2}/\pi_{\ell 2}$ given in Eq.~\eqref{eq:Vus-Kl2}. Notice, also, that our value of $\smash{|V_{us}|_{K_{\ell3}}}$ is consistent with the results from Ref.~\cite{Cirigliano:2023nol,Crivellin:2022rhw}, but with slightly larger uncertainties due to our choice of fixing the $K\to \pi$ form-factor $q^2$-shapes with lattice QCD data, instead of using the $K_{\ell3}$ differential rates that are experimentally determined.

New Physics contributions to these processes can be easily computed using, e.g., the expression provided in Ref.~\cite{Becirevic:2020rzi}, which are summarized in Appendix \ref{app:KL3BSM}. Similarly to the leptonic decays, there are useful bounds derived from lepton-flavor-universality ratios~\cite{Becirevic:2020rzi},
%%%%%%%%%%%%%%%%%
\begin{equation}
\label{eq:LFU-Kl3}
R_{K\pi}^{\mu/e}\equiv \dfrac{\mathcal{B}(K\to \pi\mu\nu)}{\mathcal{B}(K\to \pi e   \nu)}\,,
\end{equation}
%%%%%%%%%%%%%%%%
which are independent of $|V_{us}|$. The comparison between the SM predictions and the current experimental determinations is shown below~\cite{ParticleDataGroup:2024cfk},
%%%%%%%%%%%%%%%%
\begin{align}
R_{K^-\pi^0}^{\mu/e}\big{|}_{\mathrm{exp}} &=0.663(2)\,,\qquad\quad  R_{K^-\pi^0}^{\mu/e}\big{|}_{\mathrm{SM}} =0.662(3)\,,\\[0.35em]
R_{K_L\pi^-}^{\mu/e}\big{|}_{\mathrm{exp}} &=0.662(4)\,,\qquad\quad  R_{K_L\pi^-}^{\mu/e}\big{|}_{\mathrm{SM}} =0.666(2)\,, \noindent
\end{align}
%%%%%%%%%%%%%%%
which are in perfect agreement, setting stringent constraints on four-fermion interactions breaking $\mu/e$ universality~\cite{Becirevic:2020rzi}.

\subsubsection{$\beta$-decays} 

Finally, we briefly discuss the limits from neutron and nuclear $\beta$-decays. These observables allow us to extract $V_{ud}$ within the SM, showing an apparent departure from CKM unitarity when combined with the $V_{us}$ values from kaon observables discussed above~\cite{Cirigliano:2022yyo}. In this analysis, we follow closely Ref.~\cite{Falkowski:2020pma}, taking into account the most precise observables, namely the neutron lifetime, neutron angular correlations,  and {\it superallowed} $0^+ \to 0^+$ nuclear transitions.~\footnote{See Ref.~\cite{Falkowski:2020pma} for a more comprehensive analysis that included, in particular, data from mirror $\beta$ transitions. These processes have a minor impact on the EFT scenarios that we consider.} 

The neutron lifetime provides one of the cleanest determinations of $V_{ud}$ from the theory perspective, as it is not affected by nuclear effects. The SM prediction of the neutron width can be written as~\cite{Falkowski:2020pma},
%%%%%%%%%%%%%
\begin{align}
    \Gamma_n \equiv \tau_n^{-1} = (1+\delta_R^{\prime \, n})\dfrac{G_F^2 \vert V_{ud}\vert^2 m_e^5 f_V^n}{2\pi^3}\Big[g_V^2\,(1+\Delta_R^V) +3g_A^2\, (1+\Delta_R^A)\Big]\,, 
\end{align}
%%%%%%%%%%%%%
where $g_V=1$ and $g_A=1.263(10)$~\cite{FlavourLatticeAveragingGroupFLAG:2024oxs} are the vector and axial charges, $f_V^n=1.6887(1)$ is the phase-space factor~\cite{Czarnecki:2018okw}, short-distance corrections are given by $\Delta_R^{V} = 0.02467(22)$~\cite{Seng:2018yzq} and $\Delta_R^{A} = 0.02493(34)$~\cite{Gorchtein:2021fce}, whereas the long-distance radiative correction are estimated to be $\delta_R^{\prime \, n} =1.4902(2)\%$~\cite{Towner:2010zz}. Using the PDG average for the neutron lifetime, $\tau_n^{\rm exp} = 878.4(5)~\mathrm{s}$~\cite{ParticleDataGroup:2024cfk}, that is dominated by the recent results from Ref.~\cite{UCNt:2021pcg}, together with the angular observables described in Ref.~\cite{Falkowski:2020pma}, we obtain 
%%%%%%%%%%%%%
\begin{align}
|V_{ud}|_{n}=0.97441(45)\,,
\end{align}
%%%%%%%%%%%%%
which leads to $|V_{us}|_{n}=0.2248(20)$ after applying CKM unitarity. 

The usual approach adopted in the literature is to combine the above determination based on the neutron lifetime with the measurements from {\it superallowed} $0^+\to 0^+$ nuclear $\beta$ decays, which are described by a pure Fermi transition in the SM. The latter provides a more precise extraction of $|V_{ud}|$, although it requires a careful assessment of nuclear-structure-dependent effects given the current level of precision. The experimental results for the transitions are reported in terms of the correct half-life $\mathcal{F}t$, which is universal among all {\it superallowed} transitions within the SM~\cite{Hardy:2020qwl}
%%%%%%%%%%%%%%%%%
\begin{equation}
    \mathcal{F}t_i \equiv \dfrac{f_V\,(1+\delta_i)\log2}{\Gamma_i} = \bigg[ \dfrac{M_F^2\,G_F^2 \vert V_{ud}\vert^2 m_e^5(1+\Delta_R^V)}{2 \pi^3\log2} \bigg]^{-1}\,,
\end{equation}
%%%%%%%%%%%%%%%%

\noindent where 
$M_F=\sqrt{2}$ is the Fermi matrix element in the isospin limit for $0^+\to 0^+$ transitions. The $\delta_i$ factor encodes radiative corrections that are split as $1+\delta_i\equiv  (1+\delta_R^{\prime\,i})(1+\delta_{\mathrm{NS}}^{Vi}-\delta_C^{Vi})$, where $\delta_R^{\prime i}$ represents long-distance corrections, $\delta_{\mathrm{NS}}^{Vi}$ is the nuclear-structure dependent piece and $\delta_C^{Vi}$ stand for isospin-breaking corrections. We follow the approach, and use the theoretical and experimental inputs described in Ref.~\cite{Falkowski:2020pma} (cf.~also~Ref.~\cite{Cirigliano:2023nol}), which is based on the $\mathcal{F}t$ values reported in Ref.~\cite{Hardy:2020qwl}. We then obtain
%%%%%%%%%%%%%%%%%
\begin{equation}
\label{eq:Vud-beta-nuclear}
|V_{ud}|_{0^+\to 0^+} =0.97366(29)\,,
\end{equation}
%%%%%%%%%%%%%%%%
which is more precise than the determination based on the neutron lifetime, leading to $|V_{us}|_{0^+\to 0^+}=0.2280(12)$ within the SM. Finally, by combining the $0^+\to 0^+$ transitions lifetimes with neutron decay data, we obtain
%%%%%%%%%%%%%%%%%
\begin{equation}
\label{eq:Vud-beta-decay}
|V_{ud}|_{\beta} ={0.97388(24)}\,,
\end{equation}
%%%%%%%%%%%%%%%%
which can be translated into $|V_{us}|_{\beta}=0.2271(10)$ through CKM unitarity. This value is larger than the ones derived from leptonic and semileptonic kaon decays, leading to a discrepancy in the determination of the Cabibbo angle.~\footnote{We note, in particular, that the discrepancy between kaon and $\beta$-decay is more pronounced if {\it superallowed} transitions are considered, as noted before in Ref.~\cite{Cirigliano:2022yyo} and references therein.
} New Physics contributions to both types of $\beta$-decays can be computed using the expression provided in Ref.~\cite{Falkowski:2020pma} that we consider in our numerical analysis.

\subsubsection{Numerical analysis}

We consider now the flavor observables described above, which are used to simultaneously constrain the CKM input $\lambda \equiv |V_{us}|$ and the effective coefficients $\delta g_I^{q\ell}$ entering this transition, as defined in Eq.~\eqref{eq:left}. While several Wilson coefficients can potentially contribute to the $d\to u\ell\nu$ and $s\to u\ell\nu$ transitions, several constraints make it challenging to account for these discrepancies within an EFT invariant under the SM gauge symmetry $SU(3)_c\times SU(2)_L\times U(1)_Y$\,:
%%%%%%%%%%%%%%
\begin{itemize}
    \item[i)] Semileptonic operators ($\psi^4$) are constrained by $\mu/e$ universality tests in the leptonic and semileptonic ratios defined in Eqs.~\eqref{eq:LFU-pion}. Furthermore, they lead to energy-enhanced effects in $pp\to \ell \nu$ processes at high-$p_T$ that supersede low-energy bounds for specific effective coefficients~\cite{Allwicher:2022gkm,Allwicher:2022mcg}, see also Ref.~\cite{Cirigliano:2018dyk}. For instance, this is the case for the tensor ones,~$\delta g_T^{q\ell}$.
    \item[ii)] The Higgs-current operators ($\psi^2 H^2 D$) evade the above-mentioned constraints, but are subject to relevant constraints from EWPO, in particular to left-handed operators, as discussed in Sec.~\ref{sec:EWPO} (cf.~Fig.~\ref{fig:EWPOvsLHC}).
    \item[iii)] For both $\psi^4$ and $\psi^2 H^2 D$ types of operators, the ones involving left-handed quarks will induce effects in both charged- and neutral-current transitions. The latter are tightly constrained by $\Delta F=1$ and $\Delta F=2$ processes. While certain combinations of effective operators can suppress FCNCs in the down-quark sector (e.g., $\Delta m_K$), there are still constraints from charm physics that must be taken into account in these scenarios, such as $\Delta m_D$~\cite{Crivellin:2022rhw}.    
\end{itemize}
%%%%%%%%%%%%%%

\noindent The above considerations make it clear that the best possibility for accommodating these discrepancies is Higgs-current operators with right-handed quarks, i.e.~$\mathcal{O}_{Hud}$, as they evade the constraints described above. This aligns with the findings of detailed global analyses, which confirm that such a scenario provides the best description of current data, whereas other scenarios can only alleviate the observed tensions~\cite{Crivellin:2022rhw,Cirigliano:2023nol}.

\begin{figure}[!t]
\begin{center}
    \includegraphics[width=1\textwidth]{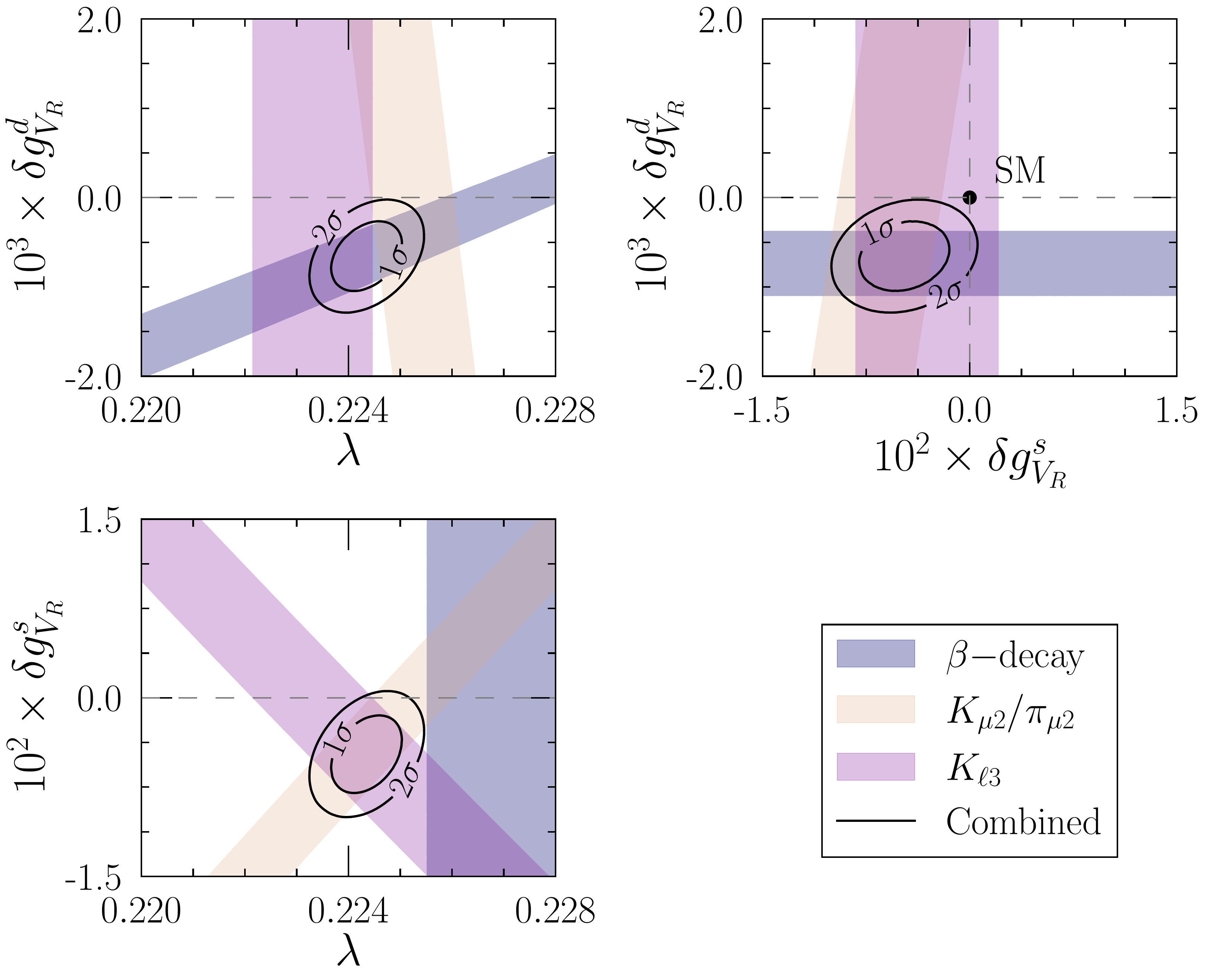}\\[0.35em]
    \caption{\small \sl Low-energy constraints on the $\lambda\equiv |V_{us}|$ and the Wilson coefficients $\smash{\delta g_{V_L}^{q} \equiv \delta g_{V_L}^{q\ell}}$ and $\smash{\delta g_{V_R}^{q} \equiv \delta g_{V_R}^{q\ell}}$ (for $\ell=e,\mu$), which are assumed to be lepton-flavor universal. The $1\sigma$ allowed regions are depicted for $K_{\mu2}/\pi_{\mu2}$ (yellow), $K_{\ell 3}$ (magenta) and $\beta$-decays (blue), with their combination to $1\sigma$ and $2\sigma$ accuracies depicted by the black lines. For the individual limits, the coefficients not displayed in the panels were set to zero in the first column, while in the panel in the second column, the value of $\lambda$ was fixed to the best-fit value from the combined fit.}
    \label{fig:flavor-fit}
\end{center}
\end{figure}

In Fig.~\ref{fig:flavor-fit}, we perform a low-energy fit to $\lbrace \lambda\,;~\delta g_{V_R}^{d }\,,~\delta g_{V_R}^{s }\rbrace$, where $\delta g_{V_R}^{q} \equiv  g_{V_R}^{q\ell}$ (with $\ell=e,\mu$), as predicted by the $\mathcal{O}_{Hud}$ operator. In each panel, we provide the individual constraints from $K_{\ell 2}/\pi_{\ell 2}$, $K_{\ell 3}$ and $\beta$-decays, as well as their combination. We confirm that such a scenario can reconcile the various observables sensitive to $\lambda$. In the following, this scenario will be used as a benchmark for comparing low- and high-energy bounds (see also Ref.~\cite{Alioli:2017ces}).

%%%%%%%%%%%%%
%%%%%%%%%%%%%

%%%%%%%%%%%%%
\begin{figure}[!t]
\begin{center}
    \includegraphics[width=0.62\textwidth]{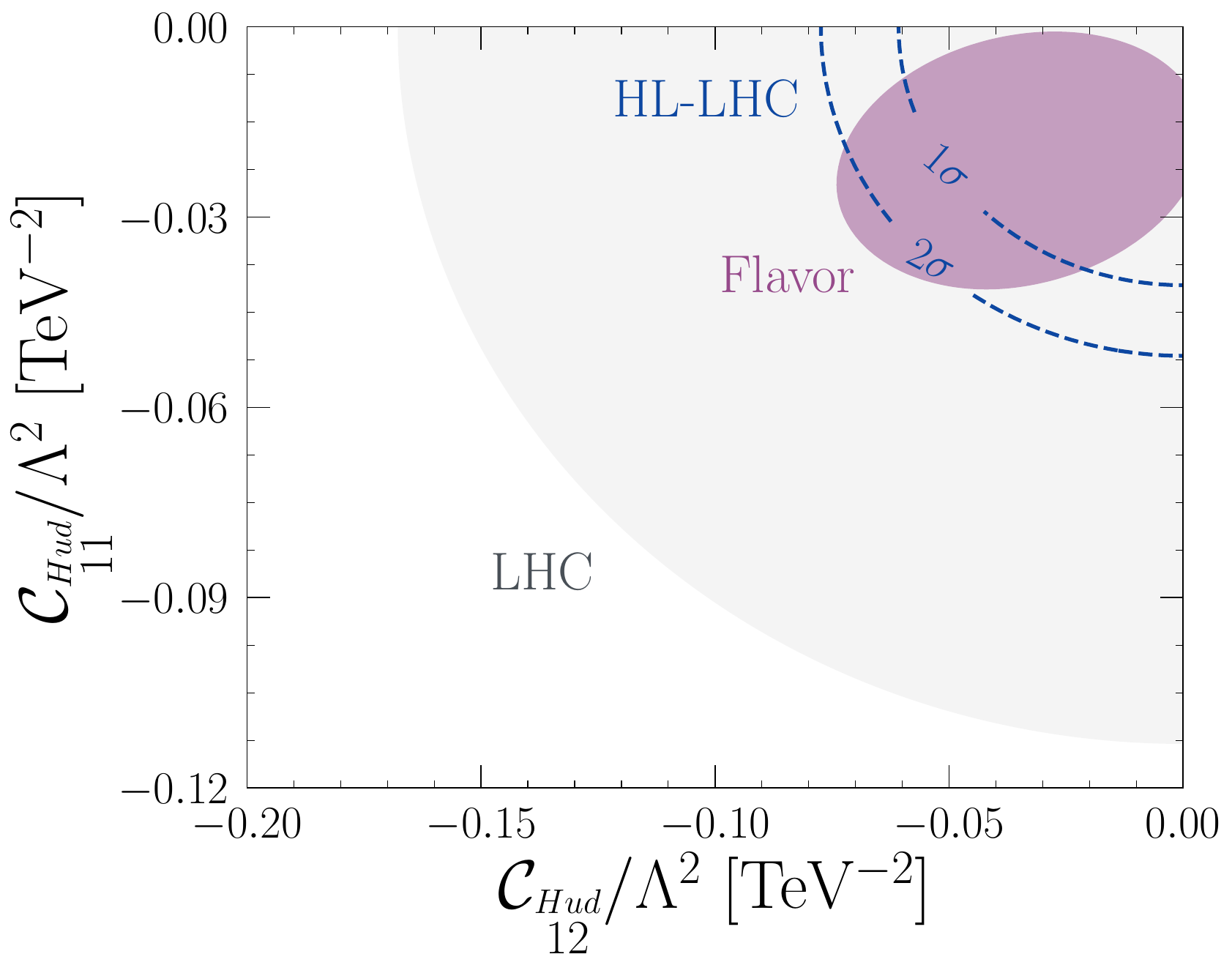}
    \caption{\small \sl Comparison of current LHC (gray) and flavor (purple) constraints at $95\%$~CL for the right-handed operators entering the $d\to u \ell\nu$ and $s\to u\ell\nu$ transitions. The projections to HL-LHC are depicted by the dashed lines, which will allow us to test the region currently allowed by low-energy bounds.}
    \label{fig:lhc-plot-VH}
\end{center}
\end{figure}
%%%%%%%%%%%%%

%%%%%%%%%%%%%%%%%%%%%%%%%%%%%%%%%%%%%%%%%%%%%%%%%%%%%%
%%%%%%%%%%%%%%%%%%%%%%%%%%%%%%%%%%%%%%%%%%%%%%%%%%%%%%
\subsection{Comparing low- and~high-energy searches}

We now compare the low-energy limits described above with the LHC constraints provided in Sec.~\ref{sec:numerical} for the scenario with right-handed currents, which is preferred by current data, as discussed above. We consider the $\mathcal{O}_{Hud}$ operators with $11$ and $12$ flavor indices, which generate the $W_\mu\,(\bar{u}_R \gamma^\mu d_R)$ and $W_\mu\,(\bar{u}_R \gamma^\mu s_R)$ interactions, respectively. We focus on the associated Higgs production, as it provides the most stringent constraint on this type of operator, cf.~Fig.~\ref{fig:limits-Wcouplings}. Notice, also, that electroweak constraints are irrelevant for these right-handed operators, as discussed in Sec.~\ref{sec:EWPO} (see Fig.~\ref{fig:EWPOvsLHC}).

The comparison between low- and high-energy bounds for this scenario is presented in Fig.~\ref{fig:lhc-plot-VH}. We find that our LHC limits are not yet competitive with flavor data for this specific quark-level transition, but that the projections to HL-LHC (with $3~\mathrm{ab}^{-1}$) will probe regions that are currently allowed by flavor data.~\footnote{The HL-LHC projections for the experimental sensitivity were taken from Ref.~\cite{ATLAS:2025wcz}.} However, we note that there are other charged-current transitions for which the LHC constraints derived in this study can already supersede flavor bounds~\cite{charm-paper}.

We stress once again that the improvement with respect to previous studies comes mostly from the differential data that have been provided in Ref.~\cite{ATLAS:2024yzu,CMS:2023vzh}, which allow us to better disentangle potential EFT signatures from the SM background at the high-energy tails. Another direction that is worth exploring in the future is the measurement of angular distributions of these processes, which is potentially available with current LHC data and has the potential to distinguish among the different EFT operators~\cite{Alioli:2017ces}.

%%%%%%%%%%%%%%%%%%%%%%%%%%%%%%%%%%%%%%%%%%%%%%%%%%%%%%
%%%%%%%%%%%%%%%%%%%%%%%%%%%%%%%%%%%%%%%%%%%%%%%%%%%%%%
\subsection{From EFTs to concrete models}

%%%%%%%%%%%%%
\begin{figure*}[!t]
    \centering
        \includegraphics[width=0.49\textwidth]{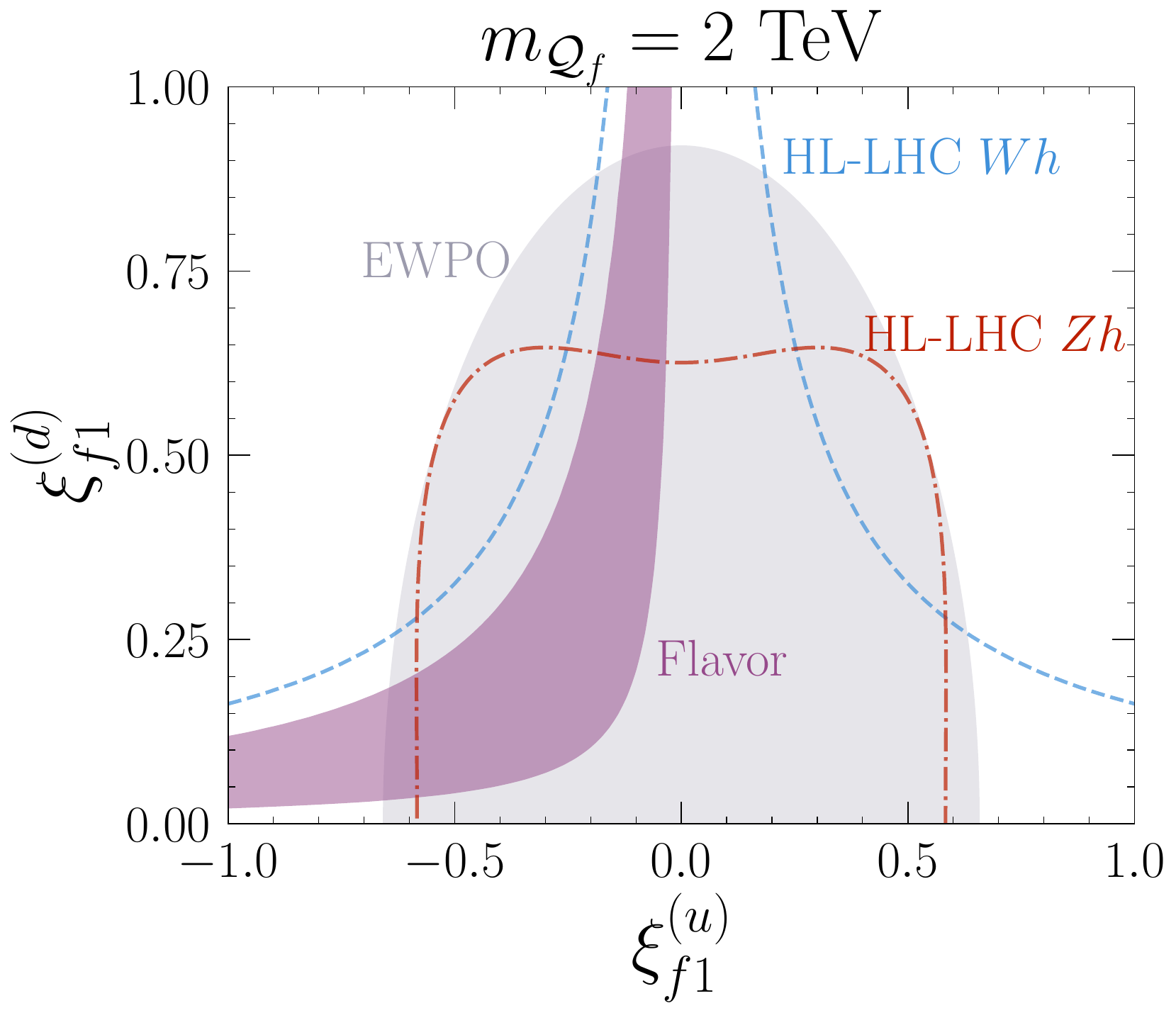}
    ~\includegraphics[width=0.49\textwidth]{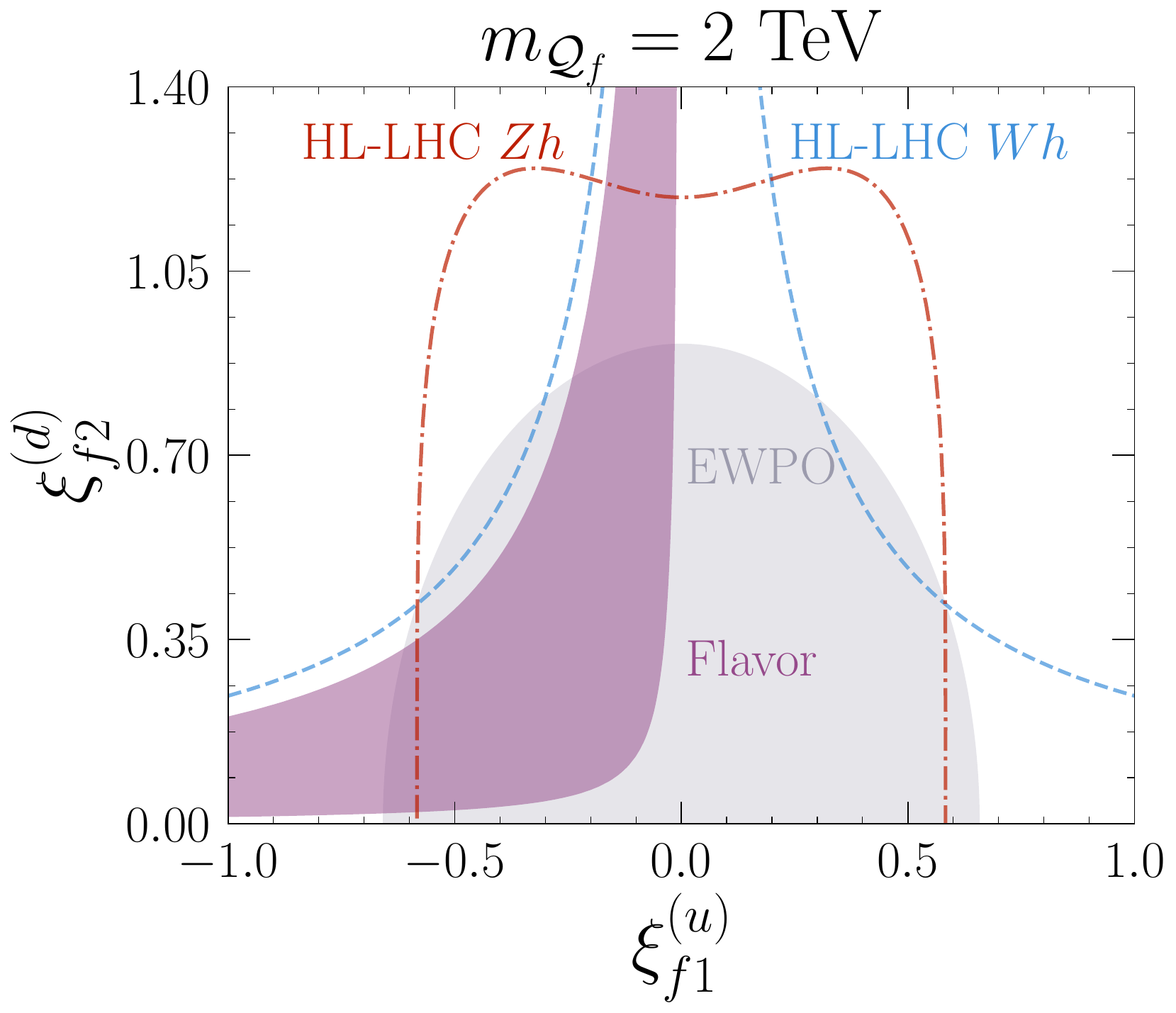}
    
    % \label{fig:limits-Wcouplings}
    \caption{\small \sl Constraints on the couplings of a vector-like quark $\mathcal{Q}\sim(\mathbf{3},\mathbf{2},1/6)$ with couplings exclusively to first-generation SM quarks (left panel), and with couplings to first-generation up-type and second-generation down-type quarks (right panel); cf.~Eq.~\eqref{eq:vlq-lag}. The vector-like quark mass is set to $m_{\mathcal{Q}}=2$~TeV and the couplings displayed are set to zero. The constraints from flavor (magenta) and electroweak (gray) data at $68\%$ CL are superimposed with our HL-LHC projections for $pp\to Wh$ (dashed blue) and $pp\to Zh$ (dashed-dotted red). See text for further details.}
    \label{fig:VLQ-limits}
\end{figure*}
%%%%%%%%%%%%%

In this Section, we extend the previous EFT analysis to concrete models that can generate the viable operators identified above. We will consider the minimal scenario identified in Ref.~\cite{Crivellin:2022rhw,Cirigliano:2023nol} (see also Ref.~\cite{Belfatto:2021jhf}), namely a model with a TeV-scale vector-like quark $\mathcal{Q}_f\sim (\mathbf{3},\mathbf{2},1/6)$, i.e., with the same quantum numbers as SM left-handed quarks, and a possible flavor index $f$. The interactions of these particles with the Higgs boson and quarks read,
%%%%%%%%%%%%%
\begin{align}
\label{eq:vlq-lag}
\mathcal{L}_{\mathrm{VLQ}} \supset -  \xi_{fi}^{(u)}\, \overline{\mathcal{Q}}_f \widetilde{H} u_i -  \xi_{fi}^{(d)}\, \overline{\mathcal{Q}}_f H d_i +\mathrm{h.c.}\,,
\end{align}
%%%%%%%%%%%%%

\noindent where the summation over the SM quark flavor index $i$ is omitted. As noted in Ref.~\cite{Crivellin:2022rhw,Cirigliano:2023nol}, the vector-like quarks should couple exclusively to first- or second-generation quarks to avoid large contributions to $K-\overline{K}$ mixing. The above Lagrangian can be matched at tree-level to the SMEFT, generating the needed dimension-six operators~\cite{Crivellin:2022rhw},
%%%%%%%%%%%%%
\begin{align}
\mathcal{C}_{\substack{Hud\\ij}}&=\dfrac{\xi_{fj}^{(d)} \xi_{fi}^{(u)\,\ast}}{ m_{\mathcal{Q}_f}^2}\,, \qquad \quad
\mathcal{C}_{\substack{Hu\\ij}}=-\dfrac{\xi_{fj}^{(u)} \xi_{fi}^{(u)\,\ast}}{2 m_{\mathcal{Q}_f}^2}\,, \qquad \quad
\mathcal{C}_{\substack{Hd\\ij}}=\dfrac{\xi_{fj}^{(d)} \xi_{fi}^{(d)\,\ast}}{2 m_{\mathcal{Q}_f}^2}\,, 
\end{align}
%%%%%%%%%%%%%

\noindent where the light-quark masses are neglected and there is an implicit summation on the vector-like quark index $f$. For concreteness, we consider in the following a single generation of vector-like quarks that couples either to $i=d$ or $i=s$.

The vector-like quark model described above generates not only the needed contribution to accommodate charged-current data, but also neutral-current interactions that can be probed by independent processes at low- and high-energies. The minimal set of couplings to reproduce low-energy data is $\lbrace \xi^{(u)}_{f1},\,\xi^{(d)}_{f1} \rbrace$ or  $\lbrace \xi^{(u)}_{f1},\,\xi^{(d)}_{f2} \rbrace$, depending whether the $d\to u \ell \nu$ or the $s\to u \ell \nu$ transition is modified~\cite{Crivellin:2022rhw}. The vector-like quark masses are conservatively set to $2$~TeV to comply with direct searches constraints~\cite{ATLAS:2024zlo}. Several independent constraints apply to the model specified above, such as electroweak precision data, as described in Sec.~\ref{sec:EWPO}. The combination of these constraints with our HL-LHC projections for $pp\to Vh$ is shown in Fig.~\ref{fig:VLQ-limits}. We find that the $pp\to Zh$ process will provide the dominant LHC constraint for this scenario, which will be competitive with electroweak data and probe viable flavor parameters in the HL-LHC phase, in particular for couplings to first-generation quarks.

\section{Summary and outlook}
\label{sec:conclusion}

In this paper, we have demonstrated that associated Higgs production with an electroweak boson, $pp\to Vh$ (with $V=W,Z$), and diboson production, $pp\to VW$, are sensitive to EFT contributions beyond the restrictive flavor assumptions previously considered in the literature, such as $U(3)^5$ symmetry and the MFV ansatz. We have shown that the limits derived from current LHC data on these processes are complementary to those extracted from electroweak and low-energy flavor data.

To this end, we have considered the complete set of dimension-six operators in the SMEFT that can induce flavor transitions in these processes and that can be generated at tree-level in weakly coupled UV scenarios. Under these assumptions, our analysis focuses on Higgs–current operators with quarks ($\psi^2 H^2 D$), which modify the couplings of the $W$- and $Z$-bosons to quarks. In Sec.~\ref{sec:helicity}, we have performed a systematic analysis of their impact on $pp \to Vh$ and $pp \to VW$, deriving the corresponding helicity amplitudes, with a general flavor structure, in the high-energy limit. In Sec.~\ref{sec:lhc}, constraints on these operators have been obtained using available Run-2 LHC data, and projections have been made for the HL-LHC phase.

In our numerical analysis, we find that associated Higgs production channels provide the most stringent constraints on Higgs-current operators at the LHC. Furthermore, we have shown that the limits from diboson production are only a factor of $\approx 2$ weaker for searches with the same integrated luminosity ($140~\mathrm{fb}^{-1}$), which can be understood from the Goldstone equivalence theorem. We stress that the differential distributions now available in experimental analyses were essential for improving upon previous studies, which relied on Higgs signal-strength data.

The limits derived in Sec.~\ref{sec:lhc} are complementary to those derived from Drell-Yan processes at high-$p_T$, which are in turn mostly sensitive to semileptonic $\psi^4$ operators, due to the energy enhancement induced by the EFT contributions \cite{Allwicher:2022gkm}, cf.~Table~\ref{tab:SMEFT-ope-energy}. Furthermore, the LHC processes considered in this paper are complementary to EWPO, as well as to various low-energy probes that can be used to test the CKM mechanism and modifications of the electroweak boson couplings to quarks. This synergy was explored in Sec.~\ref{sec:EWPO}, where we have shown, in particular, that $Vh$ production is essential to probe combinations of effective coefficients orthogonal to those constrained by electroweak data.

Finally, we have confronted our LHC constraints to flavor observables for a concrete example in Sec.~\ref{sec:illustration}, namely the apparent discrepancies in the determinations of the first row of the CKM matrix.  Indeed, the determinations of $|V_{ud}|$ and $|V_{us}|$ from $\beta$-decay and (semi)leptonic kaon decays, respectively, have mild disagreements, which could be accommodated by New Physics effects in a Higgs-current operators made of first and/or second-generation right-handed quarks~\cite{Cirigliano:2023nol,Crivellin:2022rhw}. We have shown that current constraints from LHC are not yet competitive with those derived from electroweak and flavor observables, but future HL-LHC data will allow us to probe unconstrained values of Wilson coefficients that could explain such a discrepancy. 

The examples discussed above highlight the powerful complementarity of $pp \to Vh$ and $pp \to VW$ with low-energy probes, a synergy we plan to extend to other transitions that are under experimental scrutiny in flavor experiments~\cite{charm-paper}. We also aim to incorporate the LHC constraints derived in this work into forthcoming releases of the {\tt HighPT} Mathematica package, which was developed for flavor-physics phenomenology at the LHC~\cite{Allwicher:2022mcg}. This will be a further step toward constructing a complete SMEFT likelihood based on low- and high-energy processes.

%%%%%%%%%%%%%%%%%%%%%%%%%%%%%%%%%%%%%%%%%%%%%%%%%%%%%%
%%%%%%%%%%%%%%%%%%%%%%%%%%%%%%%%%%%%%%%%%%%%%%%%%%%%%%
\section*{Acknowledgments}

 We thank G.~Gagliardi, F.~Mescia, M.~Kirk, and L.~Cadamuro for useful discussions, and L.~Allwicher for assistance with the interpretation of electroweak precision observables. This project has received support from the European Union’s Horizon 2020 research and innovation programme under the Marie Skłodowska-Curie grant agreement N$^\circ$~860881-HIDDeN and N$^\circ$~101086085-ASYMMETRY, from the IN2P3 (CNRS) Master Project HighPTflavor, from the SPRINT/CNRS agreement supported by FAPESP under Contract No.~2023/00643-0 and from the USP-COFECUB project Uc Ph194-2. M.M.~acknowledges the support of Fundação de Amparo à Pesquisa do Estado de São Paulo (FAPESP) under the grant numbers 2022/11293-8 and 2024/04246-9 while L.L. acknowledges the support of FAPESP under the grant number 2021/02283-6. OJPE is partially supported by CNPq grant number 302120/2025-4 and FAPESP grant 2019/04837-9.

\appendix

\section{$K_{\ell3}$ expressions}
\label{app:KL3BSM}

For completeness, we provide in this Appendix the expressions for $K\to\pi \ell\nu$ decays in the presence of the dimension-six operators defined in Eq.~\eqref{eq:LEFT-lag}~\cite{Becirevic:2020rzi},
%%%%%%%%%%%%%
\begin{align}
    \dfrac{\mathrm{d}\mathcal{B}}{\mathrm{d}q^2}(K\to \pi \ell \nu) &= \mathcal{N}_K(q^2)\,\lambda^{1/2}_K \Big(1-\dfrac{m_\ell^2}{q^2}\Big)^2\\[0.3em]
    &\times \Bigg\{ |g_V^{s\ell}|^2\left[\lambda_K\Big(1+\dfrac{m_\ell^2}{2q^2}\Big) |f_+(q^2)|^2 + \dfrac{3\, m_\ell^2\, m_K^4}{2\, q^2}\Big(1-\dfrac{m_\pi^2}{m_K^2}\Big)^2|f_0(q^2)|^2\nonumber\right]\\[0.3em]
    &+|g_T^{s\ell}|^2 \dfrac{8\,q^2}{(m_K+m_\pi)^2} \lambda_K\Big(1+\dfrac{2m_\ell^2}{q^2}\Big)|f_T(q^2)|^2\nonumber\\[0.3em]
    &+|g_S^{s\ell}|^2\left[ \dfrac{3\, m_K^4 \,q^2}{2\,(m_s-m_u)^2} \Big(1-\dfrac{m_\pi^2}{m_K^2}\Big)^2|f_0(q^2)|^2\right]\nonumber\\[0.3em]
    &+{\rm Re}[g_V^{s\ell}g_T^{s\ell\ast}]\, f_T(q^2) f_+(q^2) \dfrac{12\, m_\ell}{m_K+m_\pi}\lambda_K\nonumber\\[0.3em]
    &+{\rm Re}[g_V^{s\ell}g_S^{s\ell*}]\, |f_0(q^2)|^2 \dfrac{3\, m_\ell \, m_K^4}{(m_s-m_u)}\Big(1-\dfrac{m_\pi^2}{m_K^2}\Big)^2 \Bigg\}\nonumber\,,
\end{align}
%%%%%%%%%%%
where $f_T(q^2)$ is the $K \rightarrow\pi$ tensor form-factor \cite{Baum:2011rm}, and we have defined 
%%%%%%%%%%%%%%%%
\begin{equation}
\mathcal{N}_K(q^2) \equiv \tau_K \dfrac{G_F^2 |V_{us}|^2}{192\pi^3 m_K^3}  C_K^2 S_{\rm EW}\big{(}1+\delta_{\rm EM}^{K \ell}+\delta_{\rm SU(2)}^{K \pi}\big{)}\,,
\end{equation}
%%%%%%%%%%%%%%%%
and $\lambda_{K} \equiv (q^2-(m_K-m_\pi)^2)(q^2-(m_K+m_\pi)^2)$.


\begin{thebibliography}{99}

%\cite{CMS:2012qbp}
\bibitem{CMS:2012qbp}
S.~Chatrchyan \textit{et al.} [CMS],
%``Observation of a New Boson at a Mass of 125 GeV with the CMS Experiment at the LHC,''
Phys. Lett. B \textbf{716} (2012), 30-61
%doi10.1016/j.physletb.2012.08.021
[arXiv:1207.7235 [hep-ex]];
%15996 citations counted in INSPIRE as of 23 Jul 2025
%\cite{ATLAS:2012yve}
%\bibitem{ATLAS:2012yve}
G.~Aad \textit{et al.} [ATLAS],
%``Observation of a new particle in the search for the Standard Model Higgs boson with the ATLAS detector at the LHC,''
Phys. Lett. B \textbf{716} (2012), 1-29
%doi10.1016/j.physletb.2012.08.020
[arXiv:1207.7214 [hep-ex]].
%16544 citations counted in INSPIRE as of 23 Jul 2025

%\cite{Hagiwara:1986vm}
\bibitem{Hagiwara:1986vm}
K.~Hagiwara, R.~D.~Peccei, D.~Zeppenfeld and K.~Hikasa,
%``Probing the Weak Boson Sector in e+ e- ---{\ensuremath{>}} W+ W-,''
Nucl. Phys. B \textbf{282} (1987), 253-307
%doi10.1016/0550-3213(87)90685-7
%1534 citations counted in INSPIRE as of 21 Jul 2025

%\cite{Hagiwara:1993ck}
\bibitem{Hagiwara:1993ck}
K.~Hagiwara, S.~Ishihara, R.~Szalapski and D.~Zeppenfeld,
%``Low-energy effects of new interactions in the electroweak boson sector,''
Phys. Rev. D \textbf{48} (1993), 2182-2203
%doi10.1103/PhysRevD.48.2182
%772 citations counted in INSPIRE as of 21 Jul 2025

%\cite{Han:2009em}
% \bibitem{Han:2009em}
% T.~Han, D.~Krohn, L.~T.~Wang and W.~Zhu,
% %``New Physics Signals in Longitudinal Gauge Boson Scattering at the LHC,''
% JHEP \textbf{03} (2010), 082
% %doi10.1007/JHEP03(2010)082
% [arXiv:0911.3656 [hep-ph]].
%66 citations counted in INSPIRE as of 11 Jul 2025

%\cite{Degrande:2012wf}
\bibitem{Degrande:2012wf}
C.~Degrande, N.~Greiner, W.~Kilian, O.~Mattelaer, H.~Mebane, T.~Stelzer, S.~Willenbrock and C.~Zhang,
%``Effective Field Theory: A Modern Approach to Anomalous Couplings,''
Annals Phys. \textbf{335} (2013), 21-32
%doi10.1016/j.aop.2013.04.016
[arXiv:1205.4231 [hep-ph]].
%378 citations counted in INSPIRE as of 21 Jul 2025

%\cite{Corbett:2012dm}
\bibitem{Corbett:2012dm}
T.~Corbett, O.~J.~P.~Eboli, J.~Gonzalez-Fraile and M.~C.~Gonzalez-Garcia,
%``Constraining anomalous Higgs interactions,''
Phys. Rev. D \textbf{86} (2012), 075013
%doi10.1103/PhysRevD.86.075013
[arXiv:1207.1344 [hep-ph]];
%223 citations counted in INSPIRE as of 11 Jul 2025
%\cite{Corbett:2013pja}
%\bibitem{Corbett:2013pja}
T.~Corbett, O.~J.~P.~{\'E}boli, J.~Gonzalez-Fraile and M.~C.~Gonzalez-Garcia,
%``Determining Triple Gauge Boson Couplings from Higgs Data,''
Phys. Rev. Lett. \textbf{111} (2013), 011801
%doi10.1103/PhysRevLett.111.011801
[arXiv:1304.1151 [hep-ph]].
%107 citations counted in INSPIRE as of 11 Jul 2025

%\cite{Greljo:2015sla}
\bibitem{Greljo:2015sla}
A.~Greljo, G.~Isidori, J.~M.~Lindert and D.~Marzocca,
%``Pseudo-observables in electroweak Higgs production,''
Eur. Phys. J. C \textbf{76} (2016) no.3, 158
%doi10.1140/epjc/s10052-016-4000-5
[arXiv:1512.06135 [hep-ph]];
%55 citations counted in INSPIRE as of 15 Jul 2025
%\cite{Isidori:2013cga}
%\bibitem{Isidori:2013cga}
G.~Isidori and M.~Trott,
%``Higgs form factors in Associated Production,''
JHEP \textbf{02} (2014), 082
%doi:10.1007/JHEP02(2014)082
[arXiv:1307.4051 [hep-ph]].
%73 citations counted in INSPIRE as of 29 Aug 2025

%\cite{Ethier:2021ydt}
\bibitem{Ethier:2021ydt}
J.~J.~Ethier, R.~Gomez-Ambrosio, G.~Magni and J.~Rojo,
%``SMEFT analysis of vector boson scattering and diboson data from the LHC Run II,''
Eur. Phys. J. C \textbf{81} (2021) no.6, 560
%doi10.1140/epjc/s10052-021-09347-7
[arXiv:2101.03180 [hep-ph]].
%68 citations counted in INSPIRE as of 22 Jul 2025

%\cite{Butter:2016cvz}
\bibitem{Butter:2016cvz}
A.~Butter, O.~J.~P.~{\'E}boli, J.~Gonzalez-Fraile, M.~C.~Gonzalez-Garcia, T.~Plehn and M.~Rauch,
%``The Gauge-Higgs Legacy of the LHC Run I,''
JHEP \textbf{07} (2016), 152
%doi10.1007/JHEP07(2016)152
[arXiv:1604.03105 [hep-ph]];
%171 citations counted in INSPIRE as of 11 Jul 2025
%\cite{Corbett:2012ja}
%\bibitem{Corbett:2012ja}
T.~Corbett, O.~J.~P.~Eboli, J.~Gonzalez-Fraile and M.~C.~Gonzalez-Garcia,
%``Robust Determination of the Higgs Couplings: Power to the Data,''
Phys. Rev. D \textbf{87} (2013), 015022
%doi10.1103/PhysRevD.87.015022
[arXiv:1211.4580 [hep-ph]];
%219 citations counted in INSPIRE as of 11 Jul 2025
%
%\cite{Almeida:2021asy}
%\bibitem{Almeida:2021asy}
E.~d.~Almeida, A.~Alves, O.~J.~P.~{\'E}boli and M.~C.~Gonzalez-Garcia,
%``Electroweak legacy of the LHC run II,''
Phys. Rev. D \textbf{105} (2022) no.1, 013006
%doi:10.1103/PhysRevD.105.013006
[arXiv:2108.04828 [hep-ph]].
%54 citations counted in INSPIRE as of 05 Sep 2025

%\cite{Falkowski:2015jaa}
\bibitem{Falkowski:2015jaa}
A.~Falkowski, M.~Gonzalez-Alonso, A.~Greljo and D.~Marzocca,
%``Global constraints on anomalous triple gauge couplings in effective field theory approach,''
Phys. Rev. Lett. \textbf{116} (2016) no.1, 011801
%doi10.1103/PhysRevLett.116.011801
[arXiv:1508.00581 [hep-ph]];
%126 citations counted in INSPIRE as of 11 Jul 2025
%\cite{Falkowski:2016cxu}
%\bibitem{Falkowski:2016cxu}
A.~Falkowski, M.~Gonzalez-Alonso, A.~Greljo, D.~Marzocca and M.~Son,
%``Anomalous Triple Gauge Couplings in the Effective Field Theory Approach at the LHC,''
JHEP \textbf{02} (2017), 115
%doi10.1007/JHEP02(2017)115
[arXiv:1609.06312 [hep-ph]].
%142 citations counted in INSPIRE as of 11 Jul 2025

%\cite{Baglio:2017bfe}
\bibitem{Baglio:2017bfe}
J.~Baglio, S.~Dawson and I.~M.~Lewis,
%``An NLO QCD effective field theory analysis of $W^+W^-$ production at the LHC including fermionic operators,''
Phys. Rev. D \textbf{96} (2017) no.7, 073003
%doi10.1103/PhysRevD.96.073003
[arXiv:1708.03332 [hep-ph]].
%82 citations counted in INSPIRE as of 16 Jul 2025



%\cite{Franceschini:2017xkh}
\bibitem{Franceschini:2017xkh}
R.~Franceschini, G.~Panico, A.~Pomarol, F.~Riva and A.~Wulzer,
%``Electroweak Precision Tests in High-Energy Diboson Processes,''
JHEP \textbf{02} (2018), 111
%doi10.1007/JHEP02(2018)111
[arXiv:1712.01310 [hep-ph]].
%116 citations counted in INSPIRE as of 15 Jul 2025

%\cite{Grojean:2018dqj}
\bibitem{Grojean:2018dqj}
C.~Grojean, M.~Montull and M.~Riembau,
%``Diboson at the LHC vs LEP,''
JHEP \textbf{03} (2019), 020
%doi10.1007/JHEP03(2019)020
[arXiv:1810.05149 [hep-ph]].
%84 citations counted in INSPIRE as of 11 Jul 2025

%\cite{Banerjee:2018bio}
\bibitem{Banerjee:2018bio}
S.~Banerjee, C.~Englert, R.~S.~Gupta and M.~Spannowsky,
%``Probing Electroweak Precision Physics via boosted Higgs-strahlung at the LHC,''
Phys. Rev. D \textbf{98} (2018) no.9, 095012
%doi10.1103/PhysRevD.98.095012
[arXiv:1807.01796 [hep-ph]];
%71 citations counted in INSPIRE as of 11 Jul 2025
%\cite{Banerjee:2024eyo}
%\bibitem{Banerjee:2024eyo}
S.~Banerjee, D.~Reichelt and M.~Spannowsky,
%``Electroweak corrections and EFT operators in W+W- production at the LHC,''
Phys. Rev. D \textbf{110} (2024) no.11, 11
%doi10.1103/PhysRevD.110.115012
[arXiv:2406.15640 [hep-ph]].
%3 citations counted in INSPIRE as of 15 Jul 2025

%\cite{Biekotter:2014gup}
\bibitem{Biekotter:2014gup}
A.~Biek{\"o}tter, A.~Knochel, M.~Kr{\"a}mer, D.~Liu and F.~Riva,
%``Vices and virtues of Higgs effective field theories at large energy,''
Phys. Rev. D \textbf{91} (2015), 055029
%doi10.1103/PhysRevD.91.055029
[arXiv:1406.7320 [hep-ph]].
%136 citations counted in INSPIRE as of 21 Jul 2025

%\cite{Liu:2018pkg}
\bibitem{Liu:2018pkg}
D.~Liu and L.~T.~Wang,
%``Prospects for precision measurement of diboson processes in the semileptonic decay channel in future LHC runs,''
Phys. Rev. D \textbf{99} (2019) no.5, 055001
%doi10.1103/PhysRevD.99.055001
[arXiv:1804.08688 [hep-ph]].
%55 citations counted in INSPIRE as of 11 Jul 2025

%\cite{deBlas:2025xhe}
\bibitem{deBlas:2025xhe}
J.~de Blas, A.~Goncalves, V.~Miralles, L.~Reina, L.~Silvestrini and M.~Valli,
%``Constraining new physics effective interactions via a global fit of electroweak, Drell-Yan, Higgs, top, and flavour observables,''
[arXiv:2507.06191 [hep-ph]];
%3 citations counted in INSPIRE as of 25 Jul 2025
%\cite{deBlas:2016ojx}
%\bibitem{deBlas:2016ojx}
J.~de Blas, M.~Ciuchini, E.~Franco, S.~Mishima, M.~Pierini, L.~Reina and L.~Silvestrini,
%``Electroweak precision observables and Higgs-boson signal strengths in the Standard Model and beyond: present and future,''
JHEP \textbf{12} (2016), 135
%doi:10.1007/JHEP12(2016)135
[arXiv:1608.01509 [hep-ph]].
%191 citations counted in INSPIRE as of 25 Jul 2025

%\cite{Corbett:2017qgl}
\bibitem{Corbett:2017qgl}
T.~Corbett, O.~J.~P.~{\'E}boli and M.~C.~Gonzalez-Garcia,
%``Unitarity Constraints on Dimension-six Operators II: Including Fermionic Operators,''
Phys. Rev. D \textbf{96} (2017) no.3, 035006
%doi10.1103/PhysRevD.96.035006
[arXiv:1705.09294 [hep-ph]].
%60 citations counted in INSPIRE as of 17 Jul 2025

%\cite{Cornwall:1974km}
\bibitem{Cornwall:1974km}
J.~M.~Cornwall, D.~N.~Levin and G.~Tiktopoulos,
%``Derivation of Gauge Invariance from High-Energy Unitarity Bounds on the s Matrix,''
Phys. Rev. D \textbf{10} (1974), 1145
[erratum: Phys. Rev. D \textbf{11} (1975), 972]
%doi10.1103/PhysRevD.10.1145
%1424 citations counted in INSPIRE as of 22 Jul 2025

%\cite{DAmbrosio:2002vsn}
\bibitem{DAmbrosio:2002vsn}
G.~D'Ambrosio, G.~F.~Giudice, G.~Isidori and A.~Strumia,
%``Minimal flavor violation: An Effective field theory approach,''
Nucl. Phys. B \textbf{645} (2002), 155-187
%doi10.1016/S0550-3213(02)00836-2
[arXiv:hep-ph/0207036 [hep-ph]].
%1963 citations counted in INSPIRE as of 21 Jul 2025

%\cite{Barbieri:2011ci}
\bibitem{Barbieri:2011ci}
R.~Barbieri, G.~Isidori, J.~Jones-Perez, P.~Lodone and D.~M.~Straub,
%``$U(2)$ and Minimal Flavour Violation in Supersymmetry,''
Eur. Phys. J. C \textbf{71} (2011), 1725
%doi10.1140/epjc/s10052-011-1725-z
[arXiv:1105.2296 [hep-ph]].
%294 citations counted in INSPIRE as of 11 Jul 2025

%\cite{Faroughy:2020ina}
\bibitem{Faroughy:2020ina}
D.~A.~Faroughy, G.~Isidori, F.~Wilsch and K.~Yamamoto,
%``Flavour symmetries in the SMEFT,''
JHEP \textbf{08} (2020), 166
%doi10.1007/JHEP08(2020)166
[arXiv:2005.05366 [hep-ph]];
%86 citations counted in INSPIRE as of 15 Jul 2025
%\cite{Greljo:2022cah}
%\bibitem{Greljo:2022cah}
A.~Greljo, A.~Palavri{\'c} and A.~E.~Thomsen,
%``Adding Flavor to the SMEFT,''
JHEP \textbf{10} (2022), 010
%doi10.1007/JHEP10(2022)005
[arXiv:2203.09561 [hep-ph]].
%58 citations counted in INSPIRE as of 22 Jul 2025

%\cite{Buchmuller:1985jz}
\bibitem{Buchmuller:1985jz}
W.~Buchmuller and D.~Wyler,
%``Effective Lagrangian Analysis of New Interactions and Flavor Conservation,''
Nucl. Phys. B \textbf{268} (1986), 621-653
%doi10.1016/0550-3213(86)90262-2
%2643 citations counted in INSPIRE as of 23 Jul 2025

%\cite{Grzadkowski:2010es}
\bibitem{Grzadkowski:2010es}
B.~Grzadkowski, M.~Iskrzynski, M.~Misiak and J.~Rosiek,
%``Dimension-Six Terms in the Standard Model Lagrangian,''
JHEP \textbf{10} (2010), 085
%doi10.1007/JHEP10(2010)085
[arXiv:1008.4884 [hep-ph]].
%2630 citations counted in INSPIRE as of 23 Jul 2025


%\cite{Allwicher:2022gkm}
\bibitem{Allwicher:2022gkm}
L.~Allwicher, D.~A.~Faroughy, F.~Jaffredo, O.~Sumensari and F.~Wilsch,
%``Drell-Yan tails beyond the Standard Model,''
JHEP \textbf{03} (2023), 064
%doi10.1007/JHEP03(2023)064
[arXiv:2207.10714 [hep-ph]].
%93 citations counted in INSPIRE as of 23 Jul 2025

%\cite{Allwicher:2022mcg}
\bibitem{Allwicher:2022mcg}
L.~Allwicher, D.~A.~Faroughy, F.~Jaffredo, O.~Sumensari and F.~Wilsch,
%``HighPT: A tool for~
high-$p_T$
Drell-Yan tails beyond the standard model,''
Comput. Phys. Commun. \textbf{289} (2023), 108749
%doi10.1016/j.cpc.2023.108749
[arXiv:2207.10756 [hep-ph]].
%84 citations counted in INSPIRE as of 15 Jul 2025

%\cite{Grunwald:2023nli}
\bibitem{Grunwald:2023nli}
C.~Grunwald, G.~Hiller, K.~Kr{\"o}ninger and L.~Nollen,
%``More synergies from beauty, top, Z and Drell-Yan measurements in SMEFT,''
JHEP \textbf{11} (2023), 110
%doi:10.1007/JHEP11(2023)110
[arXiv:2304.12837 [hep-ph]];
%49 citations counted in INSPIRE as of 25 Jul 2025
%\cite{Bissmann:2020mfi}
%\bibitem{Bissmann:2020mfi}
S.~Bi{\ss}mann, C.~Grunwald, G.~Hiller and K.~Kr{\"o}ninger,
%``Top and Beauty synergies in SMEFT-fits at present and future colliders,''
JHEP \textbf{06} (2021), 010
%doi:10.1007/JHEP06(2021)010
[arXiv:2012.10456 [hep-ph]];
%86 citations counted in INSPIRE as of 26 Aug 2025
%\cite{Greljo:2022jac}
%\bibitem{Greljo:2022jac}
A.~Greljo, J.~Salko, A.~Smolkovi{\v{c}} and P.~Stangl,
%``Rare b decays meet high-mass Drell-Yan,''
JHEP \textbf{05} (2023), 087
%doi:10.1007/JHEP05(2023)087
[arXiv:2212.10497 [hep-ph]].
%118 citations counted in INSPIRE as of 25 Jul 2025

%\cite{Gabrielli:2024bjw}
\bibitem{Gabrielli:2024bjw}
E.~Gabrielli, L.~Marzola and K.~M{\"u}{\"u}rsepp,
%``Testing the CKM unitarity at high energy via the W+W{\ensuremath{-}} production at the LHC and future colliders,''
Phys. Lett. B \textbf{859} (2024), 139106
%doi10.1016/j.physletb.2024.139106
[arXiv:2405.14585 [hep-ph]].
%0 citations counted in INSPIRE as of 17 Jul 2025

%\cite{Cirigliano:2022yyo}
\bibitem{Cirigliano:2022yyo}
V.~Cirigliano, A.~Crivellin, M.~Hoferichter and M.~Moulson,
%``Scrutinizing CKM unitarity with a new measurement of the K{\ensuremath{\mu}}3/K{\ensuremath{\mu}}2 branching fraction,''
Phys. Lett. B \textbf{838} (2023), 137748
%doi10.1016/j.physletb.2023.137748
[arXiv:2208.11707 [hep-ph]].
%80 citations counted in INSPIRE as of 21 Jul 2025

%\cite{Alioli:2017ces}
\bibitem{Alioli:2017ces}
S.~Alioli, V.~Cirigliano, W.~Dekens, J.~de Vries and E.~Mereghetti,
%``Right-handed charged currents in the era of the Large Hadron Collider,''
JHEP \textbf{05} (2017), 086
%doi10.1007/JHEP05(2017)086
[arXiv:1703.04751 [hep-ph]].
%139 citations counted in INSPIRE as of 21 Jul 2025

%\cite{ATLAS:2024yzu}
\bibitem{ATLAS:2024yzu}
G.~Aad \textit{et al.} [ATLAS],
%``Measurements of $WH$ and $ZH$ production with Higgs boson decays into bottom quarks and direct constraints on the charm Yukawa coupling in $13\,\mathrm{TeV}$$pp$ collisions with the ATLAS detector,''
[arXiv:2410.19611 [hep-ex]].
%26 citations counted in INSPIRE as of 18 Jul 2025

%\cite{CMS:2023vzh}
\bibitem{CMS:2023vzh}
A.~Tumasyan \textit{et al.} [CMS],
%``Measurement of simplified template cross sections of the Higgs boson produced in association with W or Z bosons in the H{\textrightarrow}bb{\textasciimacron} decay channel in proton-proton collisions at s=13{\,}{\,}TeV,''
Phys. Rev. D \textbf{109} (2024) no.9, 092011
%doi10.1103/PhysRevD.109.092011
[arXiv:2312.07562 [hep-ex]].
%31 citations counted in INSPIRE as of 22 Jul 2025



%\cite{Efrati:2015eaa}
\bibitem{Efrati:2015eaa}
A.~Efrati, A.~Falkowski and Y.~Soreq,
%``Electroweak constraints on flavorful effective theories,''
JHEP \textbf{07} (2015), 018
%doi10.1007/JHEP07(2015)018
[arXiv:1503.07872 [hep-ph]].
%168 citations counted in INSPIRE as of 15 Jul 2025


%\cite{Arzt:1994gp}
\bibitem{Arzt:1994gp}
C.~Arzt, M.~B.~Einhorn and J.~Wudka,
%``Patterns of deviation from the standard model,''
Nucl. Phys. B \textbf{433} (1995), 41-66
%doi10.1016/0550-3213(94)00336-D
[arXiv:hep-ph/9405214 [hep-ph]].
%367 citations counted in INSPIRE as of 15 Jul 2025

%\cite{deBlas:2017xtg}
\bibitem{deBlas:2017xtg}
J.~de Blas, J.~C.~Criado, M.~Perez-Victoria and J.~Santiago,
%``Effective description of general extensions of the Standard Model: the complete tree-level dictionary,''
JHEP \textbf{03} (2018), 109
%doi10.1007/JHEP03(2018)109
[arXiv:1711.10391 [hep-ph]].
%282 citations counted in INSPIRE as of 18 Jul 2025

%\cite{Bobeth:2015zqa}
\bibitem{Bobeth:2015zqa}
C.~Bobeth and U.~Haisch,
%``Anomalous triple gauge couplings from $B$-meson and kaon observables,''
JHEP \textbf{09} (2015), 018
%doi10.1007/JHEP09(2015)018
[arXiv:1503.04829 [hep-ph]].
%26 citations counted in INSPIRE as of 22 Jul 2025

%\cite{Bordone:2021cca}
\bibitem{Bordone:2021cca}
M.~Bordone, A.~Greljo and D.~Marzocca,
%``Exploiting dijet resonance searches for flavor physics,''
JHEP \textbf{08} (2021), 036
%doi:10.1007/JHEP08(2021)036
[arXiv:2103.10332 [hep-ph]].
%45 citations counted in INSPIRE as of 25 Jul 2025

%\cite{Lee:1977eg}
\bibitem{Lee:1977eg}
B.~W.~Lee, C.~Quigg and H.~B.~Thacker,
%``Weak Interactions at Very High-Energies: The Role of the Higgs Boson Mass,''
Phys. Rev. D \textbf{16} (1977), 1519
%doi10.1103/PhysRevD.16.1519
%2524 citations counted in INSPIRE as of 19 Jul 2025

%\cite{Gounaris:1986cr}
\bibitem{Gounaris:1986cr}
G.~J.~Gounaris, R.~Kogerler and H.~Neufeld,
%``Relationship Between Longitudinally Polarized Vector Bosons and their Unphysical Scalar Partners,''
Phys. Rev. D \textbf{34} (1986), 3257
%doi10.1103/PhysRevD.34.3257
%339 citations counted in INSPIRE as of 11 Jul 2025

%\cite{Chanowitz:1987vj}
\bibitem{Chanowitz:1987vj}
M.~S.~Chanowitz, M.~Golden and H.~Georgi,
%``Low-Energy Theorems for Strongly Interacting W's and Z's,''
Phys. Rev. D \textbf{36} (1987), 1490;
%doi10.1103/PhysRevD.36.1490
%258 citations counted in INSPIRE as of 11 Jul 2025
%\cite{Chanowitz:1985hj}
%\bibitem{Chanowitz:1985hj}
M.~S.~Chanowitz and M.~K.~Gaillard,
%``The TeV Physics of Strongly Interacting W's and Z's,''
Nucl. Phys. B \textbf{261} (1985), 379-431
%doi10.1016/0550-3213(85)90580-2
%1301 citations counted in INSPIRE as of 11 Jul 2025

%\cite{Wulzer:2013mza}
\bibitem{Wulzer:2013mza}
A.~Wulzer,
%``An Equivalent Gauge and the Equivalence Theorem,''
Nucl. Phys. B \textbf{885} (2014), 97-126
%doi10.1016/j.nuclphysb.2014.05.021
[arXiv:1309.6055 [hep-ph]].
%45 citations counted in INSPIRE as of 11 Jul 2025


%\cite{ATLAS:2025dhf}
\bibitem{ATLAS:2025dhf}
G.~Aad \textit{et al.} [ATLAS],
%``Measurements of $W^+W^-$ production cross-sections in $pp$ collisions at $\sqrt{s}=13$ TeV with the ATLAS detector,''
[arXiv:2505.11310 [hep-ex]].
%1 citations counted in INSPIRE as of 28 Aug 2025

%\cite{ATLAS:2019rob}
\bibitem{ATLAS:2019rob}
M.~Aaboud \textit{et al.} [ATLAS],
%``Measurement of fiducial and differential $W^+W^-$ production cross-sections at $\sqrt{s}=13$  TeV with the ATLAS detector,''
Eur. Phys. J. C \textbf{79} (2019) no.10, 884
%doi10.1140/epjc/s10052-019-7371-6
[arXiv:1905.04242 [hep-ex]].
%145 citations counted in INSPIRE as of 21 Jul 2025

%\cite{CMS:2020mxy}
\bibitem{CMS:2020mxy}
A.~M.~Sirunyan \textit{et al.} [CMS],
%``W$^+$W$^-$ boson pair production in proton-proton collisions at $\sqrt{s} =$ 13 TeV,''
Phys. Rev. D \textbf{102} (2020) no.9, 092001
%doi10.1103/PhysRevD.102.092001
[arXiv:2009.00119 [hep-ex]].
%88 citations counted in INSPIRE as of 21 Jul 2025

%\cite{ATLAS:2019bsc}
% \bibitem{ATLAS:2019bsc}
% M.~Aaboud \textit{et al.} [ATLAS],
% %``Measurement of $W^{\pm}Z$ production cross sections and gauge boson polarisation in $pp$ collisions at $\sqrt{s} = 13$ TeV with the ATLAS detector,''
% Eur. Phys. J. C \textbf{79} (2019) no.6, 535
% %doi10.1140/epjc/s10052-019-7027-6
% [arXiv:1902.05759 [hep-ex]].
%213 citations counted in INSPIRE as of 22 Jul 2025

%\cite{ATLAS:2025edf}
\bibitem{ATLAS:2025edf}
G.~Aad \textit{et al.} [ATLAS],
%``Measurements and interpretations of $W^{\pm}Z$ production cross-sections in $pp$ collisions at $\sqrt{s} =$ 13 TeV with the ATLAS detector,''
[arXiv:2507.03500 [hep-ex]].
%0 citations counted in INSPIRE as of 24 Aug 2025

%\cite{CMS:2021icx}
\bibitem{CMS:2021icx}
A.~Tumasyan \textit{et al.} [CMS],
%``Measurement of the inclusive and differential WZ production cross sections, polarization angles, and triple gauge couplings in pp collisions at $ \sqrt{s} $ = 13 TeV,''
JHEP \textbf{07} (2022), 032
%doi10.1007/JHEP07(2022)032
[arXiv:2110.11231 [hep-ex]].
%97 citations counted in INSPIRE as of 15 Jul 2025

%\cite{Frederix:2018nkq}
\bibitem{Frederix:2018nkq}
R.~Frederix, S.~Frixione, V.~Hirschi, D.~Pagani, H.~S.~Shao and M.~Zaro,
%``The automation of next-to-leading order electroweak calculations,''
JHEP \textbf{07} (2018), 185
[erratum: JHEP \textbf{11} (2021), 085]
%doi10.1007/JHEP11(2021)085
[arXiv:1804.10017 [hep-ph]].
%429 citations counted in INSPIRE as of 21 Jul 2025

%\cite{Christensen:2008py}
\bibitem{Christensen:2008py}
N.~D.~Christensen and C.~Duhr,
%``FeynRules - Feynman rules made easy,''
Comput. Phys. Commun. \textbf{180} (2009), 1614-1641
%doi10.1016/j.cpc.2009.02.018
[arXiv:0806.4194 [hep-ph]];
%1120 citations counted in INSPIRE as of 23 Jul 2025
%\cite{Alloul:2013bka}
%\bibitem{Alloul:2013bka}
A.~Alloul, N.~D.~Christensen, C.~Degrande, C.~Duhr and B.~Fuks,
%``FeynRules  2.0 - A complete toolbox for tree-level phenomenology,''
Comput. Phys. Commun. \textbf{185} (2014), 2250-2300
%doi10.1016/j.cpc.2014.04.012
[arXiv:1310.1921 [hep-ph]].
%2776 citations counted in INSPIRE as of 23 Jul 2025

%\cite{Sjostrand:2007gs}
\bibitem{Sjostrand:2007gs}
T.~Sjostrand, S.~Mrenna and P.~Z.~Skands,
%``A Brief Introduction to PYTHIA 8.1,''
Comput. Phys. Commun. \textbf{178} (2008), 852-867
%doi10.1016/j.cpc.2008.01.036
[arXiv:0710.3820 [hep-ph]].
%7952 citations counted in INSPIRE as of 22 Jul 2025

%\cite{deFavereau:2013fsa}
\bibitem{deFavereau:2013fsa}
J.~de Favereau \textit{et al.} [DELPHES 3],
%``DELPHES 3, A modular framework for fast simulation of a generic collider experiment,''
JHEP \textbf{02} (2014), 057
%doi10.1007/JHEP02(2014)057
[arXiv:1307.6346 [hep-ex]].
%3319 citations counted in INSPIRE as of 23 Jul 2025

%\cite{LHCHiggsCrossSectionWorkingGroup:2016ypw}
\bibitem{LHCHiggsCrossSectionWorkingGroup:2016ypw}
D.~de Florian \textit{et al.} [LHC Higgs Cross Section Working Group],
%``Handbook of LHC Higgs Cross Sections: 4. Deciphering the Nature of the Higgs Sector,''
CERN Yellow Rep. Monogr. \textbf{2} (2017), 1-869
%doi10.23731/CYRM-2017-002
[arXiv:1610.07922 [hep-ph]].
%2570 citations counted in INSPIRE as of 23 Jul 2025

%\cite{Coy:2019rfr}
\bibitem{Coy:2019rfr}
R.~Coy, M.~Frigerio, F.~Mescia and O.~Sumensari,
%``New physics in $b\to s\ell\ell$ transitions at one loop,''
Eur. Phys. J. C \textbf{80} (2020) no.1, 52
%doi:10.1140/epjc/s10052-019-7581-y
[arXiv:1909.08567 [hep-ph]];
%27 citations counted in INSPIRE as of 27 Aug 2025
%\cite{Cornella:2018tfd}
%\bibitem{Cornella:2018tfd}
C.~Cornella, F.~Feruglio and P.~Paradisi,
%``Low-energy Effects of Lepton Flavour Universality Violation,''
JHEP \textbf{11} (2018), 012
%doi:10.1007/JHEP11(2018)012
[arXiv:1803.00945 [hep-ph]].
%34 citations counted in INSPIRE as of 27 Aug 2025

%\cite{Jenkins:2013zja}
\bibitem{Jenkins:2013zja}
E.~E.~Jenkins, A.~V.~Manohar and M.~Trott,
%``Renormalization Group Evolution of the Standard Model Dimension Six Operators I: Formalism and lambda Dependence,''
JHEP \textbf{10} (2013), 087
%doi:10.1007/JHEP10(2013)087
[arXiv:1308.2627 [hep-ph]];
%678 citations counted in INSPIRE as of 27 Aug 2025
%\cite{Jenkins:2013wua}
%\bibitem{Jenkins:2013wua}
E.~E.~Jenkins, A.~V.~Manohar and M.~Trott,
%``Renormalization Group Evolution of the Standard Model Dimension Six Operators II: Yukawa Dependence,''
JHEP \textbf{01} (2014), 035
%doi:10.1007/JHEP01(2014)035
[arXiv:1310.4838 [hep-ph]];
%691 citations counted in INSPIRE as of 27 Aug 2025
%\cite{Alonso:2013hga}
%\bibitem{Alonso:2013hga}
R.~Alonso, E.~E.~Jenkins, A.~V.~Manohar and M.~Trott,
%``Renormalization Group Evolution of the Standard Model Dimension Six Operators III: Gauge Coupling Dependence and Phenomenology,''
JHEP \textbf{04} (2014), 159
%doi:10.1007/JHEP04(2014)159
[arXiv:1312.2014 [hep-ph]].
%889 citations counted in INSPIRE as of 27 Aug 2025

%\cite{Fuentes-Martin:2020zaz}
\bibitem{Fuentes-Martin:2020zaz}
J.~Fuentes-Martin, P.~Ruiz-Femenia, A.~Vicente and J.~Virto,
%``DsixTools 2.0: The Effective Field Theory Toolkit,''
Eur. Phys. J. C \textbf{81} (2021) no.2, 167
%doi10.1140/epjc/s10052-020-08778-y
[arXiv:2010.16341 [hep-ph]];
%98 citations counted in INSPIRE as of 15 Jul 2025
%\cite{Celis:2017hod}
%\bibitem{Celis:2017hod}
A.~Celis, J.~Fuentes-Martin, A.~Vicente and J.~Virto,
%``DsixTools: The Standard Model Effective Field Theory Toolkit,''
Eur. Phys. J. C \textbf{77} (2017) no.6, 405
%doi10.1140/epjc/s10052-017-4967-6
[arXiv:1704.04504 [hep-ph]].
%163 citations counted in INSPIRE as of 15 Jul 2025

%\cite{Janot:2019oyi}
\bibitem{Janot:2019oyi}
P.~Janot and S.~Jadach,
%``Improved Bhabha cross section at LEP and the number of light neutrino species,''
Phys. Lett. B \textbf{803} (2020), 135319
%doi10.1016/j.physletb.2020.135319
[arXiv:1912.02067 [hep-ph]].
%112 citations counted in INSPIRE as of 22 Jul 2025

%\cite{Breso-Pla:2021qoe}
\bibitem{Breso-Pla:2021qoe}
V.~Bres{\'o}-Pla, A.~Falkowski and M.~Gonz{\'a}lez-Alonso,
%``A$_{FB}$ in the SMEFT: precision Z physics at the LHC,''
JHEP \textbf{08} (2021), 021
%doi10.1007/JHEP08(2021)021
[arXiv:2103.12074 [hep-ph]].
%45 citations counted in INSPIRE as of 22 Jul 2025

%\cite{deBlas:2022hdk}
\bibitem{deBlas:2022hdk}
J.~de Blas, M.~Pierini, L.~Reina and L.~Silvestrini,
%``Impact of the Recent Measurements of the Top-Quark and W-Boson Masses on Electroweak Precision Fits,''
Phys. Rev. Lett. \textbf{129} (2022) no.27, 271801
%doi10.1103/PhysRevLett.129.271801
[arXiv:2204.04204 [hep-ph]].
%199 citations counted in INSPIRE as of 14 Jul 2025

%\cite{ALEPH:2005ab}
\bibitem{ALEPH:2005ab}
S.~Schael \textit{et al.} [ALEPH, DELPHI, L3, OPAL, SLD, LEP Electroweak Working Group, SLD Electroweak Group and SLD Heavy Flavour Group],
%``Precision electroweak measurements on the $Z$ resonance,''
Phys. Rept. \textbf{427} (2006), 257-454
%doi10.1016/j.physrep.2005.12.006
[arXiv:hep-ex/0509008 [hep-ex]].
%3386 citations counted in INSPIRE as of 23 Jul 2025

%\cite{ParticleDataGroup:2024cfk}
\bibitem{ParticleDataGroup:2024cfk}
S.~Navas \textit{et al.} [Particle Data Group],
%``Review of particle physics,''
Phys. Rev. D \textbf{110} (2024) no.3, 030001
%doi10.1103/PhysRevD.110.030001
%2157 citations counted in INSPIRE as of 23 Jul 2025

%\cite{ALEPH:2013dgf}
\bibitem{ALEPH:2013dgf}
S.~Schael \textit{et al.} [ALEPH, DELPHI, L3, OPAL and LEP Electroweak],
%``Electroweak Measurements in Electron-Positron Collisions at W-Boson-Pair Energies at LEP,''
Phys. Rept. \textbf{532} (2013), 119-244
%doi10.1016/j.physrep.2013.07.004
[arXiv:1302.3415 [hep-ex]].
%925 citations counted in INSPIRE as of 21 Jul 2025


%\cite{Crivellin:2022rhw}
\bibitem{Crivellin:2022rhw}
A.~Crivellin, M.~Kirk, T.~Kitahara and F.~Mescia,
%``Global fit of modified quark couplings to EW gauge bosons and vector-like quarks in light of the Cabibbo angle anomaly,''
JHEP \textbf{03} (2023), 234
%doi10.1007/JHEP03(2023)234
[arXiv:2212.06862 [hep-ph]];
%46 citations counted in INSPIRE as of 22 Jul 2025
%\cite{Kirk:2020wdk}
%\bibitem{Kirk:2020wdk}
M.~Kirk,
%``Cabibbo anomaly versus electroweak precision tests: An exploration of extensions of the Standard Model,''
Phys. Rev. D \textbf{103} (2021) no.3, 035004
%doi10.1103/PhysRevD.103.035004
[arXiv:2008.03261 [hep-ph]].
%66 citations counted in INSPIRE as of 22 Jul 2025

%\cite{Cirigliano:2023nol}
\bibitem{Cirigliano:2023nol}
V.~Cirigliano, W.~Dekens, J.~de Vries, E.~Mereghetti and T.~Tong,
%``Anomalies in global SMEFT analyses. A case study of first-row CKM unitarity,''
JHEP \textbf{03} (2024), 033
%doi10.1007/JHEP03(2024)033
[arXiv:2311.00021 [hep-ph]].
%37 citations counted in INSPIRE as of 18 Jul 2025

%\cite{Coutinho:2019aiy}
\bibitem{Coutinho:2019aiy}
A.~M.~Coutinho, A.~Crivellin and C.~A.~Manzari,
%``Global Fit to Modified Neutrino Couplings and the Cabibbo-Angle Anomaly,''
Phys. Rev. Lett. \textbf{125} (2020) no.7, 071802
%doi:10.1103/PhysRevLett.125.071802
[arXiv:1912.08823 [hep-ph]];
%127 citations counted in INSPIRE as of 29 Aug 2025
%\cite{Crivellin:2020ebi}
%\bibitem{Crivellin:2020ebi}
A.~Crivellin, F.~Kirk, C.~A.~Manzari and M.~Montull,
%``Global Electroweak Fit and Vector-Like Leptons in Light of the Cabibbo Angle Anomaly,''
JHEP \textbf{12} (2020), 166
%doi:10.1007/JHEP12(2020)166
[arXiv:2008.01113 [hep-ph]].
%123 citations counted in INSPIRE as of 29 Aug 2025


%\cite{Belfatto:2021jhf}
\bibitem{Belfatto:2021jhf}
B.~Belfatto and Z.~Berezhiani,
%``Are the CKM anomalies induced by vector-like quarks? Limits from flavor changing and Standard Model precision tests,''
JHEP \textbf{10} (2021), 079
%doi:10.1007/JHEP10(2021)079
[arXiv:2103.05549 [hep-ph]];
%63 citations counted in INSPIRE as of 29 Aug 2025
%\cite{Branco:2021vhs}
%\bibitem{Branco:2021vhs}
G.~C.~Branco, J.~T.~Penedo, P.~M.~F.~Pereira, M.~N.~Rebelo and J.~I.~Silva-Marcos,
%``Addressing the CKM unitarity problem with a vector-like up quark,''
JHEP \textbf{07} (2021), 099
%doi:10.1007/JHEP07(2021)099
[arXiv:2103.13409 [hep-ph]];
%58 citations counted in INSPIRE as of 29 Aug 2025
%\cite{Cheung:2020vqm}
%\bibitem{Cheung:2020vqm}
K.~Cheung, W.~Y.~Keung, C.~T.~Lu and P.~Y.~Tseng,
%``Vector-like Quark Interpretation for the CKM Unitarity Violation, Excess in Higgs Signal Strength, and Bottom Quark Forward-Backward Asymmetry,''
JHEP \textbf{05} (2020), 117
%doi:10.1007/JHEP05(2020)117
[arXiv:2001.02853 [hep-ph]].
%43 citations counted in INSPIRE as of 29 Aug 2025

%\cite{FlavourLatticeAveragingGroupFLAG:2021npn}
\bibitem{FlavourLatticeAveragingGroupFLAG:2021npn}
Y.~Aoki \textit{et al.} [Flavour Lattice Averaging Group (FLAG)],
%``FLAG Review 2021,''
Eur. Phys. J. C \textbf{82} (2022) no.10, 869
%doi10.1140/epjc/s10052-022-10536-1
[arXiv:2111.09849 [hep-lat]].
%849 citations counted in INSPIRE as of 23 Jul 2025


%\cite{DiCarlo:2019thl}
\bibitem{DiCarlo:2019thl}
M.~Di Carlo, D.~Giusti, V.~Lubicz, G.~Martinelli, C.~T.~Sachrajda, F.~Sanfilippo, S.~Simula and N.~Tantalo,
%``Light-meson leptonic decay rates in lattice QCD+QED,''
Phys. Rev. D \textbf{100} (2019) no.3, 034514
%doi10.1103/PhysRevD.100.034514
[arXiv:1904.08731 [hep-lat]].
%124 citations counted in INSPIRE as of 14 Jul 2025

%\cite{Giusti:2017dwk}
\bibitem{Giusti:2017dwk}
D.~Giusti, V.~Lubicz, G.~Martinelli, C.~T.~Sachrajda, F.~Sanfilippo, S.~Simula, N.~Tantalo and C.~Tarantino,
%``First lattice calculation of the QED corrections to leptonic decay rates,''
Phys. Rev. Lett. \textbf{120} (2018) no.7, 072001
%doi10.1103/PhysRevLett.120.072001
[arXiv:1711.06537 [hep-lat]];
%105 citations counted in INSPIRE as of 14 Jul 2025
%\cite{Desiderio:2020oej}
%\bibitem{Desiderio:2020oej}
A.~Desiderio, R.~Frezzotti, M.~Garofalo, D.~Giusti, M.~Hansen, V.~Lubicz, G.~Martinelli, C.~T.~Sachrajda, F.~Sanfilippo and S.~Simula, \textit{et al.}
%``First lattice calculation of radiative leptonic decay rates of pseudoscalar mesons,''
Phys. Rev. D \textbf{103} (2021) no.1, 014502
%doi10.1103/PhysRevD.103.014502
[arXiv:2006.05358 [hep-lat]].
%59 citations counted in INSPIRE as of 11 Jul 2025

%\cite{Dowdall:2013rya}
\bibitem{Dowdall:2013rya}
R.~J.~Dowdall, C.~T.~H.~Davies, G.~P.~Lepage and C.~McNeile,
%``Vus from pi and K decay constants in full lattice QCD with physical u, d, s and c quarks,''
Phys. Rev. D \textbf{88} (2013), 074504
%doi10.1103/PhysRevD.88.074504
[arXiv:1303.1670 [hep-lat]];
%242 citations counted in INSPIRE as of 15 Jul 2025
%\cite{Bazavov:2017lyh}
%\bibitem{Bazavov:2017lyh}
A.~Bazavov, C.~Bernard, N.~Brown, C.~Detar, A.~X.~El-Khadra, E.~G{\'a}miz, S.~Gottlieb, U.~M.~Heller, J.~Komijani and A.~S.~Kronfeld, \textit{et al.}
%``$B$- and $D$-meson leptonic decay constants from four-flavor lattice QCD,''
Phys. Rev. D \textbf{98} (2018) no.7, 074512
%doi10.1103/PhysRevD.98.074512
[arXiv:1712.09262 [hep-lat]];
%324 citations counted in INSPIRE as of 20 Jul 2025
%\cite{Miller:2020xhy}
%\bibitem{Miller:2020xhy}
N.~Miller, H.~Monge-Camacho, C.~C.~Chang, B.~H{\"o}rz, E.~Rinaldi, D.~Howarth, E.~Berkowitz, D.~A.~Brantley, A.~S.~Gambhir and C.~K{\"o}rber, \textit{et al.}
%``$F_K / F_\pi$ from M{\"o}bius Domain-Wall fermions solved on gradient-flowed HISQ ensembles,''
Phys. Rev. D \textbf{102} (2020) no.3, 034507
%doi10.1103/PhysRevD.102.034507
[arXiv:2005.04795 [hep-lat]];
%46 citations counted in INSPIRE as of 15 Jul 2025
%\cite{ExtendedTwistedMass:2021qui}
%\bibitem{ExtendedTwistedMass:2021qui}
C.~Alexandrou \textit{et al.} [Extended Twisted Mass],
%``Ratio of kaon and pion leptonic decay constants with Nf=2+1+1 Wilson-clover twisted-mass fermions,''
Phys. Rev. D \textbf{104} (2021) no.7, 074520
%doi10.1103/PhysRevD.104.074520
[arXiv:2104.06747 [hep-lat]].
%55 citations counted in INSPIRE as of 15 Jul 2025

%\cite{UTfit:2022hsi}
\bibitem{UTfit:2022hsi}
M.~Bona \textit{et al.} [UTfit],
%``New UTfit Analysis of the Unitarity Triangle in the Cabibbo-Kobayashi-Maskawa scheme,''
Rend. Lincei Sci. Fis. Nat. \textbf{34} (2023), 37-57
%doi10.1007/s12210-023-01137-5
[arXiv:2212.03894 [hep-ph]].
%123 citations counted in INSPIRE as of 15 Jul 2025

%\cite{Charles:2004jd}
\bibitem{Charles:2004jd}
J.~Charles \textit{et al.} [CKMfitter Group],
%``CP violation and the CKM matrix: Assessing the impact of the asymmetric $B$ factories,''
Eur. Phys. J. C \textbf{41} (2005) no.1, 1-131
%doi10.1140/epjc/s2005-02169-1
[arXiv:hep-ph/0406184 [hep-ph]].
%2201 citations counted in INSPIRE as of 22 Jul 2025

%\cite{Bryman:2011zz}
\bibitem{Bryman:2011zz}
D.~Bryman, W.~J.~Marciano, R.~Tschirhart and T.~Yamanaka,
%``Rare kaon and pion decays: Incisive probes for new physics beyond the standard model,''
Ann. Rev. Nucl. Part. Sci. \textbf{61} (2011), 331-354
%doi10.1146/annurev-nucl-102010-130431
%60 citations counted in INSPIRE as of 11 Jul 2025

%\cite{Cirigliano:2018dyk}
\bibitem{Cirigliano:2018dyk}
V.~Cirigliano, A.~Falkowski, M.~Gonz{\'a}lez-Alonso and A.~Rodr{\'\i}guez-S{\'a}nchez,
%``Hadronic {\ensuremath{\tau}} Decays as New Physics Probes in the LHC Era,''
Phys. Rev. Lett. \textbf{122} (2019) no.22, 221801
%doi10.1103/PhysRevLett.122.221801
[arXiv:1809.01161 [hep-ph]];
%80 citations counted in INSPIRE as of 16 Jul 2025
%\cite{Cirigliano:2021yto}
%\bibitem{Cirigliano:2021yto}
V.~Cirigliano, D.~D{\'\i}az-Calder{\'o}n, A.~Falkowski, M.~Gonz{\'a}lez-Alonso and A.~Rodr{\'\i}guez-S{\'a}nchez,
%``Semileptonic tau decays beyond the Standard Model,''
JHEP \textbf{04} (2022), 152
%doi10.1007/JHEP04(2022)152
[arXiv:2112.02087 [hep-ph]];
%60 citations counted in INSPIRE as of 22 Jul 2025
%\cite{Gonzalez-Alonso:2016etj}
%\bibitem{Gonzalez-Alonso:2016etj}
M.~Gonz{\'a}lez-Alonso and J.~Martin Camalich,
%``Global Effective-Field-Theory analysis of New-Physics effects in (semi)leptonic kaon decays,''
JHEP \textbf{12} (2016), 052
%doi10.1007/JHEP12(2016)052
[arXiv:1605.07114 [hep-ph]];
%108 citations counted in INSPIRE as of 11 Jul 2025
%\cite{Cirigliano:2012ab}
%\bibitem{Cirigliano:2012ab}
V.~Cirigliano, M.~Gonzalez-Alonso and M.~L.~Graesser,
%``Non-standard Charged Current Interactions: beta decays versus the LHC,''
JHEP \textbf{02} (2013), 046
%doi10.1007/JHEP02(2013)046
[arXiv:1210.4553 [hep-ph]].
%193 citations counted in INSPIRE as of 21 Jul 2025

%\cite{FlaviaNetWorkingGrouponKaonDecays:2010lot}
\bibitem{FlaviaNetWorkingGrouponKaonDecays:2010lot}
M.~Antonelli \textit{et al.} [FlaviaNet Working Group on Kaon Decays],
%``An Evaluation of $|V_{us}|$ and precise tests of the Standard Model from world data on leptonic and semileptonic kaon decays,''
Eur. Phys. J. C \textbf{69} (2010), 399-424
%doi10.1140/epjc/s10052-010-1406-3
[arXiv:1005.2323 [hep-ph]].
%362 citations counted in INSPIRE as of 14 Jul 2025

%\cite{Marciano:1993sh}
\bibitem{Marciano:1993sh}
W.~J.~Marciano and A.~Sirlin,
%``Radiative corrections to pi(lepton 2) decays,''
Phys. Rev. Lett. \textbf{71} (1993), 3629-3632;
%doi10.1103/PhysRevLett.71.3629
%374 citations counted in INSPIRE as of 22 Jul 2025
%\cite{Sirlin:1977sv}
%\bibitem{Sirlin:1977sv}
A.~Sirlin,
%``Current Algebra Formulation of Radiative Corrections in Gauge Theories and the Universality of the Weak Interactions,''
Rev. Mod. Phys. \textbf{50} (1978), 573
[erratum: Rev. Mod. Phys. \textbf{50} (1978) no.4, 905]
%doi10.1103/RevModPhys.50.573
%460 citations counted in INSPIRE as of 11 Jul 2025

%\cite{Seng:2022wcw}
\bibitem{Seng:2022wcw}
C.~Y.~Seng, D.~Galviz, M.~Gorchtein and U.~G.~Mei{\ss}ner,
%``Complete theory of radiative corrections to K$_{ℓ3}$ decays and the V$_{us}$ update,''
JHEP \textbf{07} (2022), 071
%doi:10.1007/JHEP07(2022)071
[arXiv:2203.05217 [hep-ph]];
%44 citations counted in INSPIRE as of 09 Sep 2025
%\cite{Seng:2021nar}
%\bibitem{Seng:2021nar}
C.~Y.~Seng, D.~Galviz, W.~J.~Marciano and U.~G.~Mei{\ss}ner,
%``Update on |Vus| and |Vus/Vud| from semileptonic kaon and pion decays,''
Phys. Rev. D \textbf{105} (2022) no.1, 013005
%doi:10.1103/PhysRevD.105.013005
[arXiv:2107.14708 [hep-ph]].
%61 citations counted in INSPIRE as of 09 Sep 2025

%\cite{Carrasco:2016kpy}
\bibitem{Carrasco:2016kpy}
N.~Carrasco, P.~Lami, V.~Lubicz, L.~Riggio, S.~Simula and C.~Tarantino,
%``$K \to \pi$ semileptonic form factors with $N_f=2+1+1$ twisted mass fermions,''
Phys. Rev. D \textbf{93} (2016) no.11, 114512
%doi10.1103/PhysRevD.93.114512
[arXiv:1602.04113 [hep-lat]].
%114 citations counted in INSPIRE as of 15 Jul 2025

%\cite{FlavourLatticeAveragingGroupFLAG:2024oxs}
\bibitem{FlavourLatticeAveragingGroupFLAG:2024oxs}
Y.~Aoki \textit{et al.} [Flavour Lattice Averaging Group (FLAG)],
%``FLAG Review 2024,''
[arXiv:2411.04268 [hep-lat]].
%126 citations counted in INSPIRE as of 22 Jul 2025

%\cite{FermilabLattice:2018zqv}
\bibitem{FermilabLattice:2018zqv}
A.~Bazavov \textit{et al.} [Fermilab Lattice and MILC],
%``$|V_{us}|$ from $K_{\ell 3}$ decay and four-flavor lattice QCD,''
Phys. Rev. D \textbf{99} (2019) no.11, 114509
%doi10.1103/PhysRevD.99.114509
[arXiv:1809.02827 [hep-lat]].
%79 citations counted in INSPIRE as of 15 Jul 2025

%\cite{Moulson:2017ive}
\bibitem{Moulson:2017ive}
M.~Moulson,
%``Experimental determination of $V_{us}$ from kaon decays,''
PoS \textbf{CKM2016} (2017), 033
%doi10.22323/1.291.0033
[arXiv:1704.04104 [hep-ex]].
%69 citations counted in INSPIRE as of 15 Jul 2025

%\cite{Becirevic:2020rzi}
\bibitem{Becirevic:2020rzi}
D.~Be{\v{c}}irevi{\'c}, F.~Jaffredo, A.~Pe{\~n}uelas and O.~Sumensari,
%``New Physics effects in leptonic and semileptonic decays,''
JHEP \textbf{05} (2021), 175
%doi10.1007/JHEP05(2021)175
[arXiv:2012.09872 [hep-ph]].
%45 citations counted in INSPIRE as of 17 Jul 2025

%\cite{Falkowski:2020pma}
\bibitem{Falkowski:2020pma}
A.~Falkowski, M.~Gonz{\'a}lez-Alonso and O.~Naviliat-Cuncic,
%``Comprehensive analysis of beta decays within and beyond the Standard Model,''
JHEP \textbf{04} (2021), 126
%doi10.1007/JHEP04(2021)126
[arXiv:2010.13797 [hep-ph]].
%100 citations counted in INSPIRE as of 21 Jul 2025

%\cite{Markisch:2018ndu}
% \bibitem{Markisch:2018ndu}
% B.~M{\"a}rkisch, H.~Mest, H.~Saul, X.~Wang, H.~Abele, D.~Dubbers, M.~Klopf, A.~Petoukhov, C.~Roick and T.~Soldner, \textit{et al.}
% %``Measurement of the Weak Axial-Vector Coupling Constant in the Decay of Free Neutrons Using a Pulsed Cold Neutron Beam,''
% Phys. Rev. Lett. \textbf{122} (2019) no.24, 242501
% %doi10.1103/PhysRevLett.122.242501
% [arXiv:1812.04666 [nucl-ex]].
% %230 citations counted in INSPIRE as of 21 Jul 2025

%\cite{Czarnecki:2018okw}
\bibitem{Czarnecki:2018okw}
A.~Czarnecki, W.~J.~Marciano and A.~Sirlin,
%``Neutron Lifetime and Axial Coupling Connection,''
Phys. Rev. Lett. \textbf{120} (2018) no.20, 202002
%doi10.1103/PhysRevLett.120.202002
[arXiv:1802.01804 [hep-ph]].
%184 citations counted in INSPIRE as of 21 Jul 2025

%\cite{Seng:2018yzq}
\bibitem{Seng:2018yzq}
C.~Y.~Seng, M.~Gorchtein, H.~H.~Patel and M.~J.~Ramsey-Musolf,
%``Reduced Hadronic Uncertainty in the Determination of $V_{ud}$,''
Phys. Rev. Lett. \textbf{121} (2018) no.24, 241804
%doi10.1103/PhysRevLett.121.241804
[arXiv:1807.10197 [hep-ph]].
%256 citations counted in INSPIRE as of 15 Jul 2025

%\cite{Gorchtein:2021fce}
\bibitem{Gorchtein:2021fce}
M.~Gorchtein and C.~Y.~Seng,
%``Dispersion relation analysis of the radiative corrections to g$_{A}$ in the neutron {\ensuremath{\beta}}-decay,''
JHEP \textbf{10} (2021), 053
%doi10.1007/JHEP10(2021)053
[arXiv:2106.09185 [hep-ph]].
%35 citations counted in INSPIRE as of 21 Jul 2025

%\cite{Towner:2010zz}
\bibitem{Towner:2010zz}
I.~S.~Towner and J.~C.~Hardy,
%``The evaluation of V(ud) and its impact on the unitarity of the Cabibbo-Kobayashi-Maskawa quark-mixing matrix,''
Rept. Prog. Phys. \textbf{73} (2010), 046301
%doi10.1088/0034-4885/73/4/046301
%162 citations counted in INSPIRE as of 11 Jul 2025

%\cite{UCNt:2021pcg}
\bibitem{UCNt:2021pcg}
F.~M.~Gonzalez \textit{et al.} [UCN{\ensuremath{\tau}}],
%``Improved neutron lifetime measurement with UCN$\tau$,''
Phys. Rev. Lett. \textbf{127} (2021) no.16, 162501
%doi10.1103/PhysRevLett.127.162501
[arXiv:2106.10375 [nucl-ex]].
%161 citations counted in INSPIRE as of 21 Jul 2025

%\cite{Hardy:2020qwl}
\bibitem{Hardy:2020qwl}
J.~C.~Hardy and I.~S.~Towner,
%``Superallowed $0^+ \to 0^+$ nuclear $\beta$ decays: 2020 critical survey, with implications for V$_{ud}$ and CKM unitarity,''
Phys. Rev. C \textbf{102} (2020) no.4, 045501
%doi10.1103/PhysRevC.102.045501
%247 citations counted in INSPIRE as of 11 Jul 2025

%\cite{charm-paper}
\bibitem{charm-paper}
D.~Becirevic, M.~Martines, O.~Sumensari, S.~Rosauro-Alcaraz, \textit{In preparation}

%\cite{ATLAS:2025wcz}
\bibitem{ATLAS:2025wcz}
 [ATLAS],
%``Expected sensitivity of the ATLAS experiment to $H \to b\bar{b}$ and $H \to c\bar{c}$ decays in the $VH$ production mode at the High Luminosity LHC,''
ATL-PHYS-PUB-2025-012.
%2 citations counted in INSPIRE as of 28 Aug 2025

%\cite{ATLAS:2024zlo}
\bibitem{ATLAS:2024zlo}
G.~Aad \textit{et al.} [ATLAS],
%``Search for pair-produced vectorlike quarks coupling to light quarks in the lepton plus jets final state using 13~TeV pp collisions with the ATLAS detector,''
Phys. Rev. D \textbf{110} (2024) no.5, 052009
%doi10.1103/PhysRevD.110.052009
[arXiv:2405.19862 [hep-ex]];
%10 citations counted in INSPIRE as of 18 Jul 2025
%\cite{CMS:2024bni}
%\bibitem{CMS:2024bni}
A.~Hayrapetyan \textit{et al.} [CMS],
%``Review of searches for vector-like quarks, vector-like leptons, and heavy neutral leptons in proton{\textendash}proton collisions at {\ensuremath{\sqrt{}}}s=13 TeV at the CMS experiment,''
Phys. Rept. \textbf{1115} (2025), 570-677
%doi10.1016/j.physrep.2024.09.012
[arXiv:2405.17605 [hep-ex]].
%56 citations counted in INSPIRE as of 21 Jul 2025

%\cite{Baum:2011rm}
\bibitem{Baum:2011rm}
I.~Baum, V.~Lubicz, G.~Martinelli, L.~Orifici and S.~Simula,
%``Matrix elements of the electromagnetic operator between kaon and pion states,''
Phys. Rev. D \textbf{84} (2011), 074503
%doi:10.1103/PhysRevD.84.074503
[arXiv:1108.1021 [hep-lat]].
%67 citations counted in INSPIRE as of 05 Sep 2025



\end{thebibliography}
\end{document}